\documentclass[11pt,tightenlines,eqsecnum,floats,aps,amsmath,superscriptaddress,amssymb,nofootinbib,longbibliography]{revtex4-1}
 \usepackage{graphicx, wrapfig}
\usepackage{amssymb}
\usepackage[usenames, dvipsnames]{color}
\usepackage{diagbox}
\usepackage{mathrsfs}
\usepackage{subfigure}
\usepackage{float}
\usepackage{slashbox}
\usepackage{verbatim}
\usepackage{cancel}
\usepackage[normalem]{ulem}
\usepackage[hidelinks]{hyperref}
\hypersetup{
  colorlinks   = true, 
  urlcolor     = blue, 
  linkcolor    = blue, 
  citecolor   = blue 
}
\def\f{\frac}
\def\h{\hat}
\def\v{\vec}

\def\dd{\textrm{d}}
\def\d{\textrm{d}}

\def\pp{p_{\varphi}}
\def\dpp{\delta p_{\varphi}}
\def\dph{\delta\varphi}

\def\u{\mathfrak{A}}
\def\dd{\textrm{d}}

\def\l{\left}
\def\r{\right}

\usepackage{enumerate}

\newcommand{\ig}{\includegraphics}
\newcommand{\be}{\nopagebreak[3]\begin{equation}}
\newcommand{\ee}{\end{equation}}
\newcommand{\bfig}{\nopagebreak[3]\begin{figure}}
\newcommand{\efig}{\end{figure}}
\newcommand{\bea}{\nopagebreak[3]\begin{eqnarray}}
\newcommand{\eea}{\end{eqnarray}}

\newcommand{\bmult}{\nopagebreak[3]\begin{multline}}
\newcommand{\emult}{\end{multline}}

\newcommand{\ma}{\color{magenta}}

\begin{document}

\title{Large scale anomalies in the CMB and non-Gaussianity in bouncing cosmologies }

\author{Ivan Agullo}
\email{agullo@lsu.edu}
\author{Dimitrios Kranas}
\email{dkrana1@lsu.edu}
\affiliation{Department of Physics and Astronomy, Louisiana State University, Baton Rouge, LA 70803, U.S.A.
}
\author{V.~Sreenath}
\email{sreenath@nitk.edu.in}
\affiliation{Department of Physics, National Institute of Technology Karnataka, Surathkal, Mangalore 575025, India.}

\begin{abstract}

We propose  that several of the anomalies that have been observed at large angular scales in the CMB have a common origin in a cosmic bounce that took place before the inflationary era.  The bounce introduces a new physical scale in the problem, which breaks the almost scale invariance of inflation. As a result,  the state of scalar perturbations at the onset of inflation is no longer the Bunch-Davies vacuum, but it rather contains excitations and non-Gaussianity, which are larger for infrared modes. 
We argue that the combined effect of these excitations and the correlations  between CMB modes and longer wavelength  perturbations, can  account for the observed power suppression, for the dipolar asymmetry, and  it can also produce a preference for odd-parity correlations. The model can also alleviate the  tension in the  lensing amplitude $A_L$.  We adopt a phenomenological viewpoint by characterizing the model with a few  free parameters, rather than restricting to specific bouncing theories.
We identify the minimum set of ingredients needed for our ideas to hold, and point out examples of theories in the literature where these conditions are met. 

\end{abstract}

 \maketitle

\tableofcontents
\newpage
 
 \section{Introduction}

 The standard model of cosmology---the $\Lambda$CDM model complemented with almost scale invariant primordial curvature perturbations---provides an excellent fit to the CMB data at small and intermediate angular scales. 
 However,  observations have revealed some features at large angular scales that are in tension with it.
 The signals that have attracted most of the attention are: (i) Absence of two-point correlations, also known as power suppression; (ii) A hemispherical or dipolar asymmetry; and (iii) A bias for odd-parity correlations.

These anomalies have been identified in data from the satellites WMAP and Planck, and some are visible even in  data from COBE.  This makes it improbable that  they originate in instrumental or residual systematics.
Consequently, there is broad agreement that they are real features in the CMB. The discussion is rather whether the observed signals provide enough evidence of new physics, since their statistical significance is low, once cosmic variance is taken into account.  
More precisely, the significance of these features has been quantified by using  the so-called $p$-value, proposed by the Planck collaboration in  \cite{Ade:2013nlj}. This is the probability of obtaining a temperature map at least as extreme as the observed one, evaluated from a large number of Monte Carlo simulations of the primordial probability distribution predicted by  the $\Lambda$CDM model.  
The analysis in \cite{Ade:2015hxq,Akrami:2019bkn} associates similar $p$-values to each of the three anomalies separately, which are of the order of a fraction of per cent. It is worth emphasizing that this is the significance of producing only one of the aforementioned features from the $\Lambda$CDM model. Their collective significance can only be higher, but we are unaware of analyses of $p$-values associated with combinations of these signals.  We will also discuss a tension between the  $\Lambda$CDM model and data from Planck \cite{Akrami:2018vks}, originated from the preference of data for a value of the lensing amplitude $A_L$ larger than one.

Our goal is to construct a phenomenological model able to account for the observed features  and relate them. This model is an extension of the $\Lambda$CDM theory with an early phase of inflation, where the new ingredient is a cosmic bounce that replaces the big bang singularity (Ref. \cite{Agullo:2020fbw} contains a summary of our  ideas and results).  There exist several concrete scenarios that contain a cosmic bounce, based either on the introduction of exotic matter \cite{Khoury:2001wf,Lehners:2008vx,Cai:2007qw,Cai:2008qb,Cai:2007zv,Cai:2008qw,Lin:2010pf,Qiu:2011cy,Easson:2011zy,Brandenberger:2012zb, Raveendran:2017vfx}, a modified  theory of gravity  \cite{Shtanov:2002mb, Ijjas:2016vtq,Chamseddine:2016uef,Liu:2017puc}, or quantum gravity \cite{Ashtekar:2006rx,Ashtekar:2006wn,Ashtekar:2011ni,Agullo:2016tjh,Agullo:2013dla}. Some bouncing models face important challenges, particularly those relying on exotic matter \cite{Battefeld:2014uga}, but others are robust and compatible with existing observational constraints. We will not restrict to  any concrete theory, but rather remain as general as possible by identifying the minimum ingredients and assumptions needed to reproduce our results. This will make our findings   easily applicable to a wide range of theories.  See \cite{Ashtekar:2016pqn,Ashtekar:2016wpi,deBlas:2016puz,Ashtekar:2020gec,Xia:2014tda,Cai:2017pga,Qiu:2015nha,Cai:2014bea,Raveendran:2018yyh,Agullo:2015aba,Zhu:2017onp})
 for previous work on the relation between a cosmic bounce and individual anomalies.

In our model, the goal of the bounce is not to replace inflation---as most bouncing models do---but rather to complement it by providing a mechanism to replace the big bang singularity and to  bring the universe to a phase of slow-roll. The pre-inflationary evolution of perturbations is such that they reach the onset of inflation in a state  different from
the Bunch-Davies vacuum. Therefore, the modifications that we introduce to the $\Lambda$CDM model can be recast as a physically motivated choice for the initial conditions of inflation, and we argue that the specific initial conditions set up by a cosmic bounce can collectively account for the anomalous features in the CMB. More precisely, we argue that a cosmic bounce introduces a new physical scale in the problem that breaks the almost scale invariance of inflation and, furthermore, it introduces  non-Gaussian correlations that are confined  between   the most infra-red scales we can observe and super-horizon modes. 
Since these non-Gaussianities  involve at least one super-horizon perturbation, we cannot  observe them directly. 
But we can measure the indirect effects they induce in the CMB. 
In particular, we show that the non-Gaussianity increases the probability for certain features to appear  in individual realizations of the primordial probability distribution, like a power suppression and a dipolar asymmetry. Or in other words, primordial non-Gaussianity increases the  $p$-values of these features. 
Our model respects homogeneity and isotropy at the fundamental level, but it predicts that typical realizations look significantly more anisotropic than they would in a Gaussian universe.

This paper is organized as follows. Section \ref{sec:NGMod} provides a description of the phenomenon of non-Gaussian modulation of the power spectrum, introduced in \cite{Jeong:2012df,Lewis:2011au,Schmidt:2012ky}, and further explored in \cite{Agullo:2015aba,Adhikari:2015yya}. This is the  mechanism  on which our model rests. We will describe how the effects of the non-Gaussianity on the statistics of the CMB can be neatly encoded in the so-called Bipolar Spherical Harmonic coefficients, and show the way of computing them from the primordial non-Gaussianity. Section \ref{model} describes the details of our model,  the conditions under which an inflationary phase of the universe emerges after the bounce, a description of the power spectrum of scalar perturbations at the end of inflation, and of  non-Gaussianity. Section \ref{monopolar} is devoted to studying the monopolar modulation of the CMB.  
This section includes discussions on the power suppression of  the angular power spectrum, the lensing parameter, and the parity asymmetry. Sections \ref{dipolar} and \ref{quadrupolar} focus on anisotropic features, and describe the predictions of our model for a dipolar and quadrupolar modulation in the CMB, respectively.  Section \ref{validitypert} includes a quantitative analysis of the validity of the perturbative expansion in our model. Finally, section \ref{conclusions} contains a summary of our ideas, the assumptions in our model, and  a  discussion of our results.   We complement the content of this paper with three appendices that include some technical details omitted in the main body. 

We use units for which $\hbar=1$ and $c=1$.

 \section{Non-Gaussian modulation of the power spectrum\label{sec:NGMod}}
 
 The intuitive idea that perturbations of the gravitational potential with wavelengths  longer than the Hubble horizon at recombination do not affect the CMB, is only true in the absence of non-Gaussianity. If sub- and  super-horizon perturbations are correlated, the later can bias what we observe. The size of this effect depends on the concrete amplitude of the super-horizon modes in the vicinity of the observable patch of the universe. But these modes---as well as sub-horizon ones---are random variables with zero ``mean''. This  implies that, as we will show explicitly below, the indirect effects that super-horizon modes produce in the CMB in a non-Gaussian theory do not affect the statistical {\it mean} values of some observables, but in contrast they can affect their {\em variance}, increasing the probability of large deviations from the mean. 
 
For instance, if  non-Gaussian correlations between perturbations  in the CMB with wavenumber $\vec k_1$ and a super-horizon mode $\vec q$   depend strongly on the angle between them,\footnote{The isotropy of the probability distribution  forbids the correlations to depend on the direction of  $\vec k_1$ or $\vec q$ separately, but not on their relative direction.}  individual realizations of the primordial probability distribution would look more anisotropic than what they typically do for a Gaussian theory, and a  dipolar  modulation with large amplitude would be a feature expected in many realizations. (However, the direction of the dipole would be  random, in such a way that  the average over a large number of realizations reproduces isotropy).  Similarly, the modulation could make  the observed power spectrum to deviate from the mean value predicted by the theory  more than what is expected in the Gaussian case, so local observers would typically measure  a suppression or an enhancement of correlations. But if these observers  were aware of the existence of non-Gaussianty, they would know that such large deviations from the mean are common, and hence would not call them anomalies.  It is in this sense that primordial non-Gaussianity can solve the puzzle with the observed anomalies: not because the non-Gaussian theory predicts a mean value of the power suppression, dipolar modulation,  and parity asymmetry that agrees with the observed features, but rather because these features appear in concrete realizations with higher probability.

In the remaining of this section we analyze in quantitative terms the impact of non-Gaussianity on the probability distribution of the temperature  anisotropies of the CMB.

 \subsection{Primordial Power spectrum in presence of a spectator mode}\label{modPsp}
 We are interested in understanding the effects that correlations  between near- and super-horizon modes produce in a typical realization of the CMB (see \cite{Jeong:2012df,Lewis:2011au,Schmidt:2012ky,Agullo:2015aba,Adhikari:2015yya} for previous analyses).  To achieve this goal, we will first study the way the power spectrum of the Bardeen
 potential $\Phi_{\vec k}$ for a Fourier mode $\vec k$ that is observable in the CMB is modified by the  presence of a spectator mode $\Phi_{\vec q}$.  By spectator mode we mean a concrete realization of $\Phi_{\vec q}$.  
 
If $\Phi$  is a non-Gaussian random field, different Fourier modes are coupled, and this coupling makes it possible  
for the power spectrum of $\Phi_{\vec k}$ to be 
affected by the presence of $\Phi_{\vec q}$. To calculate this effect we will work at leading order in non-Gaussianity, and assume that the underlying probability distribution is statistically homogenous and isotropic. This type of non-Gassianity can be modeled by writing the random field $\Phi$ as a quadratic convolution of a Gaussian field $\phi$, which in position space reads \cite{Schmidt:2010gw}
\be \Phi(\vec x,t)=\phi(\vec x,t) +\frac{1}{2}\int d^3y d^3z\, F_{NL}(\vec y,\vec z)\, \phi(\vec x+\vec y,t)\, \phi(\vec x+\vec z,t)\, , \ee
and, in Fourier space 
 \be \label{Phik} \Phi_{\vec k}(t)=\phi_{\vec k}(t) +\frac{1}{2}\, \int \frac{d^3q}{(2\pi)^3}\,   f_{NL}(\vec q,\vec k-\vec q)\, \phi_{\vec q}(t)\, \phi_{\vec k-\vec q}(t)\, , \ee
where $f_{NL}$ is the Fourier transform of $F_{NL}$. The magnitude and ``shape'' of the non-Gaussianity is  encoded in the function $f_{NL}(\vec k_1,\vec k_2)$. Statistical homogeneity and isotropy constrain this function to depend only on the lengths of $\vec k_1$ and $\vec k_2$, and their relative orientation $\mu\equiv \hat k_1\cdot \hat k_2$. This is equivalent to saying that $f_{NL}(\vec k_1,\vec k_2)$ is a function of  the triangle  defined by $\vec k_1$ and $\vec k_2$, but it is independent of the orientation of this triangle: $f_{NL}(\vec k_1,\vec k_2)=f_{NL}(k_1,k_2,\mu)=f_{NL}(k_1,k_2,k_3)$, where $\vec k_3=-(\vec k_1+\vec k_2)$. Some useful properties  of $   f_{NL}(\vec k_1,\vec k_2)$ are
\bea f_{NL}(\vec k_1,\vec k_2)&=&f_{NL}(-\vec k_1,-\vec k_2)\nonumber \\
f_{NL}(\vec k_1,\vec k_2)&=&f_{NL}(\vec k_1,\vec k_3)=   f_{NL}(\vec k_2,\vec k_3)\nonumber \\
f_{NL}(\vec k_1,\vec k_3)&\in& \mathbb{R} \, .\eea
The relation between $   f_{NL}(\vec k_1,\vec k_2)$ and the bispectrum $B_{\Phi}(\vec k_1,\vec k_2)$ of  $\Phi_{\vec k}$, defined as $\langle  \Phi_{\vec k_1}\Phi_{\vec k_2} \Phi_{\vec k_3} \rangle=(2\pi)^3\, \delta(\vec k_1+\vec k_2+\vec k_3)\, B_{\Phi}(\vec k_1,\vec k_2)$, can be easily obtained using  (\ref{Phik}), and it reads 
\be \label{bisp} B_{\Phi}(\vec k_1,\vec k_2)= f_{NL}(\vec k_1,\vec k_2) \, [P_{\phi}(\vec k_1)P_{\phi}(\vec k_2)+1
\leftrightarrow 3+2
\leftrightarrow 3]\, , \ee
 where $P_{\phi}(\vec k_1$) is the  power spectrum of $\phi$, defined as \be \label{ps} \langle \phi(\vec k_1)\phi^{\star}(\vec k_2) \rangle =(2\pi)^3\, \delta(\vec k_1-\vec k_2)\, P_{\phi}(\vec k_1)\, .\ee  Hence, $f_{NL}(\vec k_1,\vec k_2)$ is the generalization of the parameter $f_{NL}$ first introduced in \cite{Komatsu:2001rj} to describe the so-called ``local'' non-Gaussianity. 
 We proceed now to compute  the two-point correlation function of the non-Gaussian field $\Phi_{\vec k}$ in  presence of a spectator mode $\Phi_{\vec q}$. Using (\ref{Phik}), we have
 \bea \label{modtp} \langle \Phi_{\vec k_1}\Phi^{\star}_{\vec k_2}\rangle |_{\Phi_{\vec q}}= \langle \phi_{\vec k_1}\phi^{\star}_{\vec k_2}\rangle &+& \frac{1}{2}\, \int \frac{d^3q'}{(2\pi)^3}\, f_{NL}(\vec{q'} ,\vec k_1-\vec{q'})\, \langle\phi_{\vec{q'}}\, \phi_{\vec k_1-\vec{q'}}\, \phi^{\star}_{\vec k_2}\rangle \nonumber \\ &+& \frac{1}{2}\, \int \frac{d^3q'}{(2\pi)^3}\, f_{NL}(\vec{q'},\vec k_2-\vec{q'})\, \langle\phi_{\vec k_1}\, \phi^{\star}_{\vec{q'}}\, \phi^{\star}_{\vec k_2-\vec{q'}}\rangle+\mathcal{O}(f_{NL}^2).
  \eea
Because we are not averaging over the spectator mode, we  must take $\phi_{\vec{q'}}$ out of the statistical average. We are then left with  two-point functions inside the integrals, and by using  (\ref{ps}) and the properties of $f_{NL}$, the previous expression reduces to 
  \bea \label{ngmod} \langle \Phi_{\vec k_1}\Phi^{\star}_{\vec k_2}\rangle |_{\Phi_{\vec q}}&=& (2\pi)^3\, \delta(\vec k_1-\vec k_2)\, P_{\phi}(\vec k_1)\nonumber \\&+&   f_{NL}(\vec k_1, - \vec k_2)\, \frac{1}{2}\, \big(P_{\phi}(\vec k_1)+P_{\phi}(\vec k_2)\big)\, \phi_{\vec q}+\cdots \, ,\eea
where $\vec q$ is constrained to be $\vec q= \vec k_1-\vec k_2$, otherwise the second term in the right hand side vanishes. (We have also used that, because $\phi(\vec x)$ is real,  $\phi^{\star}_{\vec q}=\phi_{-\vec q}$.) The first term in the right hand side is the familiar power spectrum, while the second  is the modification that the presence of the spectator mode $\phi_{\vec q}$ induces in the  two-point function of  $\Phi_{\vec k_1}$ and $\Phi_{\vec k_2}$.  It is proportional to both, the amplitude of the spectator mode  and the intensity of the correlations $f_{NL}(\vec k_1,-\vec k_2)$. We will refer to this  contribution as the non-Gaussian modulation of the two-point function. 

 Expression (\ref{ngmod}) contains two additional important messages. On the one hand, it tells us that only the mode $\vec q= \vec k_1-\vec k_2$ can affect the two-point function of the modes $\vec k_1$ and $\vec k_2$. This is expected from the underlying statistical homogeneity, which implies that only triples $(-\vec k_1,\vec k_2,\vec q)$ that close a triangle can be correlated among themselves. 
 On the other hand, the fact that $\vec q\neq 0$ implies that  this mode can affect $ \langle \Phi_{\vec k_1}\Phi^{\star}_{\vec k_2}\rangle$  only for $\vec k_1 \neq \vec k_2$. In other words, the non-Gaussian modulation  can only modify the  ``non-diagonal'' part of the two-point function. 
 But recall  that statistical homogeneity forces the two-point function $\langle \Phi_{\vec k_1}\Phi^{\star}_{\vec k_2}\rangle$ to be  proportional to $\delta(\vec k_1-\vec k_2)$, i.e.\ to be diagonal. Hence, (\ref{ngmod}) tells us that the presence of a spectator mode breaks homogeneity. This is not a surprise either, since it is obvious that the mere existence of a concrete realization of a  mode $\phi_{\vec q}$ with a finite wavelength  breaks homogeneity. The fact that our model respects statistical  homogeneity at the fundamental level  becomes manifest if we  take average also on $\phi_{\vec q}$, and take into account that   $\langle \phi_{\vec q}\rangle =0$. Therefore, the second term in (\ref{ngmod}) should  be understood as the modulation that  non-Gaussian correlations produce for a {\em concrete} realization of the mode $\phi_{\vec q}$.  In a typical realization, we expect $\phi_{\vec q}$ to take values of the order of the square root of its power spectrum. Then, substituting this value in (\ref{ngmod}) we obtain the expected size of the non-diagonal terms  in a typical realization.  In the standard inflationary paradigm $f_{NL}$ is small  at 
all scales, and therefore these terms can be neglected. But if the non-Gaussianities were large for some wavenumbers, they could leave an imprint in the CMB angular power spectrum, as we now discuss.

 \subsection{Non-Gaussian modulation of the  temperature covariance matrix}\label{covmaxmod}
 
 Next, we compute the temperature two-point function in angular 
 multipole space from (\ref{ngmod}). The  angular multipoles $a_{\ell m}$ are defined from the temperature $T(\hat n)$ of the CMB in the direction $\hat n$ as
\be \delta T(\hat n)\equiv \frac{ T(\hat n)-\bar T}{\bar T}=\sum_{\ell,m}a_{\ell m}\, Y_{\ell m}(\hat n)\, , \ee 
 where $\bar T$ is the mean temperature. The coefficients $a_{\ell m}$ are  related to the Bardeen potential $ \Phi_{\vec k}$  by
 \be \label{alm}  a_{\ell m}=4\pi \int \frac{d^3k}{(2\pi)^3}\, (-i)^{\ell}\, \Delta_{\ell}(k)\, Y^*_{\ell m}(\hat k)\, \Phi_{\vec k} \, , \ee 
where  $ \Delta_{\ell}(k)$ are the temperature radiation transfer functions. From this expression, the covariance matrix $ \langle a_{\ell m} a^{\star}_{\ell' m'}\rangle$ can be written in terms of the two-point function of the Bardeen potential as
\be \label{aaa} \langle a_{\ell m} a^{\star}_{\ell' m'}\rangle= (4\pi)^2\int \frac{d^3k_1}{(2\pi)^3}\int \frac{d^3k_2}{(2\pi)^3}\, (-i)^{\ell-\ell'}\, \Delta_{\ell}(k_1)\, \Delta_{\ell'}(k_2)\, Y^*_{\ell m}(\hat k_1)\, Y_{\ell' m'}(\hat k_2)\, \langle \Phi_{\vec k_1}\Phi^{\star}_{\vec k_2}\rangle |_{\Phi_{\vec q}}\, .\ee
In order to write this expression in a more useful form, we first recall that $f_{NL}(\vec k_1, - \vec k_2)$ can be written as a function of $k_1$, $q$ and  $\mu=\vec k_1\cdot \vec q$, and define the angular multipole moments of  $f_{NL}(k_1,q,\mu)$ as
\be \label{defGL} G_L(k_1,q)\equiv \int_{-1}^1d\mu\, f_{NL}(k_1,q,\mu)\, P_L(\mu)\, , \ee
where $ P_L(\mu)$ are Legendre polynomials. Equivalently, 
\be f_{NL}(k_1,q,\mu)=\sum_L G_L(k_1,q)\, \frac{2L+1}{2}\, P_L(\mu)=\,2\pi\, \sum_{LM} G_L(k_1,q)\,Y_{L M}(\hat k_1)Y^{\star}_{LM}(\hat q) \, . \ee
We also expand $\phi_{\vec q}$ in spherical harmonics  $\phi_{\vec q}=\sum_{L'M'} \phi_{L'M'}(q)\, Y_{L'M'}(\hat q)$. With this,  (\ref{aaa}) reduces to
 \bea \label{covmod} \langle a_{\ell m} a^*_{\ell' m'}\rangle= C_{\ell}\, \delta_{\ell\ell'}\delta_{m m'}&+&\frac{2}{(2\pi)^3}\,  \int d^3k_1\,{dq\, q^2} \, (-i)^{\ell-\ell'}\,\Delta_{\ell}(k_1)\,  \Delta_{\ell'}(k_2)\, \Big(P_{\phi}(k_1)+P_{\phi}(k_2)\Big) \nonumber\\  &\times &  \, \sum_{L,M} G_L(k_1,q)\, \phi_{LM}(q)\, Y^{\star}_{\ell m}(\hat k_1)Y_{\ell' m'}(\hat k_2)Y_{L M}(\hat k_1)\, , \eea
 where we have replaced the integral in $\vec k_2$ by an integral in $\vec q$, and  performed the integration in the direction of $\vec q$. 
The first term contains the familiar angular power spectrum $C_{\ell}\equiv \frac{2}{\pi}\int dk\, k^2\, \, \Delta^2_{\ell}(k)\, P_{\phi}(k)$, while 
the second term describes the non-Gaussian modulation caused by the presence of the spectator mode $\phi_{\vec q}$.

In order to evaluate this expression in the scenario of interest for this paper, we  now introduce the following approximation. We will assume that the non-Gaussianity only correlates modes with very different wavelengths, 
i.e., we will assume that $f_{NL}(k_1,q,\mu)$ is large only for $q <  k_1$.  As we will see, this is in fact the case for the model we consider in this paper (see Appendix \ref{approx} for details). Under this approximation, the second term in the previous equation is dominated by configurations for which $\vec k_2\approx \vec k_1$. With this, the covariance matrix can be written as
 \be \label{abiposh} \langle a_{\ell m} a^*_{\ell' m'}\rangle= C_{\ell}\, \delta_{\ell\ell'}\delta_{m m'} +(-1)^{m'}\,\sum_{LM} A^{LM}_{\ell\ell'}\,  C^{LM}_{\ell m \ell' -m'}\,, \ee
where  $C^{LM}_{\ell m \ell' m'}$ are Clebsch-Gordan coefficients, and 
\bea  \label{valueBipoSH} A^{LM}_{\ell\ell'} &=&\frac{4}{(2\pi)^3}\,  \int dk_1\, k_1^2\,{dq\, q^2} \, (-i)^{\ell-\ell'}\,\Delta_{\ell}(k_1)\,  \Delta_{\ell'}(k_1)\, P_{\phi}(k_1) \, \,  G_L(k_1,q)\, \phi_{LM}(q)\, \nonumber \\ &\times& C^{L0}_{\ell 0 \ell' 0}\,  \sqrt{\frac{(2\ell+1)(2\ell'+1)}{4\pi\, (2L+1)}}. \eea
Interestingly, looking at expression (\ref{abiposh}) we identify  $A^{LM}_{\ell\ell'}$ with the well-known Bipolar Spherical Harmonic (BipoSH) coefficients, commonly used to characterize the  correlation functions in the CMB.  Appendix \ref{BipoSH} contains a brief summary of the definition and physical interpretation of the BipoSH coefficients. They  organize the information about the modulation in an efficient manner: the labels $L$ and $M$ indicate the ``shape'' of the modulation, while $\ell,\ell'$ account for a possible variation of the modulation amplitude at different scales in the CMB.  For example, for $L=0$, $A^{LM}_{\ell\ell'}$ introduces a monopolar (spherically symmetric) modulation, i.e.\ a shift of the value of $C_{\ell}$ to $C_{\ell}+(-1)^{\ell}\, A_{\ell\ell}^{00}/{\sqrt{2\ell+1}}$. In contrast,  for $L>0$,  $A^{LM}_{\ell\ell'}$ produces an {\em anisotropic} modulation of the CMB, with angular distribution characterized by $L$ (i.e. dipolar for $L=1$, quadrupolar for $L=2$, etc.). A non-zero value of $A^{LM}_{\ell\ell'}$ for any $L>0$ implies that angular multipoles $\ell$ and $\ell'=\ell+L$ are correlated, something that is excluded for a statistically  isotropic  CMB. On the other hand, if these coefficients vary with $\ell$ and $\ell'$ 
 we  say we have a scale-dependent modulation. This will be in fact the case  for the model we consider in this paper.

Therefore, we have shown that the non-Gaussian modulation produces anisotropies in the CMB that can be quantified by the value of the BipoSH coefficients given by expression  (\ref{valueBipoSH}). Note that the ``shape'' of these anisotropies originates from  the  angular multipole components $G_L(k_1,q)$ of the function  $f_{NL}(k_1,q,\mu)$, i.e. it is  the particular dependence on $\mu$ (the cosine of the angle between $\vec k_1$ and $\vec q $) of the primordial  non-Gaussianity  that determines whether the modulation is mostly monopolar, dipolar, or something more complicated.  On the other hand, the  dependence of the BipoSH coefficients on $\ell$ and $\ell'$ originates from  the dependence of  $G_L(k_1,q)$ on $k_1$ and $q$.

Before we conclude this section, note that because the BipoSH coefficients  depend  on the random variable $\phi_{\vec q}$, we cannot predict the concrete value of $A^{L M}_{\ell,m,\ell',m'}$  on the sky. The best we can do is to compute their statistical root mean square $\sqrt{\langle |A^{L M}_{\ell,m,\ell',m'}|^2\rangle}$, which 
measures the expected value of this amplitude in a typical CMB. Obviously, this variance will depend on the mean square value of the spectator mode $\phi_{\vec q}$, i.e.\ its power spectrum $P_{\phi}$.  Using
 \be \langle \phi_{LM}(q) \phi_{L'M'}^*(q') \rangle =(2\pi)^3\, \delta_{L L'} \delta_{M M'}\delta(q-q')\,\frac{1}{q^2}\,  P_{\phi}(q)\, , \ee
  we arrive at%
\be \label{varalpha}    \sqrt{\langle|A^{L M}_{\ell \ell' }|^2\rangle}= \left[\frac{1}{2\pi} \, \int dq\, q^2 \, P_{\phi}(q) \, |\mathcal{C}_{\ell \ell'}^L(q)|^2\, \right]^{1/2} \times C^{L0}_{\ell 0 \ell' 0}\,  \sqrt{\frac{(2\ell+1)(2\ell'+1)}{4\pi\, (2L+1)}}\, , \ee
where we have defined
 \be \label{Ctilde}  \mathcal{C}_{\ell \ell'}^{L}(q)\equiv \frac{2}{\pi}\int dk_1 \, k_1^2 \, (i)^{\ell-\ell'} \, \Delta_{\ell}(k_1)\Delta_{\ell'}(k_1) \, P_{\phi}(k_1)\, G_L(k_1,q) \, .\ee
Note that the right hand side of (\ref{varalpha})  does not depend on $M$, as expected from  the underlying statistical isotropy. We will use  expression (\ref{varalpha}) to make predictions for  the CMB, after introducing the details of our model in the next section.

\section{The model: a cosmic bounce followed by a phase of  inflation}\label{model}

The scenario  considered in this work is a simple extension of the standard $\Lambda$CDM model with inflation, on which we add a cosmic bounce 
before the inflationary era. The bounce  modifies the initial state of scalar and tensor perturbations at the beginning of inflation, by a state that is not the commonly assumed Bunch-Davies vacuum, but it rather contains excitations and non-Guassianity relative to the vacuum. These are the ``memories'' that perturbations retain from the pre-inflationary history of the universe.

 As we already mentioned in the introduction, one peculiarity of the model proposed here is that the bounce is not meant to replace inflation, as it is more common in the literature, but rather to complement it---the bounce  substitutes the big bang singularity and  drives the universe to an inflationary phase.  Having a phase of inflation after the bounce changes significantly the set of wavelengths that become observable in the CMB and, in particular, makes it possible that the details of the bounce become observable, as we argue below.  In subsection \ref{sec:3.1} we review the way an inflationary phase appears after the bounce, provided a scalar field with potential $V(\varphi)$ dominates the matter sector  at some instant after the bounce. We will then discuss the concrete imprints that a pre-inflationary bounce  leaves on the power spectrum of primordial perturbations in subsection  \ref{powspec}, and finally we will describe the predictions for non-Gaussianity. We will keep our description as generic as possible, and avoid making assumptions about the mechanism that causes the bounce.

More concretely, we will use the following parameterization of the scale factor around  the time of the bounce

\be \label{an} a(t)=a_B\, \left(1+ b\,  t^2\right)^n\, ,\ee
with $b$ a constant. It is straightforward to check that $b$ determines the value of the Ricci curvature scalar $R(t)$  at the bounce, namely $R_B=12\, n\, b$. Therefore, a bounce in this family is characterized by two parameters, $R_B$ and $n$, which codify the new physics causing the bounce. We will make physical  predictions for different values of these parameters. 
Although one could think of other shapes of a bouncing scale factor, expression (\ref{an}) is generic enough for our purposes. 
In order for inflation to occur, the new physics producing the bounce must  lose relative relevance as the universe expands, in such a way that the potential energy of the scalar field eventually dominates. This is the scenario on which our model is built.

As an example, this paradigm  can be obtained from  a modified Friedmann equation of the form
\be \label{modF} H^2=\frac{\kappa}{3} \, \rho \, \left(1-\frac{\rho}{\rho_{\rm max}}\right)\, ,\ee
where $\kappa=8\pi G$, and $\rho_{\rm max}$ is a constant. The second term in parentheses is absent in general relativity, and it makes the universe bounce when the matter energy density $\rho$ reaches its maximum value $\rho_{\rm max}$.  If the matter sector is made of a scalar field, the solutions to  the above equation---complemented with the equation of motion of the  scalar field---are of the form (\ref{an}), with $n\approx 1/6$ around the time of the bounce.\footnote{This concrete value of $n$ appears when the potential energy of the  scalar field at the bounce is small relative to its kinetic energy, otherwise $n$ is slightly different.} After the bounce the new term in the Friedmann equation quickly loses relevance,  and a phase of inflation occurs. 
Equation (\ref{modF}) has been derived in the context of loop quantum cosmology \cite{Ashtekar:2011ni,Agullo:2016tjh}, where the deviations from classical general relativity originate from quantum effects, and it also emerges in  higher-derivative scalar-tensor theories \cite{Chamseddine:2016uef,Liu:2017puc}.

\subsection{Inflation in a bouncing scenario} \label{sec:3.1}

The conditions under which an inflationary phase appears in a bouncing scenario have been studied in  detail in the literature \cite{Ashtekar:2009mm,Ashtekar:2011rm,Bolliet:2017czc,Bonga:2015kaa,Bonga:2015xna,Zhu:2017jew}, and we  provide here only a brief summary of these results. Although these analyses have been derived for a specific bouncing model, the conclusions are generic. 

For inflation to take place, one needs a scalar field $\varphi$ with an inflationary potential $V(\varphi)$ in the matter sector  that dominates the dynamics far away from the bounce. We assume the spacetime geometry to be spatially flat, homogeneous and isotropic, although the analysis can be generalized to include anisotropies \cite{Gupt:2013swa,Agullo:2020wur,1784621} and other features.  Under these circumstances, the answer is that almost every dynamical trajectory finds an inflationary phase (see  \cite{Ashtekar:2011rm} for a mathematically precise statement). In other words, slow-roll inflation is a dynamical attractor. This is not surprising, and it is well-known to be the case in general relativity. The statement here is that the presence of a bounce does not change the standard  inflationary attractor mechanism.   

One can get some intuition about the origin of this attractor mechanism by considering the following qualitative argument. First of all, we must keep in mind that the energy density of the scalar field at the bounce  $\rho(t_B)= \dot \varphi^2(t_B)/2 + V(\varphi(t_B))$ is finite. If the value of the field at the bounce $\varphi(t_B)$ is small (the field is near the bottom of the potential),  the velocity $\dot \varphi(t_B)$ must be large (kinetic dominated bounce). The field will then quickly climb the potential up, slowing down until eventually $\dot \varphi$ will vanish and the potential energy dominates over the kinetic term. The field will  then start rolling down, and slow-roll inflation will begin. In contrast, if $\varphi(t_B)$ is large at the bounce, then we are even closer to the conditions required for inflation, and slow-roll will start earlier. Namely, if the field is initially moving up the potential, it will soon stop, and will start rolling down while the universe inflates. If the field is moving down, inflation will start right away. For intermediate values of $\varphi(t_B)$ the situation is also in between, and inflation starts not too long after the bounce. An inflationary  phase is difficult to avoid. As mentioned before, this simple qualitative statements have been confirmed in detailed numerical simulations   \cite{Ashtekar:2009mm,Ashtekar:2011rm,Bolliet:2017czc,Bonga:2015kaa,Bonga:2015xna,Zhu:2017jew}. Details, such as the duration of the pre-inflationary and inflationary phases, depend on the form of the potential and the bouncing scale factor, but the global aspects of the dynamics are quite universal. 

One interesting aspect is that the duration of the inflationary phase is longer the larger the  value of the field  $\varphi(t_B)$ at the bounce  is. This is to say, if the potential energy of the scalar field is large at the bounce, the inflationary phase ends up being significantly  longer. This is because inflation would begin when the field is higher up in the potential,  since the energy lost by Hubble friction is smaller. Such situations will not be of interest for us since, for them, inflation is so long that all effects imprinted in perturbations by the bounce will be red-shifted out of our present Hubble radius. Therefore, we will focus from now on bounces for which the potential energy of the scalar field is small relative to its kinetic energy.\footnote{Notice that this is a restriction only on the relative size of the two contributions to the energy of the scalar field, and not on the dominant contribution to the gravitational field equations. In other words, close to the bounce there may---and should---exist other contributions that dominate and cause the bounce, either in the matter sector or of pure gravitational origin. Hence, the form of the scale factor $a(t)$ around the time of the bounce is not  tied to the properties of the scalar field. The potential energy of the scalar field is, on the other hand, assumed to take over and dominate at later times, as explained above.}

\subsection{The power spectrum}\label{powspec}

 A cosmic bounce introduces a new physical scale in the problem. This scale can be defined from the value of the Ricci curvature scalar at the bounce, $R_B$.  If a bounce took place before the inflationary era,  we expect this  new scale to be imprinted in the primordial  power spectrum. 

We will focus on scalar perturbations defined in terms of the comoving curvature perturbations in Fourier space $\mathcal{R}_k$, where $k\equiv |\vec k|$, which is related to the Bardeen potential by $\mathcal{R}_k=-\frac{5}{3} \Phi_{k}$. It is  convenient to define the re-scaled variable  $v_k(\eta)\equiv z(\eta)\, \mathcal{R}_k(\eta)$, where $z=a \, \frac{\dot \varphi}{H}$, whose dynamics in conformal time $\eta$ is dictated by 
\be \label{chieq} v_k''(\eta)+ \left(k^2-f(\eta)\right )v_k(\eta)=0\, , \ee
with $f(\eta)\equiv a^2(\eta)\, (\f{R}{6}-\u)$, where $R(\eta)$ is the Ricci  curvature scalar  and  $\u(\eta)$ is a time dependent effective potential given by  $\u=a^2 [V(\varphi)\,r - 2V_\varphi(\varphi)\sqrt{r} + V_{\varphi\varphi}(\varphi)]$, where $r=3\dot \varphi^2\,\frac{8\pi G}{\rho}$,  and $V_{\varphi}(\varphi)\equiv \dd V(\varphi)/\dd\varphi$ (see e.g.\ \cite{Agullo:2013ai,Agullo:2017eyh} for a derivation of this potential in general relativity).   Equation (\ref{chieq}) has the same mathematical form as in general relativity.  Therefore, we are assuming that the new physics that causes the bounce does not modify substantially the form of this equation, except for the fact that the evolution of the functions $a(\eta)$ and  $\varphi(\eta)$ is different.

 It turns out that  for the  bounces we are interested in---for which $\dot \varphi^2\ll V(\varphi)$ at the bounce, in order to avoid an excessively long inflationary phase---the effective potential $\u$ becomes much smaller than $\f{R}{6}$ at the bounce and during the entire pre-inflationary phase, and consequently  $f(\eta)\approx a^2\f{R}{6}$ at any time around the bounce and before inflation. This happens because, on the one hand, the parameter $r$ in $\u$, which is proportional to the ratio between the kinetic and total energy density of $\varphi$, is close to one.
And, on the other hand, because $V(\varphi)$ and its derivatives are much smaller than $R$. We have checked this explicitly for a quadratic potential and for the Starobinski potential \cite{Barrow:1988xi,Barrow:1988xh,PhysRevD.39.3159,Starobinsky:2001xq}, with the coefficient in these potentials obtained from the Planck normalization \cite{Aghanim:2019ame}. %
This fact simplifies the analysis significantly, since the pre-inflationary evolution of cosmological perturbations becomes independent of the shape of the potential $V(\varphi)$---the potential acquires relevance only during inflation. This universality will allow us to make generic statements without having to commit to a specific form of  $V(\varphi)$.

In the scenario  we are considering, the evolution of perturbations begins in the far past before the bounce. There, all the wavenumbers of interest for the CMB are in the adiabatic regime, i.e.\  $k/a\gg R$. Then, equation (\ref{chieq}) tells us that they oscillate as positive frequency modes in conformal time, $v_k(\eta)\approx 1/\sqrt{2 k}\, e^{-ik\eta}$. In other words, perturbations begin the evolution in an adiabatic vacuum well before the bounce.  We are interested in computing the primordial power spectrum\footnote{Note that we denote with a calligraphic $\mathcal{P}_{\mathcal{R}}$ the dimensionless power spectrum; it is related to  ${P}_{\mathcal{R}}(k)$, defined in the previous section, by 
$\mathcal{P}_{\mathcal{R}}(k)=\frac{k^3}{2\pi^2}\,{P}_{\mathcal{R}}(k)$.} $\mathcal{P}_{\mathcal{R}}(k)=\frac{k^3}{2\pi^2}\, \left(\frac{H}{{ a} \dot \varphi}|v_k|\right)^2$ evaluated at the end of inflation. This quantity encodes the information of the evolution of perturbations across both, the bounce and the inflationary era. In order to gain intuition, we would like to isolate from $\mathcal{P}_{\mathcal{R}}(k)$  the contribution coming from the pre-inflationary evolution. This information is contained in the ratio $\mathcal{P}_{\mathcal{R}}(k)/\mathcal{P}^{\rm BD}_{\mathcal{R}}(k)$, where $\mathcal{P}^{\rm BD}_{\mathcal{R}}(k)$ is the almost scale-invariant power spectrum produced by inflation alone, i.e.\ obtained by using the Bunch-Davies vacuum as initial state for perturbations at the onset of inflation. Furthermore, this ratio can be written as 
\be \mathcal{P}_{\mathcal{R}}(k)/P^{\rm BD}_{\mathcal{R}}(k)=|\alpha_k+\beta_k|^2\, , \ee
where $\alpha_k$ and $\beta_k$ are the Bogoliubov coefficients that relate the modes  $v_k(\eta)$ that begin as positive frequency exponentials in the past before the bounce, and the modes $v^{BD}_k(\eta)$ that define the Bunch-Davies vacuum during inflation\footnote{During inflation, these modes are approximated at leading order in the slow-roll parameters $\epsilon=-\dot H/H^2$ and $\delta=2\ddot H/(2\dot H H)$ by $v^{BD}_k(\eta)=  e^{i(\pi/4+\pi/2 \mu)} \sqrt{|\eta|\pi/4z^2}\, H^{(1)}_{\mu}(k|\eta|)$, where $H^{(1)}_{\mu}$ is a Hankel function and $\mu=3/2+2\epsilon+\delta$.} 
\be v_k(\eta) =\alpha_k\,  v^{BD}_k(\eta)+\beta_k\, v^{BD}_k(\eta)^{\star}\, . \ee
The coefficients $\alpha_k$ and $\beta_k$ encode the information of the pre-inflationary evolution of scalar perturbations.   
Hence, to understand the contribution of the bounce to the primordial power spectrum, we need to compute $|\alpha_k+\beta_k|^2$ for bounces described by (\ref{an}), for different values of $R_B$ and $n$.

To gain some intuition, we begin by considering a bounce with $R_B=10^{-2}$ (in Planck units) and $n=1/6$, and plot $|\alpha_k+\beta_k|^2$  in Figure \ref{LQCAB}. This  figure is obtained by solving numerically equation (\ref{chieq}) from some time before the bounce until just before inflation starts (see figure caption for more details), and it  contains the main features we want to analyze in the rest of this section.
\renewcommand{\thempfootnote}{\arabic{mpfootnote}}
\bfig
 \ig[width=.7\textwidth]{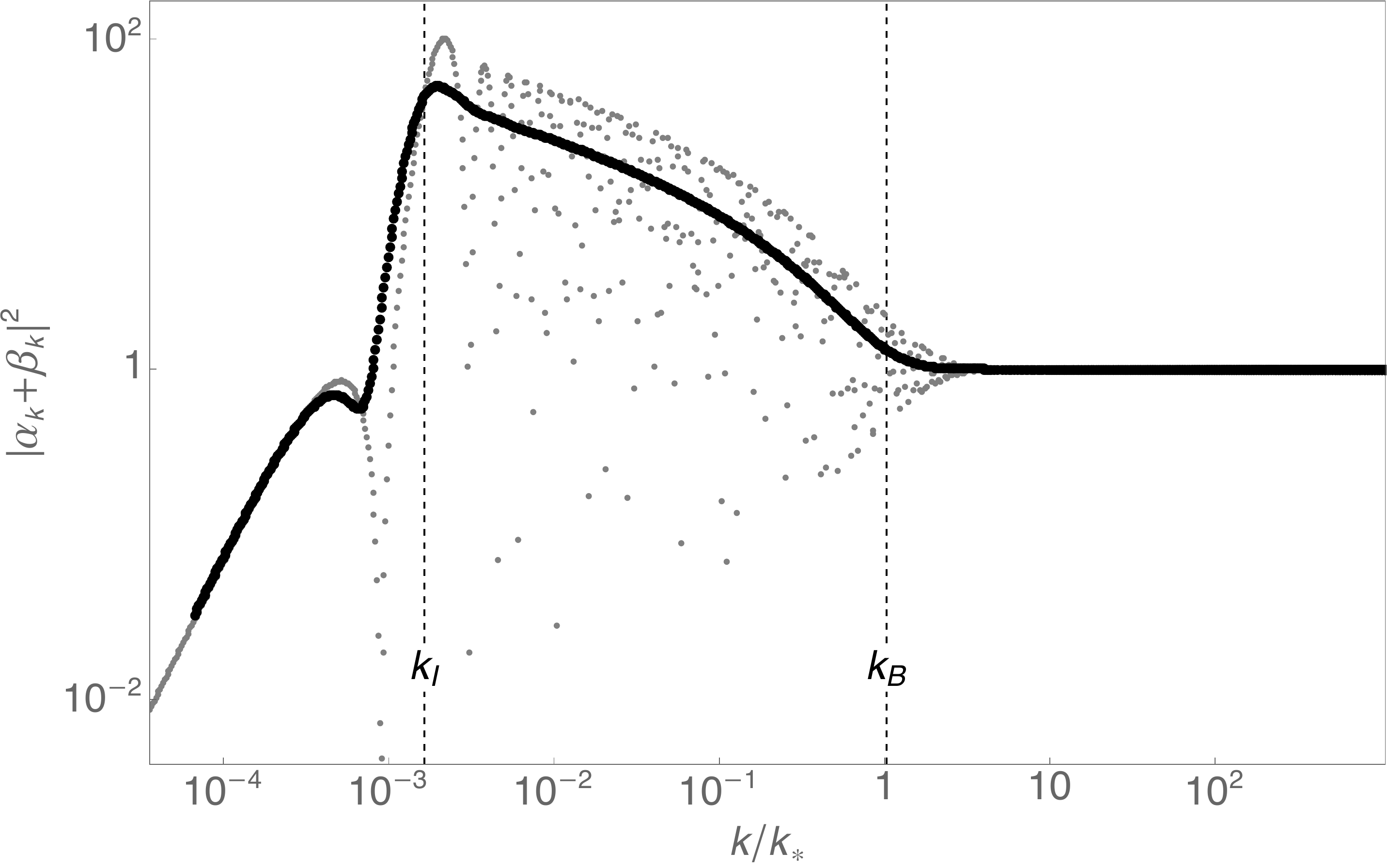}
\caption[]{$|\alpha_k+\beta_k|^2$ versus $k$ for a bounce with $R_B=10^{-2}$ in Planck units, and $n=1/6$.  Gray points indicate the value of $|\alpha_k+\beta_k|^2$ for a set of individual $k$'s, while the black solid line shows the average of the gray points, obtained by binning them in a sufficiently small window. This plot is obtained by solving numerically the differential equation  (\ref{chieq}) with the scale factor 
(\ref{an}) and initial conditions corresponding to positive frequency exponentials.\footnotemark[6] \, We have started the evolution $30000$ Planck times before the bounce, but choosing an earlier time  does not modify the results. We have evolved the modes $v_k(\eta)$ until a time a bit before inflation starts, around   $30000$ Planck times after the bounce.  As explained above, the inflaton potential $V(\varphi)$ can be neglected in this calculation, for the family of bounces we are interested in.}\label{LQCAB}
\efig
First of all,  $|\alpha_k+\beta_k|^2$ oscillates with $k$ around a  mean value, which is indicated by the black line in  Figure \ref{LQCAB}. This is not surprising, since oscillations  appear quite generically when new physics is added to the simplest inflationary scenario.  
On the other hand, we see three  distinct regions in  Figure \ref{LQCAB}, that we have separated by  two special values of $k$, namely $k_B$ and $k_I$.  The first one, $k_B$, is defined as $k_B\equiv a_B\, \sqrt{R_B/6}$, where $R_B$ is the Ricci scalar at the bounce. This wavenumber informs us about the value of the space-time curvature at the bounce, and hence $k_B$ defines the ``scale''  at which the bounce takes place. 
The other scale, $k_I$, refers to the inflationary era, and it is defined as $k_I\equiv 2\pi\,  a(\eta_I)\, \sqrt{R_I/6}$, where $R_I$ is the value of the Ricci scalar evaluated at the beginning of inflation $\eta_I$. 
The \footnotetext{This corresponds to choosing a vacuum of zeroth adiabatic order. A higher adiabatic order vacuum can be defined following  \cite{Parker:2009uva,Agullo:2013ai,Agullo:2014ica}. However, it would produce differences in physical observables negligibly small.  For the sake of simplicity, we work here with a vacuum of zeroth adiabatic order.} three distinct regimes in Figure \ref{LQCAB} are:

\begin{enumerate}
\item  $k>k_B$: we see that  $|\alpha_k+\beta_k|^2\approx 1$  approaches the Bunch-Davies power spectrum $\mathcal{P}^{\rm BD}_{\mathcal{R}}(k)$.
\item  $k_I<k<k_B$: the mean value of $|\alpha_k+\beta_k|^2$  behaves as $\propto k^{-0.7}$. This corresponds to a red-tilted spectrum (more power on infra-red scales). 
\item  $k<k_I$: the mean value of $|\alpha_k+\beta_k|^2$  scales as $\sim k^2$.
\end{enumerate}

We now discuss the physical origin of each of these regions, and its dependence on the peculiarities of the bounce.

\begin{enumerate}
\item  Region $k>k_B$. These are modes that at the bounce are more ultraviolet than the  scale $k_B$. This implies that the term $k^2/a^2$ in the wave equation (\ref{chieq}) dominates over $-\frac{R}{6}$, and hence the solutions for these modes are simply positive frequency exponentials $\sim e^{-ik\eta}$. This remains true during the bounce, the pre-inflationary phase, until the onset of slow-roll. At that time these modes are indistinguishable from the modes that define the Bunch-Davies vacuum (recall that the limit of the Bunch-Davies modes  when $k|\eta|\gg 1$ is a positive frequency exponential). Hence, for these modes, the Bogoliubov coefficients  are just $\alpha_k\approx 1$ and $\beta_k \approx 0$, and  $|\alpha_k+\beta_k|^2\approx 1$. In simple words, modes with wavenumbers $k>k_B$ are so ultraviolet that they are not affected by the bounce or the pre-inflationary evolution, and reach the onset of inflation in the Bunch-Davies vacuum. 
Therefore, this region of Figure \ref{LQCAB}  does not depend on the details of the bounce, and it knows only about inflationary physics. 

\item Region $k_I>k>k_B$. The  enhancement observed in Figure \ref{LQCAB} for this range of wavenumbers is entirely produced by the evolution of modes across the bounce. During  a time interval around the bounce, the effective frequency in equation (\ref{chieq}), namely $ \sqrt{k^2-f(\eta)}$, becomes imaginary, making the amplitudes of the modes to grow during that period, resulting in  $|\alpha_k+\beta_k|^2$  approximately proportional to $k^q$ with ${q=-0.7}$ for the concrete bounce we are using.  
An interesting question is whether this value of $q$ is universal, or  it depends on the details of the bounce. To  investigate this question, we first compute numerically  the value of  $|\alpha_k+\beta_k|^2$  (averaged on small bins $\Delta k$ to eliminate the unimportant oscillations) for  $k>k_I$  for different types of bounces, as parameterized by (\ref{an}) with $n$ ranging from  $n=1/4$ to $n=1/7$. As discussed below, this is the most interesting range. The results appear  in Figs.\ \ref{alpha+beta} for $R_B=10^{-2}$ in Planck units. First of all, we see that for all values of $n$ the 
 averaged value of  $|\alpha_k+\beta_k|^2$ is enhanced (i.e. larger than one), and this means that $\mathcal{P}_{\mathcal{R}}(k)$ is larger than $\mathcal{P}^{\rm BD}_{\mathcal{R}}(k)$. Hence, the bounce  {\em enhances} the  power spectrum. We find that  $|\alpha_k+\beta_k|^2$ behaves approximately as a power law also for other values of $n$, and we observe that the tilt $q$ is more negative for larger values of $n$. 
 Table \ref{n&q} shows the approximate values of $q$ for different values of $n$. On the other hand, the value of $q$  is quite insensitive to $R_B$; we observe that $q$ changes only in a few per cent when $R_B$ changes in three orders of magnitude.

 \bfig[htp]
 \centering
\includegraphics[width=.7\textwidth]{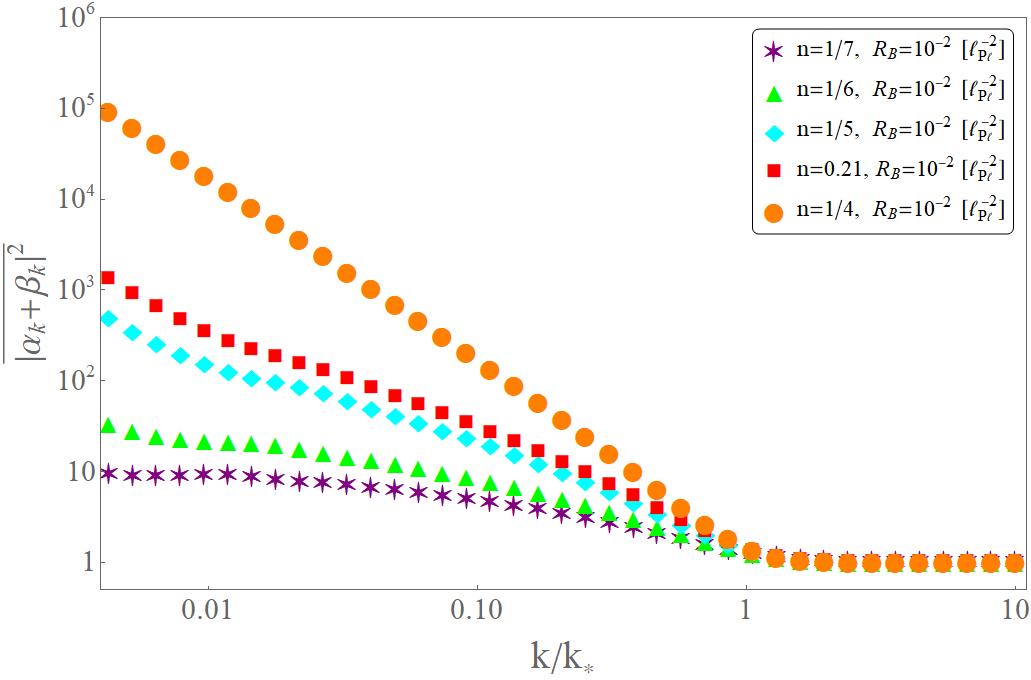} 
\caption{$|\alpha_k+\beta_k|^2$ versus $k$ for different values   $n$ and  $R_B=10^{-2}$ in Planck units.   
 While the slope of $|\alpha_k+\beta_k|^2$ varies with $n$, it is quite insensitive to $R_B$.}\label{alpha+beta}
\efig

\begin{table}
\caption{Values of the tilt $q$ for different values of $n$. The value of $q$ is quite insensitive to the choice of $R_B$.  These numbers are obtained by adjusting the numerical result to a power law. } 
\centering 
\begin{tabular}{|c|c|} 
\hline 
$n$ & $q$\\ [0.5ex] 
\hline \hline 
1/4 & -2\\ 
0.21&-1.24\\
1/5 & -1.1\\
1/6 & -0.7\\
1/7 & -0.5\\
 [1ex] 
\hline 
\end{tabular}
\label{n&q} 
\end{table}

The physical origin of the relation between  $q$ and $n$ can be understood as follows. In Figure \ref{fvseta} we plot the function $f(\eta)\approx a^2(\eta)\frac{R}{6}$ that appears in the effective frequency in equation (\ref{chieq}) for different values of $n$. We see that $f(\eta)$ has the approximate form of a Gaussian, whose width {\em increases} with $n$. For large $n$ the Gaussian becomes wider, and it becomes exactly  flat for $n=1/2$. Now, recall that when the effective frequency  $ \sqrt{k^2-f(\eta)}$  becomes imaginary, the amplitude of the mode $v_k$ grows exponentially, and this happens around the time of the bounce when $f(\eta)$ is maximum. The wider the shape of $f(\eta)$ is, the longer the effective frequency remains imaginary around the bounce, and consequently the larger the enhancement is. Hence, we expect a stronger enhancement for larger $n$ or, in other words, the tilt $q$ is expected to take more negative values the larger $n$ is. This trend is confirmed in our numerical simulations (see Table \ref{n&q}).

\bfig
 \ig[width=.7\textwidth]{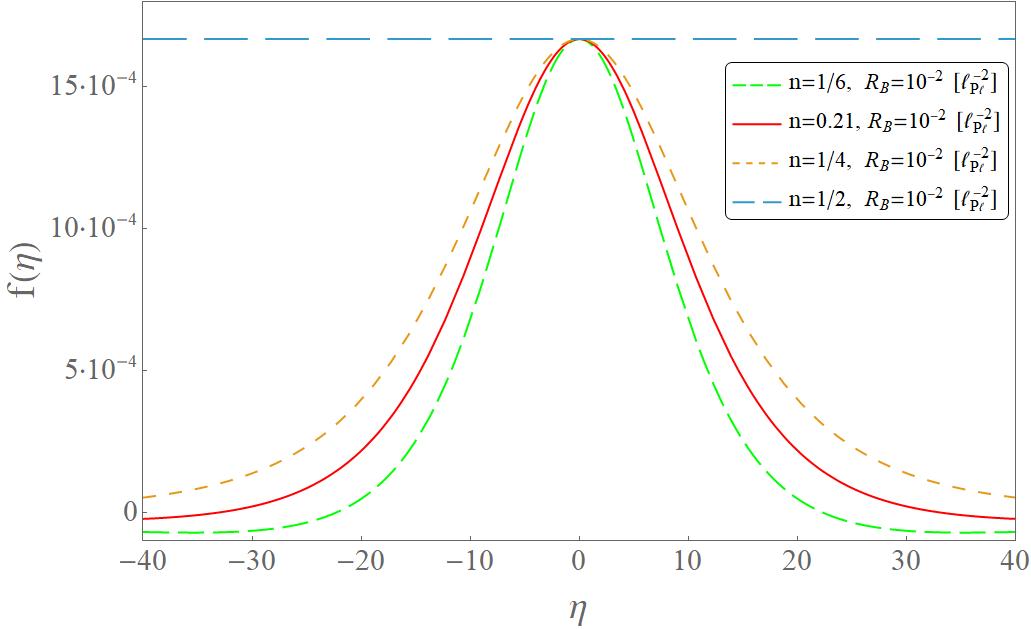}
\caption{Plot of the function $f(\eta)\approx a^2(\eta)\frac{R}{6}$, versus conformal time around the bounce, for $R_B=0.01$ and different values of $n$. The plot shows that  $f(\eta)$ has a Gaussian-like shape, whose width is bigger the larger $n$ is. This is the reason why $q$ is more negative for larger values of $n$.}\label{fvseta}
\efig

For values of $n$ close to $n=1/2$ the enhancement in $|\alpha_k+\beta_k|^2$ is so large that the power spectrum $\mathcal{P}_{\mathcal{R}}(k)$ becomes of order one and perturbation theory breaks down. Hence, from now on we will restrict to values of $n$ for which there is some enhancement, but not too large, so that perturbation theory remains under control.  This happens for $n\in [1/7,1/4]$, although we will argue below that $n=1/4$ is also problematic in our model. The particular value $n=1/6$ arises in  higher derivatives scalar-tensor theories \cite{Chamseddine:2016uef,Liu:2017puc} and loop quantum cosmology \cite{Ashtekar:2011ni,Agullo:2016tjh}.

\item Region $k<k_I$. The power spectrum $\mathcal{P}_{\mathcal{R}}(k)$ becomes blue tilted for  very infrared modes $k<k_I$. More concretely, we find   $\mathcal{P}_{\mathcal{R}}(k)\propto k^2$. Furthermore, this concrete power law appears to be universal, in the sense that it does not depend on the details of the bounce. This can be  understood from the following qualitative argument. These are modes that during the entire evolution, including both the  bounce and inflationary era, are out of the curvature radius. 
For these wavenumbers  the comoving curvature perturbations $\mathcal{R}_{k}$ remain frozen (constant in time) during the entire post-bounce evolution, and this makes the $k$-dependence of $\mathcal{R}_{k}$ to be given simply by  $\mathcal{R}_{k} \propto \frac{1}{\sqrt{k}}$, and $\dot{\mathcal{R}}_{k} \approx 0$. This remains true until the end of inflation. On the other hand, the Bunch-Davies modes during inflation are almost scale invariant on super-Hubble scales: $ \mathcal{R}^{\rm BD}_{k} \propto  k^{-3/2+\mathcal{O}(\epsilon)}$, $\dot{\mathcal{R}}^{\rm BD}_{k} \approx 0$, where $\epsilon$ is the first slow-roll parameter. Hence, the Bogoliubov coefficients that relate both sets of modes 
must satisfy $|\alpha_k+\beta_k|\propto k$ for $k<k_I$. 

\end{enumerate}%

The ratio between the two relevant scales in the problem, $k_I/k_B$, is given by $k_I/k_B=2\pi \frac{a(\eta_I)}{a_B}\sqrt{\frac{R_I}{R_B}}$. Furthermore,  the  ratio $a(\eta_I)/a_B$ can be written  in terms of $R_B/R_I$, by the following argument.  For a large portion of time between the bounce and the onset of inflation, the universe is dominated by the kinetic energy of the scalar field, and the  equation  of state during that phase is  $p=w\, \rho$ with $w\approx 1$. This in turns implies that $\rho(t)\propto a(t)^{1/6}$ during that period. From this, we obtain the approximate expression $a(\eta_I)/a_B\approx (\rho_B/\rho(\eta_I))^{1/6}=(R_B/R_I)^{1/6}$, which implies  $k_I/k_B\approx 2\pi (R_I/R_B)^{1/3}$---we have used that $R=\kappa (1-3 w)\rho$. The approximation is due to the fact that  it is not strictly true that the kinetic energy of the scalar field dominates during the entire pre-inflationary phase, particularly close to the bounce and right before inflation.  
However, the kinetic dominated phase is generally longer, and one can get a reasonable  estimate for $a_B/a(\eta_I)$ out of it. 
On the other hand, although there are no direct measurements of the scale of inflation $R_I$, the Planck satellite in combination with Bicep2/Keck provides the upper bound $R_I(t_{k*})<3.1\times 10^{-10}$ in Planck units, where  $t_{k*}$ is the time around which observable wavenumbers exit the Hubble radius during inflation \cite{Akrami:2018odb}. At the beginning of inflation, one  expects $R_I$ to be slightly larger, so we will use $R_I =5\times  10^{-10}$  in Planck units.

Another important scale in Figure \ref{LQCAB} is the location of the {\em observable window} in the $k$-axis. This window is made of the set of wavenumbers that we can directly observe in the CMB, and it is  approximately  given by $k\in [k_*/10,1000k_*]$, where $k_*$ is a reference scale whose physical value today is $k_*/a_{\rm today}=0.002\, {\rm Mpc}^{-1}$. In Figure \ref{LQCAB} the modifications that the bounce imprints in perturbations  appear for  $k  \lesssim k_B$. It is not difficult to understand that the value of the ratio $k_B/k_*$ depends crucially on the amount of expansion accumulated after the bounce. If that expansion is very large, then the physical scale $k_B$ will experience a large red-shift during the cosmic evolution, and today it will be very infrared compared to $k_*$ (i.e. $k_B/k_*\ll1$), and consequently unobservable. In terms of wavelengths, $\lambda_B=2\pi/k_B$ would be a super-horizon mode today. On the contrary, if the   expansion accumulated after the bounce is not too large, we would have  $k_B/k_*\approx 1$, and part of the enhancement produced by the bounce would be visible in  the CMB. On the other hand, and as we described before, the amount of expansion  after the bounce is larger for potential dominated bounces. Then, the conclusion is that  the CMB may contain imprints from the bounce only if the potential energy of the field $\varphi$ is not too large at the time of the bounce, otherwise these effects  are red-shifted  away of the observable universe. Therefore, since the goal of this paper is to understand what are the signatures of a bounce in the CMB, and whether they can account for the observed anomalies, from now on  we will assume that the value of the potential energy at the bounce is  small enough to produce  $k_B/k_* \approx 1$. This condition can be easily implemented in our model by {\em adjusting the value of the total number of e-folds after the bounce so  $k_B/k_*=1$}. This makes the effects caused by the bounce appear only at large angular scales in the CMB, in such a way that we reproduce an almost-scale invariant power spectrum for $k>k_*$ that is in agreement with observations. 
Concrete  theories may come with arguments to support or disfavor this choice. For instance, in loop quantum cosmology, the arguments proposed in \cite{Ashtekar:2016wpi} pick up the configuration that produces $k_B/k_*\approx 1$.

To summarize, we model different types of  bounces by expression  (\ref{an}) for the scale factor, and assume that at later times the universe becomes dominated by a scalar field. The free parameters in this model are $n$ and the value of the Ricci scalar at the bounce $R_B$. We restrict to  $n\in [1/4,1/7]$, where many bouncing models in the literature belong to. The number of e-folds of expansion between the bounce and the end of inflation is also a free parameter, but we fix it in such a way that the effects of the bounce in the CMB appear for  $k  \lesssim k_*$.  This implies that the kinetic energy of the scalar field is larger than its potential energy at the bounce. There are two relevant scales in the problem, the scale of the bounce $k_B$ and the inflationary scale $k_I$. The ratio  $k_B/k_I$ is determined by the ratio of the spacetime Ricci scalar at the bounce and at the onset of inflation. The primordial power spectrum resulting from a bounce within this family has three characteristic regions, as shown in Figure \ref{PS}, separated by $k_B$ and $k_I$, and in each of them $P_{\mathcal{R}}(k)$ is well approximated by a power law.  The details of the bounce are encoded in the value of $k_B$ and the tilt $q$, and the later takes negative values for all the scenarios we have considered---i.e. the bounce enhances the primordial power spectrum (see Table \ref{n&q}). However, this enhancement does not extend to arbitrarily small wavenumbers, but rather $\mathcal{P}_{\mathcal{R}}(k)$ reaches a maximum around $k\approx k_I$ and decreases for smaller $k$. On the other hand, the  inflationary potential only affects the region $k>k_B$, and it dictates the value of  $n_s$, i.e.\ the spectral index for $k>k_B$. Our general analysis is in agreement with exact calculations obtained in concrete bouncing models, as for instance in loop quantum cosmology (see e.g. \cite{Agullo:2013ai,Agullo:2015tca}). In the rest of the paper we will use the form of $P_{\mathcal{R}}(k)$ shown in Figure \ref{PS} to compute the effects of the bounce in the CMB.

\bfig
 \ig[width=.7\textwidth]{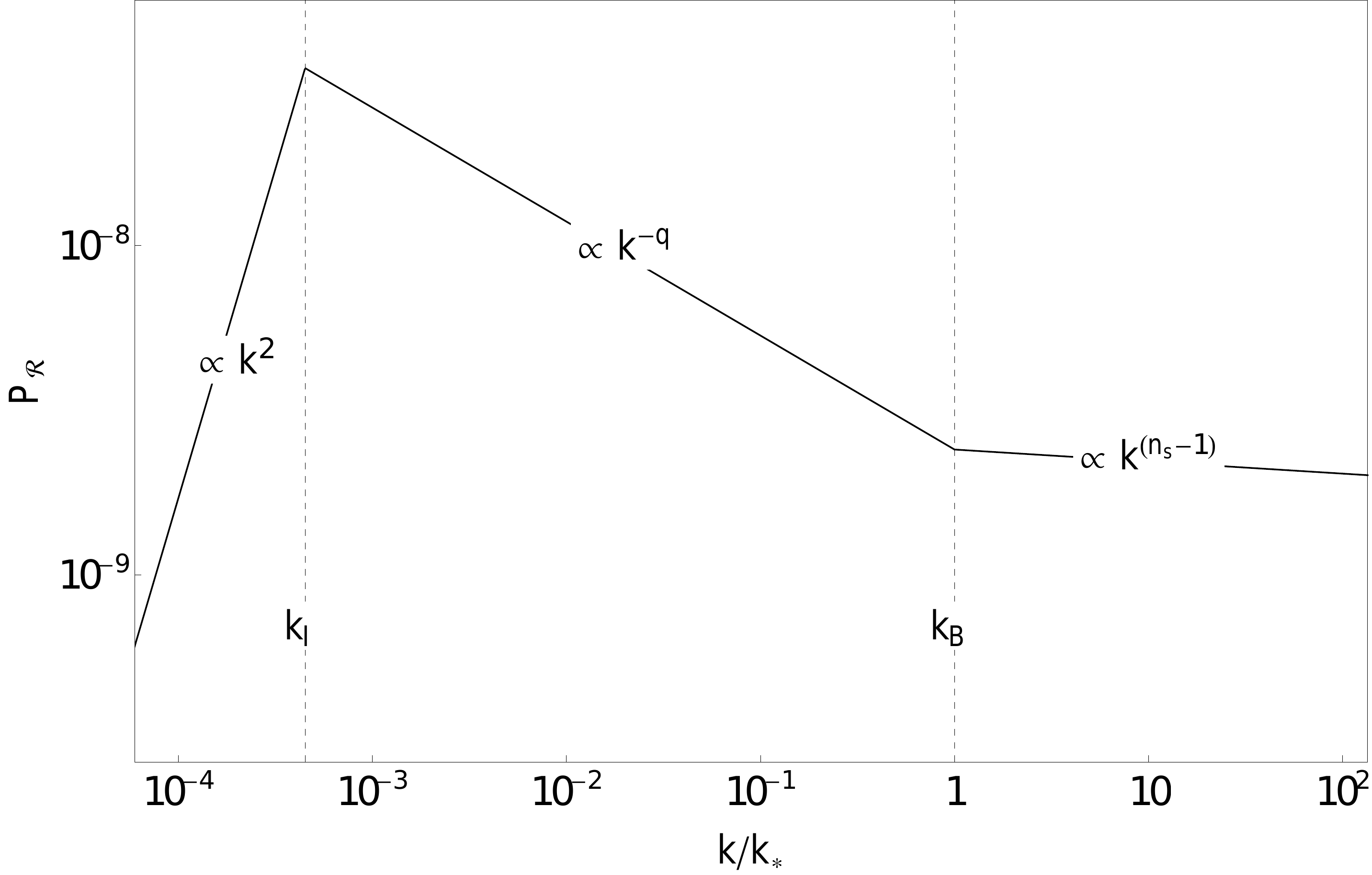}
\caption{Approximation of the scalar power spectrum for the family of bounces considered in this paper, for which a phase of slow-roll inflation follows the bounce.  We use the values for the amplitude $A_s$ and the spectral index $n_s$ for $k\geq k_*$ extracted from Planck data   \cite{Akrami:2018odb} (see section \ref{angcorr}). The value of the tilt $q$ depends on the details of the bounce, and they are given in Table \ref{n&q}.} \label{PS}
\efig

\subsection{The bispectrum}\label{bispectrum}
The goal of this section is to estimate the ``shape" of the non-Gaussianity in our model; i.e., 
the form of the  function $f_{NL}(k_1,k_2,k_3)$. The exact calculation of $f_{NL}(k_1,k_2,q)$ is not possible unless we specify the details of the  theory that produces the bounce. However, the goal of this section is not to obtain such a detailed and exact calculation, but rather to derive the generic features of $f_{NL}(k_1,k_2,k_3)$ in presence of a bounce. 
We will use a simple argument, introduced in \cite{Agullo:2017eyh}, which does not require knowledge of the details of the dynamical evolution. In spite of its simplicity and the use of crude approximations,  this argument has been proven to reproduce remarkably well the form of $f_{NL}(k_1,k_2,k_3)$ in the bouncing scenarios where an exact calculation is available \cite{Agullo:2017eyh}.  The result of this section, which we prove below, is that  a bounce that occurs before the inflationary era introduces a contribution to primordial non-Gaussianity of the form 

 \be \label{fNL}  f_{NL}(k_1,k_2,k_3) \approx \mathfrak{f}_{_{\rm NL}} \ e^{- \alpha/k_{\rm B}  \, (k_1+k_2+k_3)}\,  ,\ee 
where $\mathfrak{f}_{_{\rm NL}}$ is a constant. Since the three wavenumbers $\vec k_1$, $\vec k_2$, $\vec k_3$ must form a triangle, $k_2$ can be written in terms of $k_1$, $k_3$ and $\mu=\hat k_1\cdot \hat k_3$, as $k_2=k_1\, \sqrt{1+\frac{k_3^2}{k_1^2}+2\,\mu \frac{k_3}{k_1}}$. The amplitude  $\mathfrak{f}_{_{\rm NL}}$ depends on the details of the model, and we will leave it as a free parameter. In the example of loop quantum cosmology, where the bounce takes place at the Planck scale, it takes values of order  $\mathfrak{f}_{_{\rm NL}} \sim 10^3$. This form of $f_{NL}(k_1,k_2,k_3) $ is strongly scale-dependent, exponential in fact, and the scale dependence is determined from the details of the bounce, which are encoded in the constant $\alpha/k_{\rm B}$. Here $k_B$ is the scale of the bounce introduced in the previous section,  $\alpha=\sqrt{\frac{n\, \pi} 2}\,  \frac{\Gamma[1-n]}{\Gamma[3/2-n]}$, where $n$ defines the scale factor (\ref{an}), and $\Gamma[x]$ is the Gamma function. The most interesting property of this form of $ f_{NL}$ is that it makes the CMB very close to Gaussian for large wavenumbers $k_i\gtrsim k_B$, therefore satisfying observational constraints, while at the same time it allows  strong non-Gaussian correlations between smaller wavenumbers. In particular, it predicts strong correlations between the smallest wavenumbers we can observe in the CMB and super-horizon modes. The rest of this section is devoted to justifying the approximation (\ref{fNL}).

To obtain $  f_{NL}(k_1,k_2,k_3)$, we  must compute the bispectrum $B_{\Phi}(\vec k_1,\vec k_2)$ defined in  equation  (\ref{bisp}).  As it is well known (see e.g. \cite{Maldacena:2002vr,Sreenath:2014nca}).
$B_{\Phi}(\vec k_1,\vec k_2)$ can be computed in the quantum theory by using time dependent perturbation theory. Truncating the time evolution operator at leading order in perturbations, one obtains 
\be B_{\mathcal{R}}(\vec k_1,\vec k_2)=-i/\hbar\, \int^{\eta}  d\eta' \langle 0|\Big[\hat{\mathcal{R}}^I_{\vec k_1}{\mathcal{R}}^I_{\vec k_2}{\mathcal{R}}^I_{\vec k_3},\mathcal{H}^I_{\rm Int}\Big]|0 \rangle\, + \mathcal{O}(\mathcal{H}_{\rm Int}^2)\, .
 \ee
where the superscript $I$ indicates operators in the interaction picture, and  $\mathcal{H}^I_{\rm Int}$ is the Hamiltonian describing self-interaction between primordial scalar perturbations.  This Hamiltonian needs to be derived from the gravitational theory one is using in the early universe, and in   Appendix \ref{pertsham}  we provide the expression one obtains from general relativity. The previous integral takes the form
\bea \label{1int} B_{\mathcal{R}}(\vec k_1,\vec k_2)&=& \frac{v_{{k}_1(\eta_f)}}{a(\eta_f)}\frac{v_{{k}_2(\eta_f)}}{{a(\eta_f)}}\frac{v_{{k}_3(\eta_f)}}{a(\eta_f)} \int_{\eta_i}^{\eta_f} \d\eta\, \Big[ f_1\, \frac{v^*_{{k}_1}}{a}\frac{v^*_{{k}_2}}{{a}}\frac{v^*_{{k}_3}}{a} 
+f_2\, \frac{v^*_{{k}_1}}{a}\frac{v^*_{{k}_2}}{a}\frac{d}{d\eta}\left(\frac{v^*_{{k}_3}}{a}\right) \nonumber \\
&+& f_3 \, \frac{v^*_{{k}_1}}{a}\, \frac{d}{d\eta}\left(\frac{v^*_{{k}_2}}{a}\right)\, \frac{d}{d\eta}\left(\frac{v^*_{{k}_3}}{a}\right)+(\v k_1 \leftrightarrow \v k_3)+ (\v k_2 \leftrightarrow \v k_3) \Big] \, ,
 \eea
where $f_1( \eta)$, $f_2(\eta)$ and $f_3(\eta)$ are  functions of the background degrees of freedom $a(\eta)$, $\varphi(\eta)$, and their conjugate momenta $\pi_a(\eta)$ and $p_{\varphi}(\eta)$, also given in Appendix \ref{pertsham} for general relativity. The range of the previous integral extends from an initial time $\eta_i$ before the bounce to the end of inflation $\eta_f$. The contribution of the inflationary epoch to this integral is known to be at the order of the slow-roll parameters. We are  interested here in the contribution from an interval around the bounce.  An estimate  of this contribution can be obtained by approximating the mode functions during the time of the bounce by positive frequency exponentials $v_{{k}}\sim e^{-i k\eta}$. This is an excellent approximation for $k\gtrsim k_{\rm B}$. The contribution of the bounce to the primordial non-Gaussianity is then given by an integral of the form

\be \label{2int} B_{\mathcal{R}}(\vec k_1,\vec k_2)= \int_{-\eta_0}^{\eta_0} \d\eta \, g(k_i,\eta)\, e^{i (k_1+k_2+k_3)\, \eta}\approx \int_{-\infty}^{\infty} \d\eta \ g(k_i,\eta)\, e^{i (k_1+k_2+k_3)\, \eta} \, W(\eta,\Delta), \ee
where we have restricted the integral to some time interval $\eta_0$ before and after the bounce (the bounce takes place at $\eta=0$). In this expression $k_t \equiv  k_1+k_2+k_3$;  $g(k_i,\eta)$ is a combination of the functions $f_i$'s that can be  easily read from (\ref{1int}). And in the last equality we have extended the limits of the integral to $\pm\infty$ by introducing a window function  $W(\eta,\Delta \eta)$ that  is equal  to zero for $|\eta|> \eta_0$, equals one for $|\eta|< \eta_0$, and smoothly interpolates between both values.  Its concrete form  will be unimportant for our purposes.

With the integral written in this form, Cauchy's integral theorem tells us that the right hand side of \eqref{2int} is equal to  $2\pi i$ times  the sum of the residues of the poles of $g(k_i,\eta)$ that have {\em positive imaginary part}---since we must close the integration contour  in the upper complex plane. %
Each of these residues is proportional to $e^{i (k_1+k_2+k_3)\, \eta_p}$, where $\eta_p$ denotes a pole of $g(k_i,\eta)$. Hence, the real part of each pole contributes to the oscillatory behavior of the integral as a function of $(k_1+k_2+k_3)$, while the imaginary part introduces an exponentially decreasing factor.  Now, out of the four background functions that appear in $g(k_i,\eta)$, 
the scale factor $a(\eta)$ is the only one having a {\em minimum} at the bounce, and consequently it is the pole of $1/a(\eta)$ at the bounce that dominates the integral. This pole can be estimated by expanding $a(\eta)$ near  the bounce as $a(\eta)\approx a_B+ \frac{1}{2} \, a''(\eta_B)\, \eta^2+...$, from where we see that $\eta_p=\pm i\, \sqrt{\frac{2a(\eta_B)}{a''(\eta_B)}}$ (only the positive pole contributes to our integral). Furthermore, recall that the Ricci scalar at the bounce is $R_B=6\, a''(\eta_B)/a(\eta_B)$, so the location of this pole and the value of its residue is dictated by $R_B$.

One can obtain a more accurate expression of the pole and its residue by working  with the exact form of the scale factor (\ref{an}), rather than truncating the Taylor expansion of $a(\eta)$. The result is that the pole of $a^{-1}$ with positive imaginary part is $t_p=i/\sqrt{R_B/(2n)}=i \sqrt{2n/3}\, \frac{1}{k_B}$, when written in cosmic time,\footnote{The relation between $t$ and $\eta$ is given by a hypergeometric function $\eta=\int_0^t  a(t')^{-1}\, dt'=t\, a_B^{-1} \, \sqrt{R_B/(2n)}\, _2F_1[n,\f{1}{2},\f{3}{2},-R_B/(2n)\, t^2]$.} and
\be \label{pole} \eta_p=i \, \alpha/k_B \, , \, \ \ {\rm where} \  \ \alpha=\sqrt{\frac{n\, \pi}{ 2}}\,  \frac{\Gamma[1-n]}{\Gamma[3/2-n]}\, , \ee
 in conformal time. The concrete value of the residue at this pole and its dependence on the three wavenumbers require knowledge of  the function $g(k_i,\eta)$, but its magnitude is always dominated by the exponential factor $e^{i (k_1+k_2+k_3)\, \eta_p}$. Therefore, this  argument tells us that  the  bounce produces a contribution to $f_{NL}(k_1,k_2,k_3)$ whose overall dependence on $k_i$ is given by
%
 \be  f_{NL}(k_1,k_2,k_3) = \mathfrak{f}_{_{\rm NL}} \, e^{- \alpha  (k_1+k_2+k_3)/k_{\rm B}}\,  .\ee 

    \begin{table}
\caption{Values of the tilt $\alpha$  of ${f}_{_{\rm NL}}$ for different $n$'s. } 
\centering 
\begin{tabular}{|c|c|} 
\hline 
$n$ & $\alpha$\\ 
\hline \hline 
1/4 & 0.85\\
0.21 &0.75 \\
1/5 & 0.73 \\
1/6 & 0.65\\
1/7 & 0.59 \\
\hline 
\end{tabular}
\label{alphan} 
\end{table}

As mentioned above, we have checked that this simple argument captures remarkably well the overall form of $ f_{NL}(k_1,k_2,k_3)$  in a concrete bouncing scenario  \cite{Agullo:2017eyh}. On the other hand, our argument does not capture other finer details of  that ${f}_{NL} $ could have, like oscillating components on the top of the exponential behavior.  We will neglect these possible oscillations, so the results derived in the next sections for the effects in the CMB should be understood as upper bounds, since the effects we describe could be partially reduced by the presence of such oscillations. On the other hand, our argument breaks down for very infrared wavenumbers $k_i\ll k_B$. In that regime $ f_{NL}(k_1,k_2,k_3)$  is expected to become small, for the same reason as the power spectrum does, namely because extremely infrared modes are not excited either by the bounce nor by inflation. This expectation is indeed borne out in concrete models  \cite{Agullo:2017eyh}. However, it will not be necessary to work out these details here, since the fact that the power spectrum also becomes very small for very infrared (super-horizon) scales acts as an effective infrared cut-off, making the value of the non-Gaussianity at these scales unimportant. Our approximations are trustable for wavenumbers $k$ in the range $\sim [k_I,k_B]$ and these are precisely the values that  are responsible for the effects we explore in this paper.
 
\section{Monopolar modulation \label{monopolar}}

\begin{figure}[htp]
\centering
\includegraphics[width=.45\textwidth]{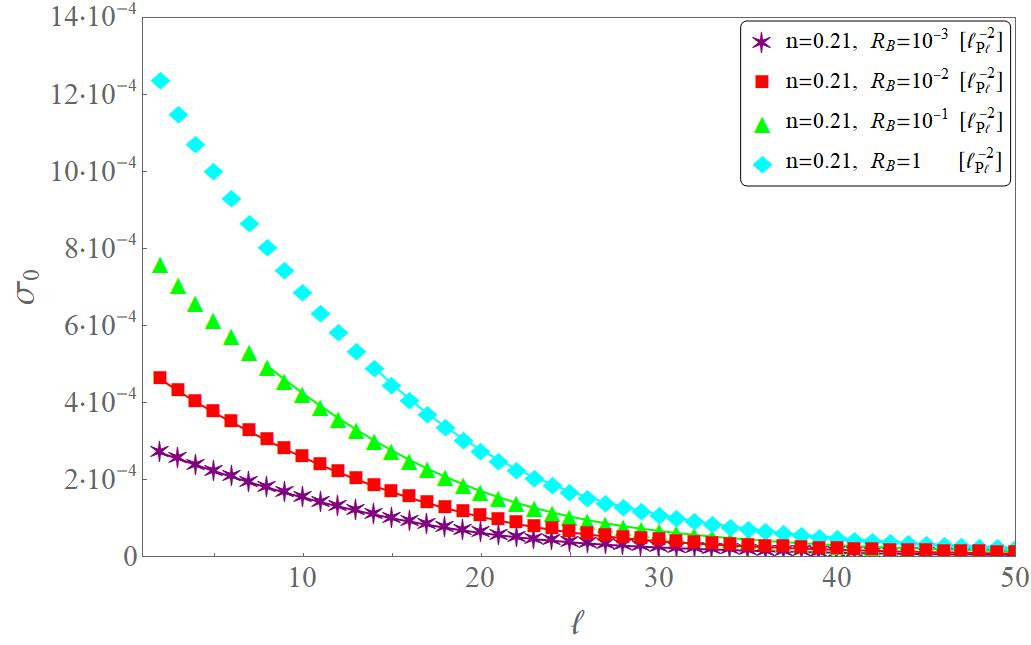}\quad
\includegraphics[width=.45\textwidth]{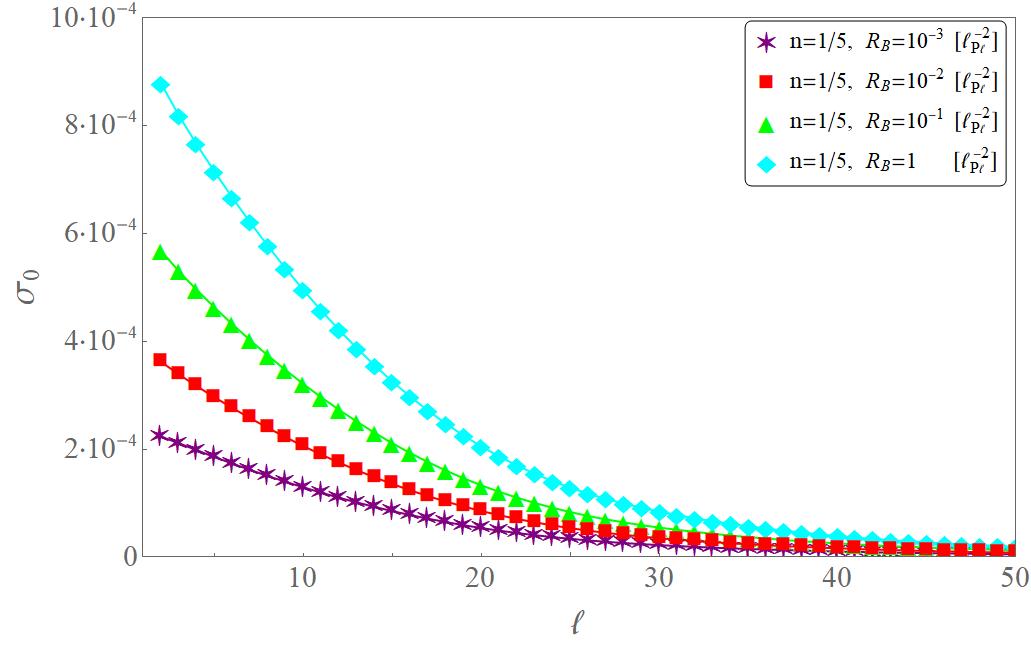}\quad
\medskip
\includegraphics[width=.45\textwidth]{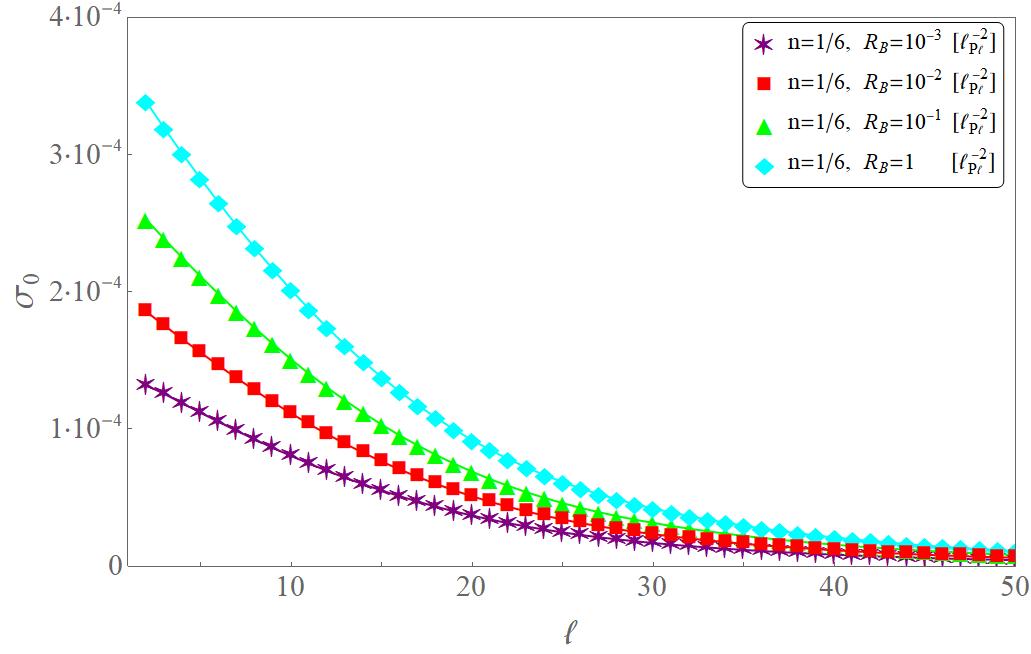}
\includegraphics[width=.45\textwidth]{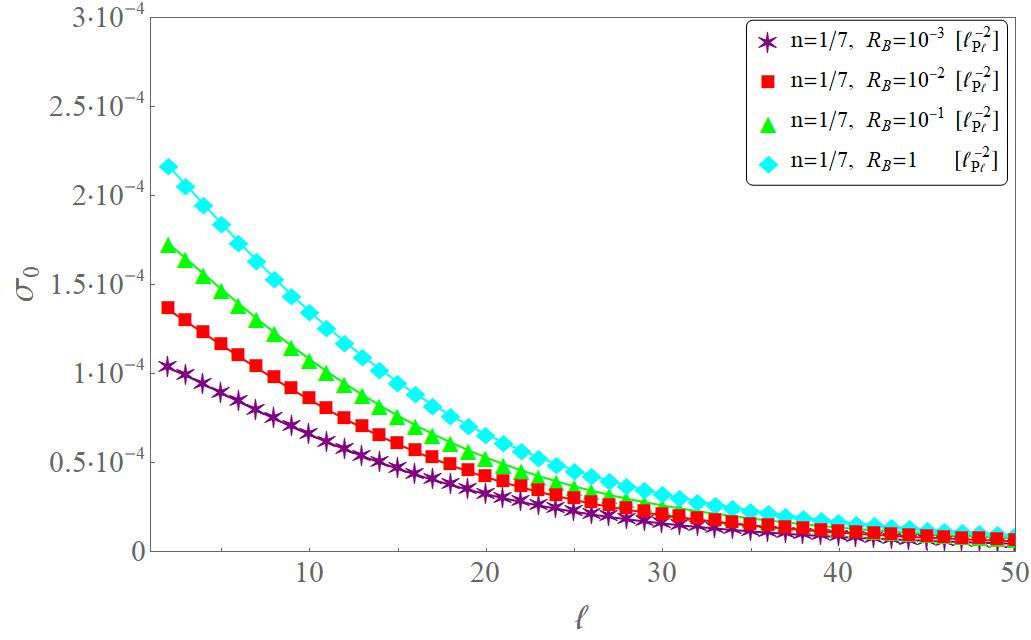}\quad
\caption{ Results for   the variance  $\sigma_0(\ell)$ of the monopolar modulation  in our model for $\mathfrak{f}_{_{\rm NL}}=1$, and different values of $n$ and $R_B$. These figures  show that   $\sigma_0(\ell)$ is strongly scale dependent.} 
\label{fig:sigma0}
\end{figure}

The goal of this section is to evaluate  the square mean value of the BipoSH coefficients with $L=0$, i.e.\  $A^{00}_{\ell\ell}$, and to discuss the effects they produce on the angular power spectrum.  If any of the coefficients $A^{00}_{\ell\ell}$ are different from zero in our local universe, then the observed angular power spectrum will be modulated as 
(see the discussion following equation (\ref{valueBipoSH}))
\be \label{Cobs} C_{\ell}^{ \rm mod} = C_{\ell}\, \left(1+\frac{1}{C_{\ell}}(-1)^{\ell}\, \frac{A^{00}_{\ell\ell}}{\sqrt{2\ell+1}}\right) \, , \ee
where $C_{\ell}$ is the ``bare'' (i.e.\ Gaussian) angular power spectrum.  Note that $A^{00}_{\ell\ell}$ can be either positive or negative, and consequently $C^{\rm mod}_{\ell}$ can be  enhanced or suppressed relative to $C_{\ell}$. In all scenarios discussed in this paper, the magnitude of the modulating term $|\frac{1}{C_{\ell}}(-1)^{\ell}\, \frac{A^{00}_{\ell\ell}}{\sqrt{2\ell+1}}|$ remains smaller than one.  Furthermore, because $A^{00}_{\ell\ell}$ could---and actually does---depend on $\ell$, the magnitude of this modulation can vary for different angular scales in the CMB. As we already discussed earlier, we cannot predict the exact value of $A^{00}_{\ell\ell}$, since it depends on the stochastic primordial perturbations. But we can compute the square mean value of the modulation
\be \sigma_0^2(\ell)\equiv \frac{1}{C^2_{\ell}}\, \frac{\langle |A^{00}_{\ell\ell}|^2\rangle}{2\ell+1} \, .\ee
Using the results of section \ref{sec:NGMod}, and in particular equation (\ref{varalpha}), we find
\be \label{singma0variance} \sigma^2_0(\ell)= \frac{1}{C^2_{\ell}}\, \frac{1}{8\pi^2}\,  \int dq\, q^2 \, P_{\phi}(q) \, |\mathcal{C}_{\ell \ell}^0(q)|^2\, ,\ee
where $\mathcal{C}_{\ell \ell'}^L(q)$, defined in (\ref{Ctilde}), contains the information about the non-Gaussianity. 
If this variance turns out to be large, an enhancement or suppression of the observed  power spectrum would happen with high probability. We plot in Figure \ref{fig:sigma0} the results for the variance $ \sigma_0(\ell)$ for different values of  $n$ and $R_B$. We have used $\mathfrak{f}_{_{\rm NL}}= 1$ in these plots (recall $\sigma_0(\ell)$ is proportional to  $\mathfrak{f}_{_{\rm NL}}$). %
We see  that  $\sigma_0(\ell)$ has a strong dependence on $\ell$, becoming smaller as $\ell$ increases. This implies that the non-Gaussian modulation will affect mostly low CMB multipoles.  This is of course a consequence of the scale-dependence of both, the power spectrum and the non-Gaussianity in our model.

The $\ell$-dependence in  $\sigma_0$  can be approximated by an exponential fall-off\footnote{Although this approximation works well for  $n=1/6$ and $1/7$ (per cent error), for  $n=1/4$ the error is larger, reaching  10 per cent for large $\ell$. But this will not be a problem for us, since in our computations we will rather use the exact numerical results. The only goal of this approximation is to provide a simple understanding of the way $\sigma_0(\ell)$ depends on $\ell$.}, $\sigma_0(\ell)=\underline{\sigma}_0\, e^{-\delta_n \, \ell}$, where  $\delta_n$ depends on $n$.\footnote{Actually, $\delta_n$  also depends on $R_B$ but in a much milder manner, with variation at the per cent  level or less when $R_B$ changes by two orders of magnitude.} Table \ref{deltan} shows the values of $\delta_n$ for several values of $n$.  We see that $\delta_n$ decreases for smaller  $n$. This is indeed expected, since  the tilt of both, the power spectrum and the non-Gaussianity,  parameterized by $q$ and $\alpha$, respectively, also decrease with $n$, as shown in  Tables \ref{n&q} and \ref{alphan}. The amplitude of the exponential fall-off can be directly read from Figure \ref{fig:sigma0}.

   \begin{table}
\caption{Values of the $\delta_n$  obtained by fitting the function  $\sigma_0(\ell)$ to an exponential $\sigma_0(\ell)=\underline{\sigma}_0\, e^{-\delta_n \, \ell}$. The amplitudes $\underline{\sigma}_0$ can be directly read from Figure \ref{fig:sigma0}} 
\centering 
\begin{tabular}{|c|c|} 
\hline 
$n$ & $\delta_n$\\ 
\hline \hline 
1/4 & 0.088\\
0.21 &0.082 \\
1/5 & 0.080 \\
1/6 & 0.072\\
1/7 & 0.066 \\
\hline 
\end{tabular}
\label{deltan} %
\end{table}

\subsection{Power suppression}

Observations from WMAP \cite{Bennett:2003bz} have revealed strong evidence of a lack of 2-point correlations at angular separations larger than approximately $60^{\circ}$  \cite{Efstathiou:2003tv}. This feature was  already  noticed by COBE one decade earlier \cite{Hinshaw:1996ut}, and it has been confirmed again in all data releases by Planck \cite{Ade:2013kta,Ade:2015hxq,Akrami:2019bkn} with similar statistical significance. See \cite{Copi:2006tu,Copi:2013cya,Schwarz:2015cma} for a detailed discussion.   This lack of correlations is particularly evident in real space;  the observed two-point angular correlation function\footnote{The relation between $C(\theta)$ and the angular power spectrum is $ C(\theta)=\sum_{\ell} \frac{2\ell+1}{4\pi}\, C_{\ell}\, P_{\ell}(\cos \theta)$, where $P_{\ell}(x)$ are Legendre polynomials.}  $C(\theta)=\langle \delta T(\hat n)\delta T(\hat n')\rangle$, where $\cos \theta=\hat n\cdot \hat n'$,  lies close to zero for angles between $60^{\circ}$ and $170^{\circ}$, in sharp contrast with what is expected from $\Lambda$CDM (see e.g.\ figure 2 of \cite{Ade:2015hxq}, and Figure \ref{modClfixRB} below).  In order to quantify the tension between data and  the $\Lambda$CDM model, different groups \cite{Copi:2013cya, Schwarz:2015cma,Ade:2015hxq,Akrami:2019bkn} have used the estimator proposed in \cite{Spergel:2003cb}, based on $S_{1/2}$ defined as
\be S_{1/2}=\int_{-1}^{1/2} \big(C(\theta)\big)^2\, d(\cos\theta) \, . \ee
This quantity  measures the total amount of correlations for angular separations in the range   $\theta \in  [60^{\circ},180^{\circ}]$. The  value of $S_{1/2}$ expected from  $\Lambda$CDM  is $S_{1/2}\approx  45000\, \mu K^4$, which is significantly larger than the observed one, which lies around $S^{\rm obs}_{1/2}\approx 1500\, \mu K^4$. (It is reported in \cite{Schwarz:2015cma} that the value of $S^{\rm obs}_{1/2}$ varies between 1887 and 911, depending on the data set and the details of the mask used.) 
The $p$-values of this observation within the $\Lambda$CDM model has been found to be consistently  below $1\%$ \cite{Ade:2015hxq,Akrami:2019bkn}. Reference \cite{Schwarz:2015cma} reports  a $p$-value  $\leq 0.5\%$, and points out that the analysis by Planck  cannot resolve values below $0.2\%$.\footnote{The $p$-value is defined in slightly different ways in the literature. We  follow the definition used in \cite{Copi:2006tu,Copi:2013cya,Schwarz:2015cma}.  On the other hand, the Planck satellite team \cite{Ade:2015hxq,Akrami:2019bkn} defines it as the number of simulations expressed in per cent  with a value of $S_{1/2}$ {\em larger}  than the observed one. Therefore, a $p$-value of $99\%$ according to \cite{Ade:2015hxq,Akrami:2019bkn} corresponds to $1\%$ in \cite{Copi:2006tu,Copi:2013cya,Schwarz:2015cma}.}

Using the results of the previous subsection, we evaluate now what is the value of the amplitude of the primordial non-Gaussianity $\mathfrak{f}_{_{\rm NL}}$ that makes the probability  of measuring $S^{\rm obs}_{1/2}\leq1500\, \mu K^4$ in our model approximately equal to $20\%$. We estimate this probability by assuming that the amplitude of the monopolar modulation follows a Gaussian probability distribution with zero mean and variance given in (\ref{singma0variance}). This assumption is reasonable since, although each BipoSH coefficient $A^{00}_{\ell\ell}$   arises from a large number of stochastic degrees of freedom, some of which are non-Gaussian,  deviations from Gaussianity  in the statistics of $A^{00}_{\ell\ell}$ are second order in the primordial non-Gaussianity, and can be neglected. We consider neither instrumental nor other observational effects in our calculation of probabilities. However, our estimation suffices  to understand whether our model can account for the observed suppression. We report the results in Table \ref{T1}.

\begin{table}
\centering
    \begin{tabular}{|l||c| c|c|c|c|c|}
    \hline
    \hline
      \diagbox{$R_B$}{$n$} & $1/4$ & 0.21&  $1/5$& $1/6$ & $1/7$ \\
    \hline \hline
$\ \ \ \ \ 1\, \ell^{-2}_{P\ell}$ &  - & 959&  1334& $3326$ &5031 \\ \hline
$10^{-1}\, \ell^{-2}_{P\ell}$ &  - & 1560 &  2065& $4454$ &6298 \\ \hline
$10^{-2}\, \ell^{-2}_{P\ell}$ & -  & 2573 & 3238& 6066& 8024 \\ \hline
$10^{-3}\, \ell^{-2}_{P\ell}$ & - & 4372 &5234  & 8518&10530  \\
    \hline
    \hline
    \end{tabular}
\caption{Values of the amplitude of the primordial Bispectrum $\mathfrak{f}_{_{NL}}$ that  make the probability of obtaining   $S^{\rm obs}_{1/2}\leq 1500\, \mu K^4$ in our CMB equal to $20\%$. The results are shown for different values of the parameters $n$ and $R_B$, the later expressed in Planck units. For $n\leq1/4$, the non-Gaussian modulation cannot produce the observed suppression.
\label{T1}}
\end{table}

Overall, our calculations show that for values of $\mathfrak{f}_{_{\rm NL}}$ of the order $10^3$ our model is able to make the observed suppression a  common feature in a typical CMB, with probabilities of the order of $20\%$, even though the bare (unmodulated) power spectrum is enhanced relative to the $\Lambda$CDM result. 
If we look at the details, we can see that the required $\mathfrak{f}_{_{\rm NL}}$ varies with $R_B$ and $n$. We  find that in order to produce the desired suppression, $\mathfrak{f}_{_{\rm NL}}$ needs to be larger the  smaller  the curvature at the bounce $R_B$ is. This is because the closer $R_B$ is to the inflationary scale $R_I$ the lesser  modes $k$ can be accommodated between $k_B$ and $k_I$. Since these are the modes whose bare power spectrum is enhanced, small $R_B$ translates to less power in infrared modes, and  consequently to a larger value of $\mathfrak{f}_{_{\rm NL}}$ to compensate. On the other hand, we observe that  $\mathfrak{f}_{_{\rm NL}}$ needs to be larger the smaller $n$ is. This can be understood by looking at Table \ref{n&q}: the tilt $q$ of the bare power spectrum  decreases with $n$, and this again translates to less power in infrared modes that needs to be compensated by  $\mathfrak{f}_{_{\rm NL}}$. We also find that for $n=1/4$   there is no choice of $\mathfrak{f}_{_{\rm NL}}$  able to account for the observed suppression. This is because the tilt $q$ is so large that the non-Gaussian modulation is unable to compensate the enhancement of the bare power spectrum to produce the observed suppression. 

The values of $\mathfrak{f}_{_{\rm NL}}$ shown in Table \ref{T1} are  several orders of magnitude larger than the ones predicted by slow-roll inflation alone. 
 But notice that, on the one hand, this is precisely the order of magnitude found in concrete bouncing models in which a phase of inflation  takes place at or near the Planck scale  before the inflationary era \cite{Agullo:2017eyh}. This shows that there  exist concrete models that reproduce the requirements we have found in this section. On the other hand, this large value of $\mathfrak{f}_{_{\rm NL}}$ does not conflict with observational constraints, since the non-Gaussian correlations are restricted mostly to super-horizon modes, and therefore they cannot be directly observed. However, a large value of $\mathfrak{f}_{_{\rm NL}}$ raises concerns about the validity of  perturbation theory. This is discussed in detail in section \ref{validitypert}, where we show that the perturbative expansion is under control for the values of $\mathfrak{f}_{_{\rm NL}}$ shown in Table \ref{T1}.

\subsection{Angular correlation functions and cosmological parameters}\label{angcorr} 

\begin{figure}[htp]
\centering
\includegraphics[width=.485\textwidth]{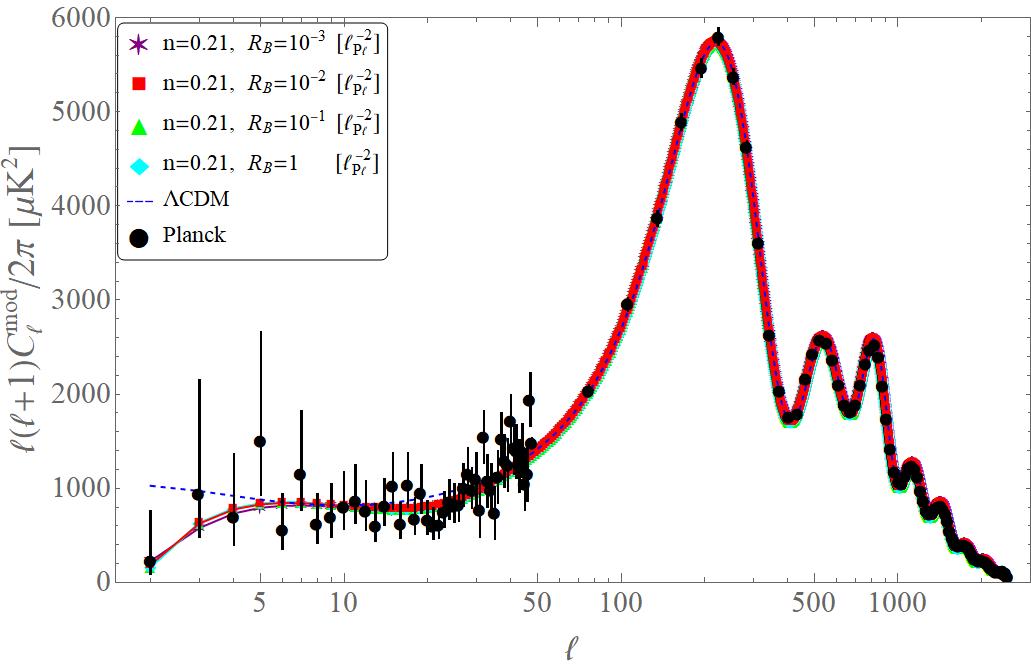}\quad
\includegraphics[width=.485\textwidth]{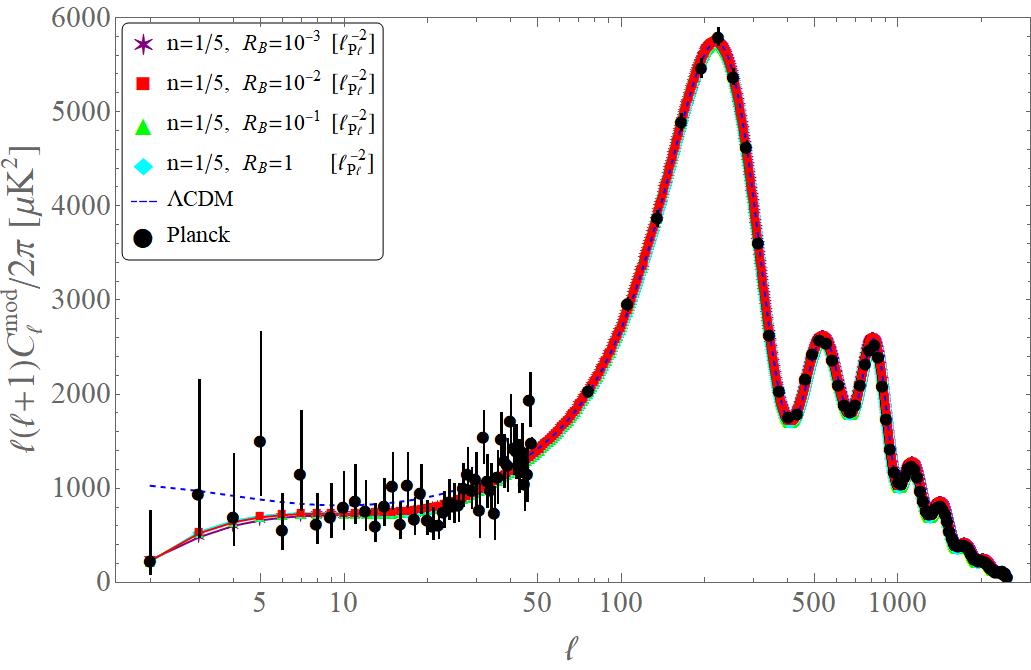}\quad
\medskip
\includegraphics[width=.485\textwidth]{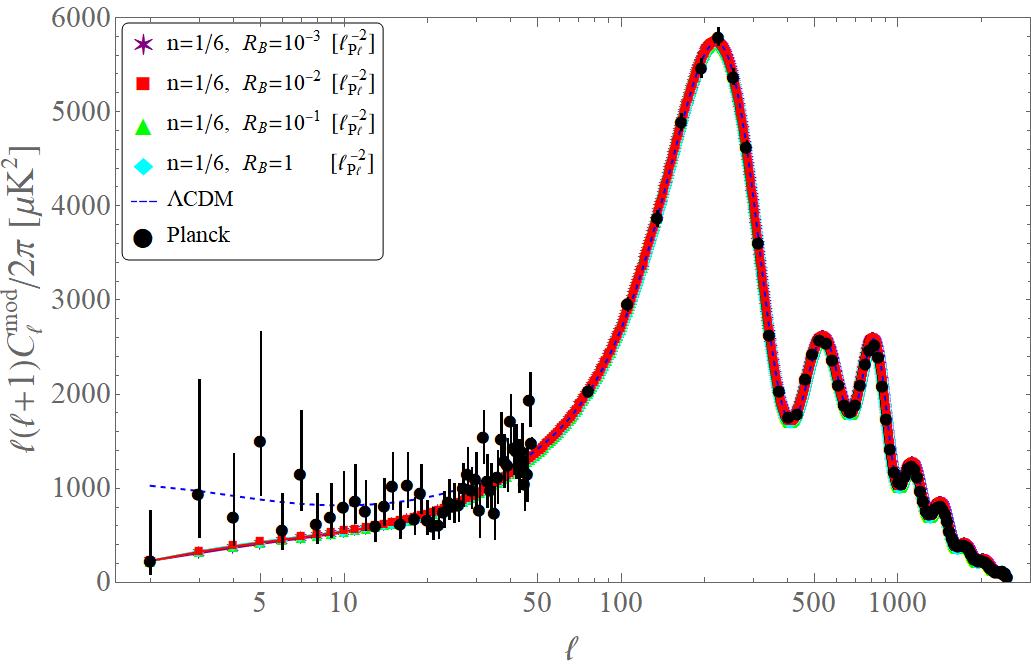}
\includegraphics[width=.485\textwidth]{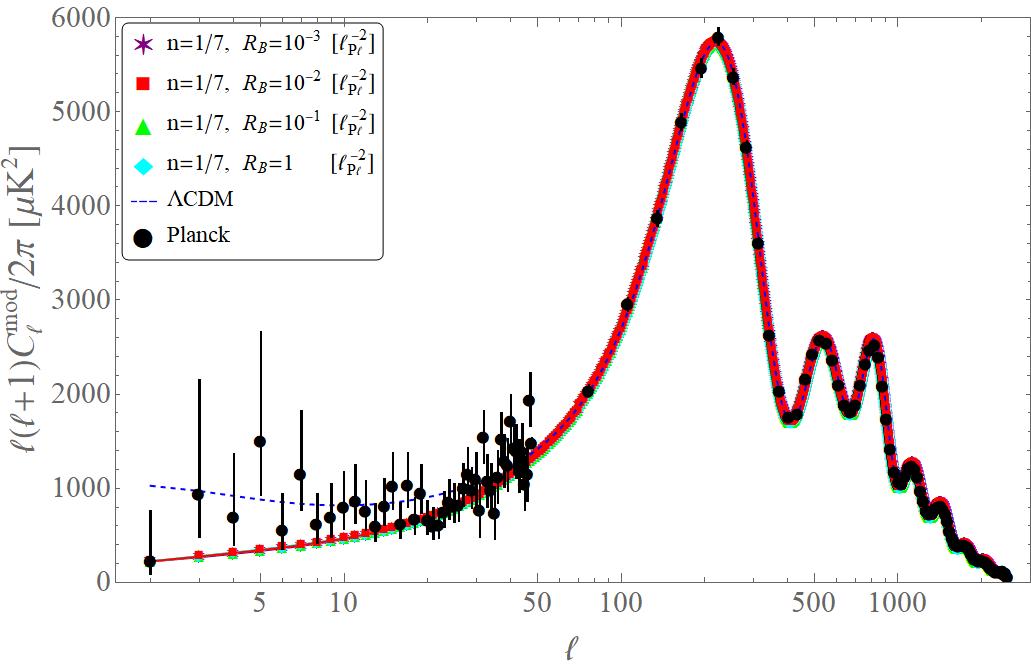}\quad

\caption{Modulated angular power spectrum obtained from our model for different values of $n$ and $R_B$, the later in Planck units. We have used the best-fit values for the cosmological parameters.  The result of the $\Lambda$CDM model with  almost scale invariant primordial perturbations, and data from Planck, are also shown for comparison.}
\label{modCl}
\end{figure}

\begin{figure}[htp]
\centering
\includegraphics[width=.485\textwidth]{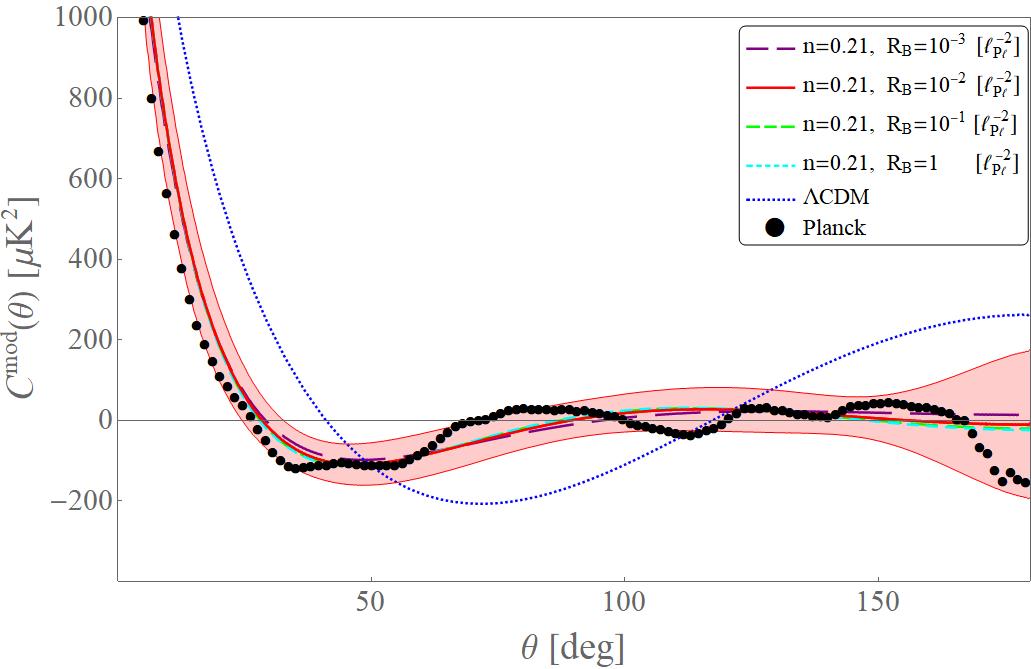}\quad
\includegraphics[width=.485\textwidth]{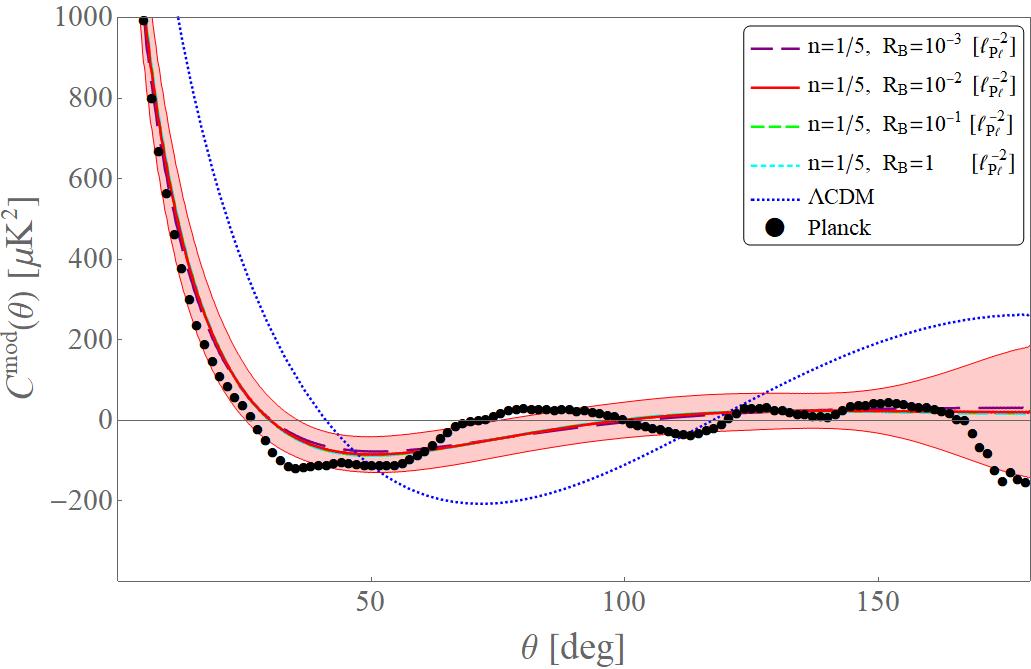}\quad
\medskip
\includegraphics[width=.485\textwidth]{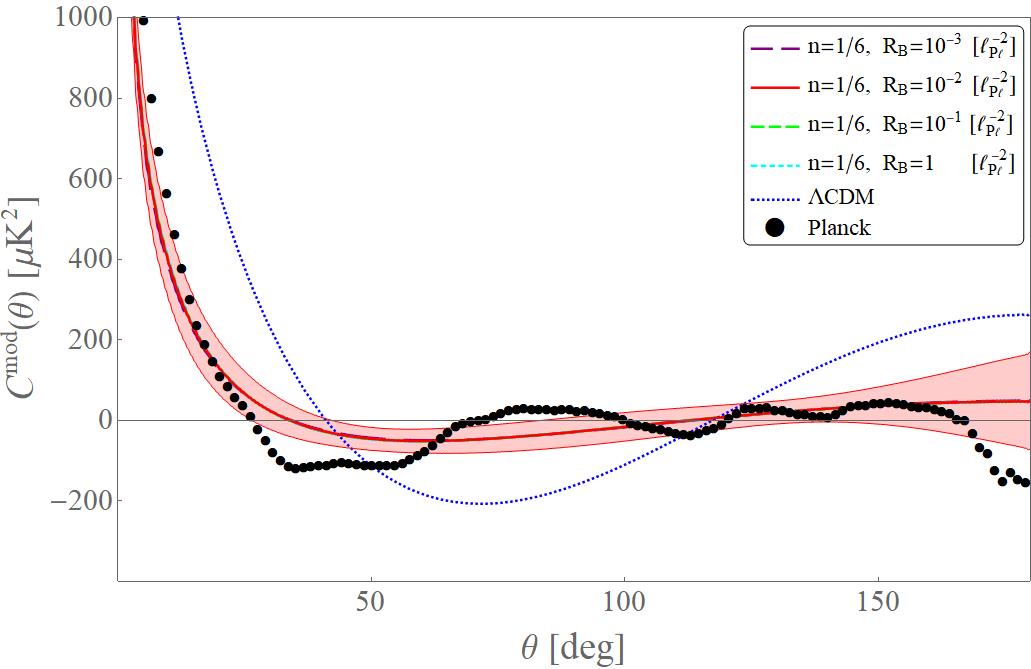}
\includegraphics[width=.485\textwidth]{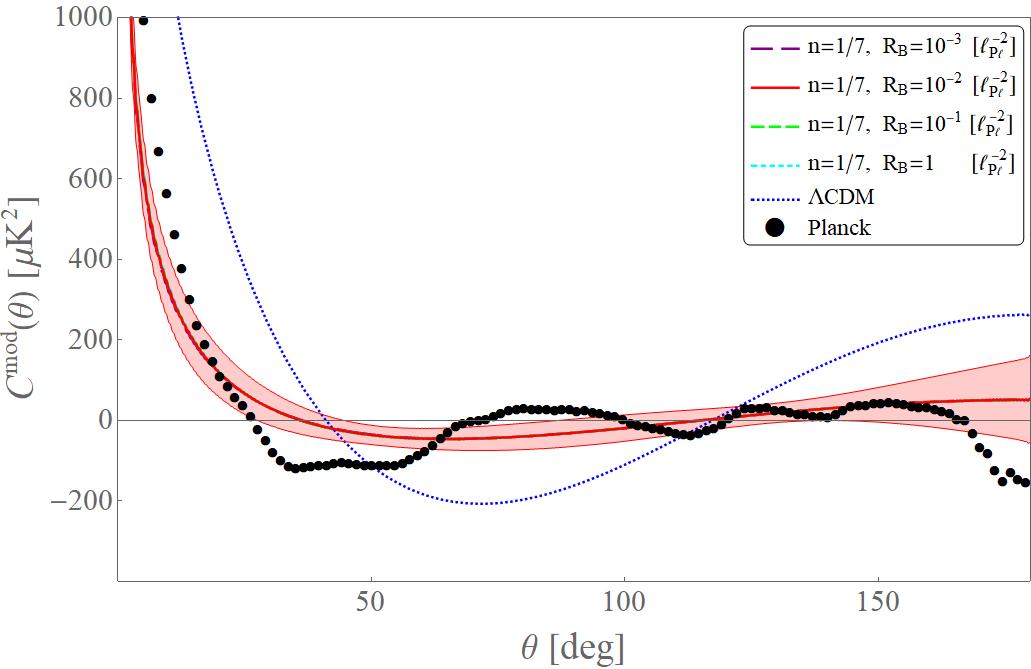}\quad

\caption{Angular two-point correlation function $C^{\rm mod}(\theta)$ computed  from $C^{\rm mod}_{\rm \ell}$ in our model for different values of $n$ and $R_B$. The two-point function of the $\Lambda$CDM model with  almost scale invariant primordial perturbations, and data from Planck, are also shown for comparison. We have added the one-standard deviation cosmic variance (shadowed region) associated to one of the curves in our model, namely the curve with $R_B=10^{-2}\ell_{\rm PL}$, in order to facilitate a meaningful comparison with data. Since for a fixed $n$ our model produces almost indistinguishable  results when $R_B$ is varied, it is unnecessary to include the cosmic variance for other values of $R_B$. Notice that all curves---except the one corresponding to the $\Lambda$CDM model---produce a $S_{1/2}=1500\, \mu K^4$.}
\label{modCtheta}
\end{figure}

From now on we will focus on   realizations of the probability distribution of our model for which  $S_{1/2}= 1500\, \mu K^4$, and we will study other properties of these realizations. First, we plot in Figures \ref{modCl} and \ref{modCtheta} the  angular power spectrum $C_{\ell}^{ \rm mod}$ and the associated angular  two-point function $C^{ \rm mod}(\theta)$ for different values of $n$ and $R_B$. We also show in these plots the predictions of the $\Lambda$CDM model with the standard ansatz of an almost scale invariant primordial power spectrum, as well as data from Planck \cite{Akrami:2019bkn}. It is important to keep in mind that all curves obtained from our model in Figures \ref{modCl} and \ref{modCtheta} reproduce the same value of $S_{1/2}$, regardless of $n$ and $R_B$, but they differ  in the details. 
So these plots contain additional information about the power spectrum at low multipoles that is not captured by $S_{1/2}$.

\begin{figure}[htp]
\centering
\includegraphics[width=.485\textwidth]{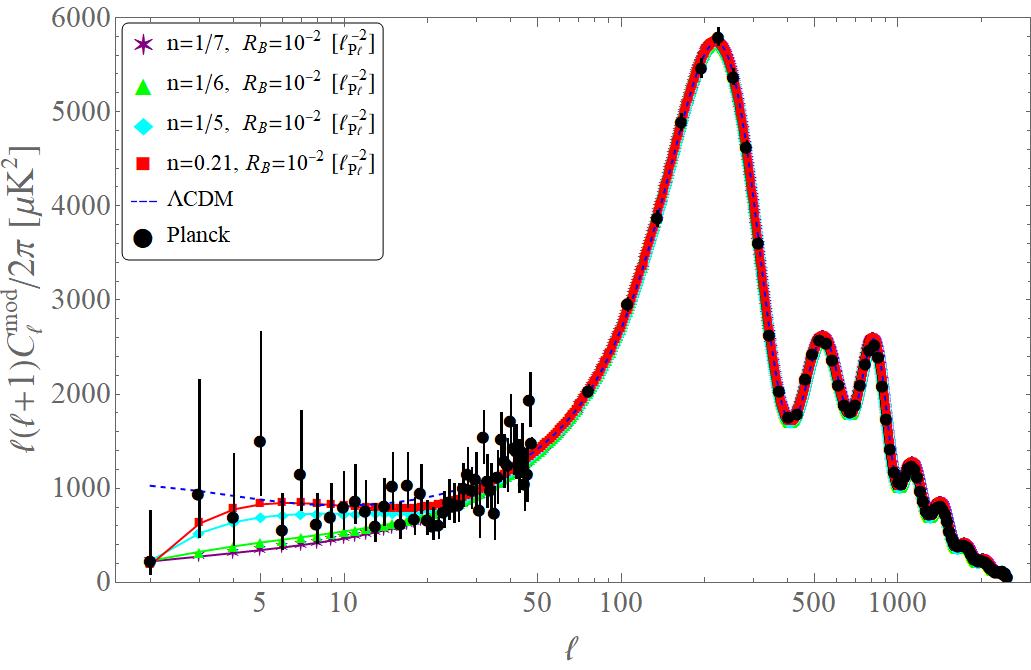}\quad
\includegraphics[width=.485\textwidth]{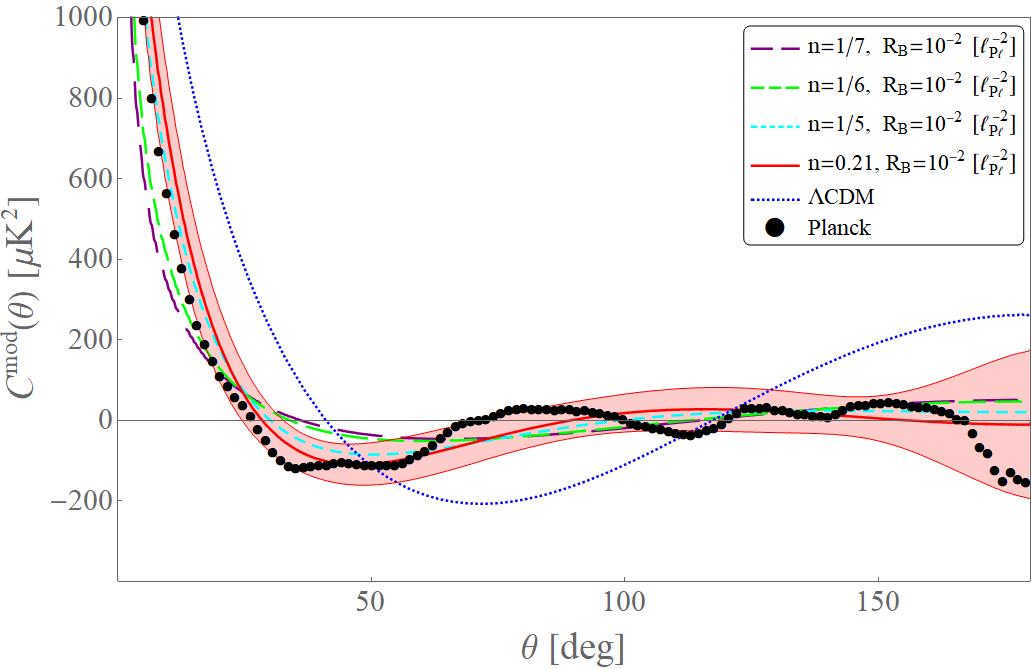}\quad
\caption{Modulated angular power spectrum $C_{\ell}^{ \rm mod}$ and two-point correlation function $C^{\rm mod}_2(\theta)$ for   different values of $n$  and $R_B=10^{-2}\ell_{\rm PL}$. The curves for the $\Lambda$CDM model  and data from Planck are also shown for comparison. The shadowed region indicates the one-standard deviation cosmic variance for $n=0.21$ (we do not include the cosmic variance for other values of $n$ to avoid an over-crowded figure). This figure shows  the way details of the power spectrum vary with $n$. Large values of $n$ (e.g.\ $n=0.21$) do better in reproducing the details of data---in fact the curve with $n=0.21$ produces a surprisingly good agreement with data at all scales.} 
\label{modClfixRB}
\end{figure}

We first observe that, although the results of our model do not change appreciably with $R_B$, they do with $n$. To emphasize this last point, we plot in Figure \ref{modClfixRB} $C_{\ell}^{\rm mod}$ and  $C^{ \rm mod}(\theta)$  for different values $n$, and fixed $R_B$. We observe that the largest values of $n$ we have considered ($n=0.21$ and 
 $n=1/5$) produce a remarkably good qualitative agreement with data. Namely,  $C^{\rm mod}_{\ell}$ in these models nicely reproduces the ``valley'' in the data around $\ell\approx 25$, and the strong suppression at $\ell \lesssim 5$. Similarly, these large values of $n$ also produce a nice qualitative agreement with data for $C^{\rm mod}(\theta)$. Smaller $n$'s  ($n=1/6$ and $n=1/7$) do not seem to reproduce as well some of the features observed in data.

\begin{figure}[htp]
\centering
\includegraphics[width=.8\textwidth]{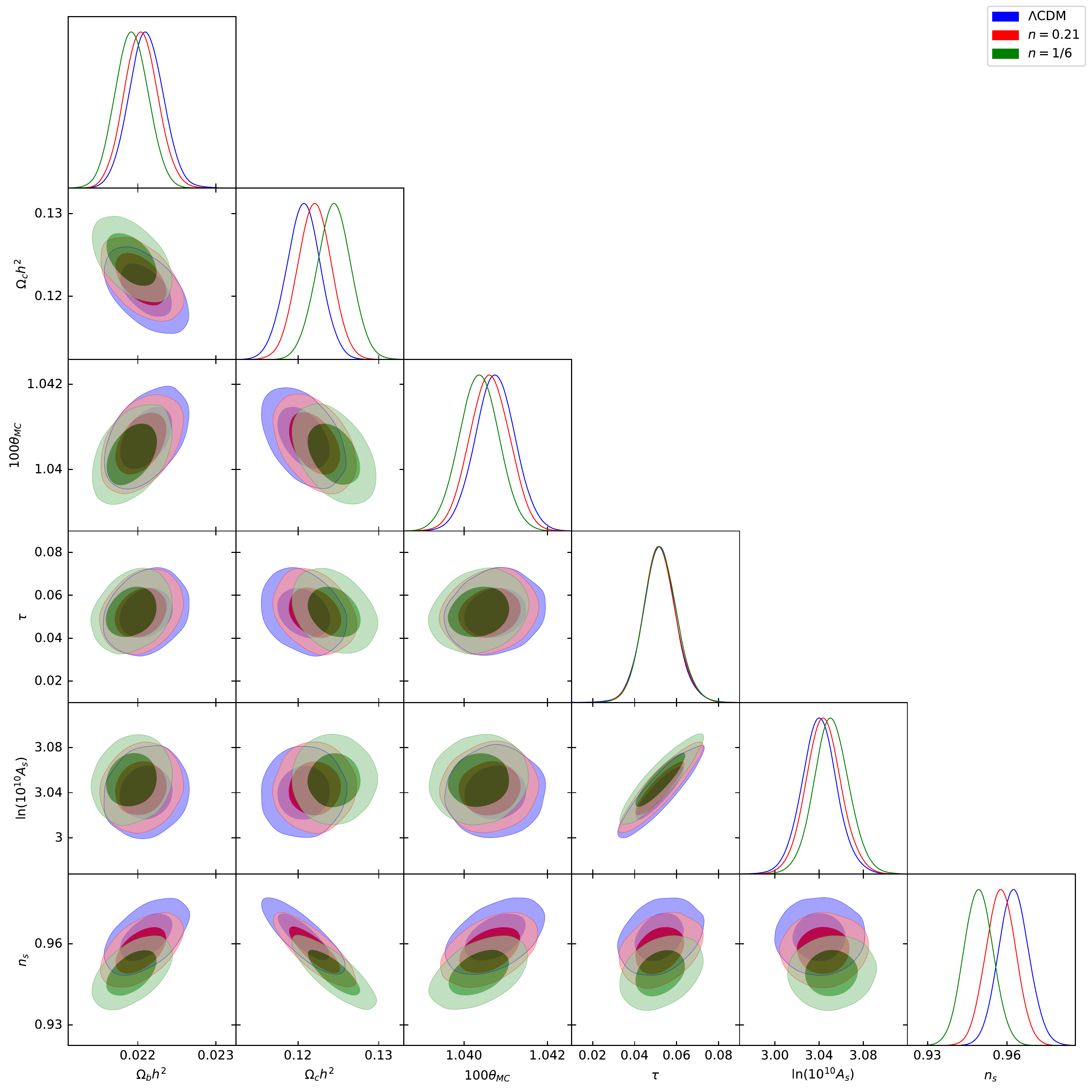}
\caption{Joint-probability distribution of the six cosmological parameters, {\it viz.}\ $\Omega_b\,h^2$, $\Omega_c h^2$, $100\theta_{MC}$, $\tau$, ${\rm{ln}}(10^{10} A_s)$ and $n_s$, for 
the $\Lambda$CDM model and two bouncing scenarios with $n=0.21$ and $n=1/6$, and $R_B=10^{-2}$ in Planck units.}
\label{tri6p}
\end{figure}

In order to make the discussion more quantitative, we have carried out a Markov chain Monte Carlo (MCMC) analysis (we use  the CosmoMC software \cite{Lewis:2002ah}), using Planck TT and low-$\ell$ $EE$ data \cite{Aghanim:2019ame}, where 
we vary the six standard cosmological parameters \cite{Weinberg_2008}, namely $\Omega_b$, $\Omega_c$, $\theta_{MC}$, $\tau$, $A_s$ and $n_s$---along with other nuisance parameters. We do not treat $n$, $R_B$ and $\mathfrak{f}_{_{\rm NL}}$ as free parameters in this likelihood analysis; they are rather thought of as fixed parameters that must be predicted by individual theories. 
 The results of this analysis are summarized in Table \ref{table:mcmc1} and Figure \ref{tri6p},  for two representative  bouncing models corresponding to $n = 0.21$ and $n = 1/6$. We also include the analysis corresponding to the $\Lambda$CDM model for comparison.   Table \ref{table:mcmc1} contains the mean and standard deviation of the marginalized posterior distribution of the six parameters.  The results show that the impact of the non-Gaussian modulation in the cosmological parameters is modest, since changes are of the order of a percent or less. More precisely, we find that the best-fit values  for $\Omega_b$, $\Omega_c$, $\theta_{MC}$, $\tau$, $A_s$ and $n_s$ for  $n=0.21$ differ from the $\Lambda$CDM  values by $0.3\%$, $1.1\%$, $0.01\%$, $0.25\%$, $0.13\%$ and $0.51\%$, respectively; while for $n=1/6$ we obtain $0.84\%$, $3.1\%$, $0.04\%$, $0.82\%$, $0.36\%$, $1.4\%$. The small change relative to $\Lambda$CDM is due to the fact that the differences introduced by our model are restricted to multipoles $\ell \lesssim 50 $. 
The discrepancies between the two models with  $n = 0.21$ and $n = 1/6$ arise from  the different way they produce the power suppression---while $C^{\rm mod}_{\ell}$ for $n = 0.21$ follow the data ``up and down'' quite well, the model with $n = 1/6$ produces a ``monotonic'' suppression. 

This difference is clearly manifested in the value of $\chi^2$  for the best fit values. While for $n=0.21$ we find a significant improvement of $\Delta\chi^2 = -6.4$ relative to $\Lambda$CDM, for $n=1/6$ we obtain $\chi^2$ worsens by $22.6$. Therefore, this analysis teaches us a valuable lesson: while all models discussed in the previous sub-subsection are able to reproduce the observed value of  $S_{1/2}$, they differ on the details, and large values of $n$ are significantly more favored by current data. Hence it is not enough to just focus  on $S_{1/2}$.

\begin{table}[]
    \centering
    \begin{tabular}{|c|c|c|c|}
    \hline
    Parameters & Standard model  & $n\,=\,0.21$ & n\, =\, 1/6\\
    \hline
    $\Omega_b h^2$ &  $0.0221 \pm 0.00022$  & 
    $0.0220 \pm 0.00022$ & 
    $0.02192 \pm 0.00021$
    \\ 
    \hline
    $\Omega_c h^2$ & $0.1207 \pm 0.0021$  & 
    $0.1221 \pm 0.0021$ & 
    $0.1245 \pm 0.0021$
    \\ 
    \hline
    $100\theta_{MC}$ & $1.0407 \pm 0.00048$ & 
    $1.0406 \pm 0.00048$ & 
    $1.0403 \pm 0.00048$
    \\ 
    \hline
    $\tau$ & $0.0519 \pm 0.0081$  & 
    $0.052 \pm 0.0080$ & 
    $0.0523 \pm 0.0080$
    \\ 
    \hline
    ${\rm{ln}}(10^{10} A_s)$ & $3.0401 \pm 0.0163$ & 
    $ 3.0441 \pm 0.0161$ & 
    $3.0512 \pm 0.0161$
    \\ 
    \hline
    $n_s$ &  $0.9626 \pm 0.0058$ & 
    $0.9577 \pm  0.0056$ & 
    $0.9493 \pm 0.0056$
    \\ 
    \hline
    \end{tabular}
    \caption{The mean and standard deviation of the posterior distribution of the six standard parameters in the $\Lambda$CDM model with almost scale invariant primordial density perturbations, and the two representative  bouncing models corresponding to $n = 0.21$ and $n = 1/6$, and $R_B=10^{-2}$ in Planck units. The differences in the mean values between the three columns are equal  to or smaller 
     than a few percent.}
    \label{table:mcmc1}
\end{table}

\subsection{Lensing parameter}
The lensing parameter $A_L$ scales the lensing spectrum. $A_L\, =\, 0$ removes the effects of lensing from the calculations, while $A_L = 1$  corresponds to the standard lensing  occurring in the universe. This parameter 
 was introduced
to provide a consistency test for cosmological models\cite{Calabrese:2008rt}. Namely, 
if $A_L$ is left as a free parameter, the best fit to data should be compatible with one.  The Planck collaboration \cite{Akrami:2018vks} has pointed out that  $A_L=1$ is more than two standard deviations away from the best-fit value obtained for the $\Lambda$CDM model. 

\begin{figure}[htp]
\centering
\includegraphics[width=.7\textwidth]{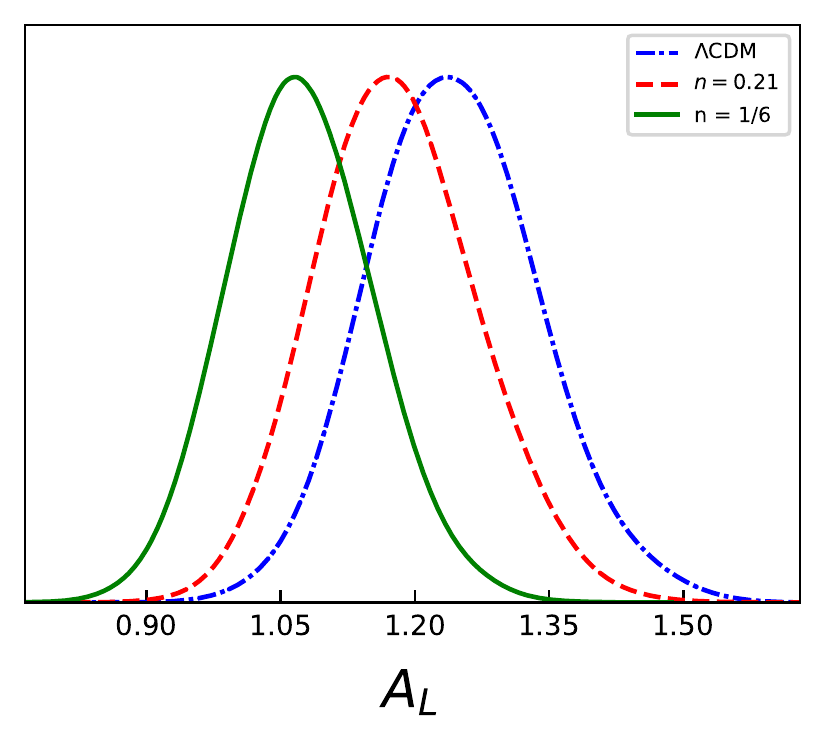}
\caption{Marginalized posterior probability distribution for the lensing amplitude $A_L$ for different models. As we can see, the mean values of the lensing amplitude are smaller for the bouncing models, in such a way that  $A_L= 1$ falls within 2-$\sigma$ of the posterior distribution for $n=0.21$, and within 1-$\sigma$ for $n=1/6$. 
\label{lensing}}
\end{figure}

The goal of this section is to extend the analysis of the previous section by including $A_L$ as a free parameter in our model. The  results of our MCMC analysis are shown in Table  \ref{table:mcmc2}, again for two representative bouncing models with $n=0.21$ and $n=1/6$, and $R_B=10^{-2}$ in Planck units. We find that, when the lensing parameter $A_L$ is included in the analysis, the best-fit values of six standard parameters change with respect to the values obtained in the previous subsection. This is not surprising, since similar changes also occur in the same analysis for the $\Lambda$CDM model. 
The most interesting result we find  is that the best fit for $A_L$ in our model is appreciably  smaller than in $\Lambda$CDM (see Figure \ref{lensing}). Namely, in $\Lambda$CDM one obtains $A_L=1.244 \pm 0.0961$. As mentioned before, this is 
 more than 2-$\sigma$ away from the desired  value  $A_L\,=\,1$ \cite{Aghanim:2018eyx}. In our model we obtain $A_L=1.179 \pm 0.0919$ for $n=0.21$,  and $A_L=1.072\pm 0.0826$ for $n = 1/6$. 
 These values are around  $5\%$ and $15\%$ away from the $\Lambda$CDM, and they bring the value $A_L=1$ within two and one standard deviations, respectively. In this sense, the tension of the $\Lambda$CDM model related to the lensing amplitude gets alleviated in our model, and the motivation to introduce spatial curvature \cite{DiValentino:2019qzk}---which is the source of several other tensions that produce a potential ``crisis in cosmlogy''---gets diluted. 
 
 The result we have found  for the lensing amplitude is a particular case of a general argument described in \cite{Ashtekar:2020gec} to relate the suppression of power at large angular scales and a lower value of $A_L$. This is because, as  shown in the previous section, the power suppression  causes an increase in the best-fit value of $A_s$. 
Since the value of $A_L$ is anti-correlated with $A_s$, a power suppression translates to a decrease in the value of $A_L$. 
 Similar changes in other parameters, such as decrease in mean values of $n_s$ and $\Omega_b h^2$, also  contribute to bringing $A_L$ closer to one.

\begin{table}[]
    \centering
    \begin{tabular}{|c|c|c|c|}
    \hline
    Parameters & Standard model  & $n\,=\,0.21$ & n\, =\, 1/6\\
    \hline
    $\Omega_b h^2$ & $0.0226 \pm 0.00029$  & 
    $0.0224 \pm 0.00028$ & 
    $0.02208 \pm 0.00028$
    \\ 
    \hline
    $\Omega_c h^2$ & $0.1166 \pm 0.0025$  & 
    $0.1189 \pm 0.0025$ & 
    $0.1230 \pm 0.0026$
    \\ 
    \hline
    $100\theta_{MC}$ & $1.0414 \pm 0.00052$ & 
    $1.0411 \pm 0.00053$ & 
    $1.0406 \pm 0.00053$
    \\ 
    \hline
    $\tau$ &  $0.0499 \pm 0.0085$ & 
    $0.0505 \pm 0.0085$  & 
    $0.0520 \pm 0.0080$
    \\ 
    \hline
    $A_L$ &  $1.244 \pm 0.0961$  & 
    $1.179 \pm 0.0919$  & 
    $1.072 \pm  0.0826$
    \\ 
    \hline
    ${\rm{ln}}(10^{10} A_s)$ & $3.027 \pm 0.0179$ &
    $3.034 \pm 0.0178$ & 
    $3.047 \pm 0.0169$
    \\ 
    \hline
    $n_s$ & $0.9742 \pm 0.0070$   & 
    $0.9664 \pm 0.0071$  & 
    $0.9530 \pm 0.0070$
    \\ 
    \hline
    \end{tabular}
    \caption{The mean and standard deviation of the marginalized posterior distribution of the seven parameters (six standard parameters and $A_L$) in the $\Lambda$CDM  model and the two bouncing models corresponding to $n = 0.21$ and $n = 1/6$. }
    \label{table:mcmc2}
\end{table}

\subsection{Parity asymmetry}\label{parity}

The $\Lambda$CDM model predicts a parity neutral universe. The parity of the primordial probability distribution of density perturbations can be tested by observing the CMB at very large angular scales ($2\leq \ell \leq 50$). This is the so-called Sachs-Wolfe plateau, which corresponds to long wavelengths that entered the Hubble radius later, and for which the primordial spectrum have not been significantly altered by late-time physics. However, an odd-parity preference has been observed in WMAP data \cite{Land:2005ad,Land:2005jq}, and it has been confirmed by Planck \cite{Ade:2015hxq}. This asymmetry has been quantified using the following estimator:

\be \label{RTT} R^{TT}(\ell_{\rm max})=\frac{D_+(\ell_{\rm max})}{D_-(\ell_{\rm max})}\, \ee
where $D_{+,-}(\ell_{\rm max})$ measures the power spectrum in even (+) or odd (-) multipoles up to $\ell_{\rm max}$. More precisely, 
\be \label{Dpm} D_{+,-}(\ell_{\rm max})=\frac{1}{\ell^{+,-}_{\rm tot}}\sum^{+,-}_{\ell=2,\ell_{\rm max}}\frac{\ell (\ell+1)}{2\pi}C_{\ell}\, , \ee
where 
$\ell^{+,-}_{\rm tot}$ is the total number of even (+) or odd (-) multiples included in the sum. The ratio $R^{TT}(\ell_{\rm max})$ reconstructed from CMB data for $\ell_{\rm max}\in [3,50]$  shows a clear odd-parity preference at large angular scales when compared to what is expected from the $\Lambda$CDM model (see \cite{Land:2005ad,Land:2005jq} for details of the observations). 

\begin{figure}[htp]
\centering
\includegraphics[width=.45\textwidth]{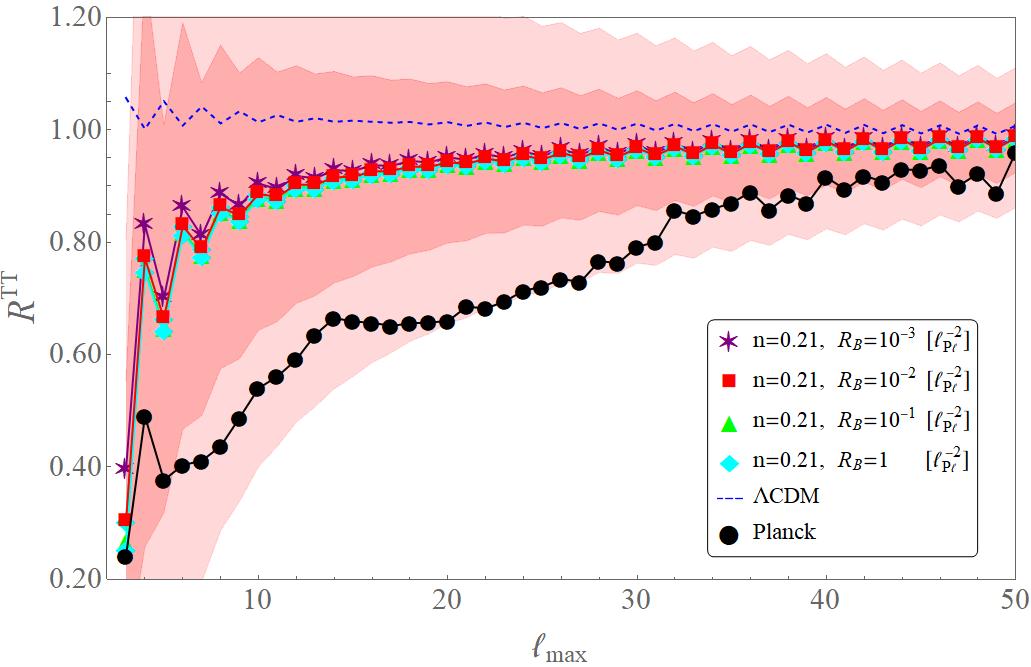}\quad
\includegraphics[width=.45\textwidth]{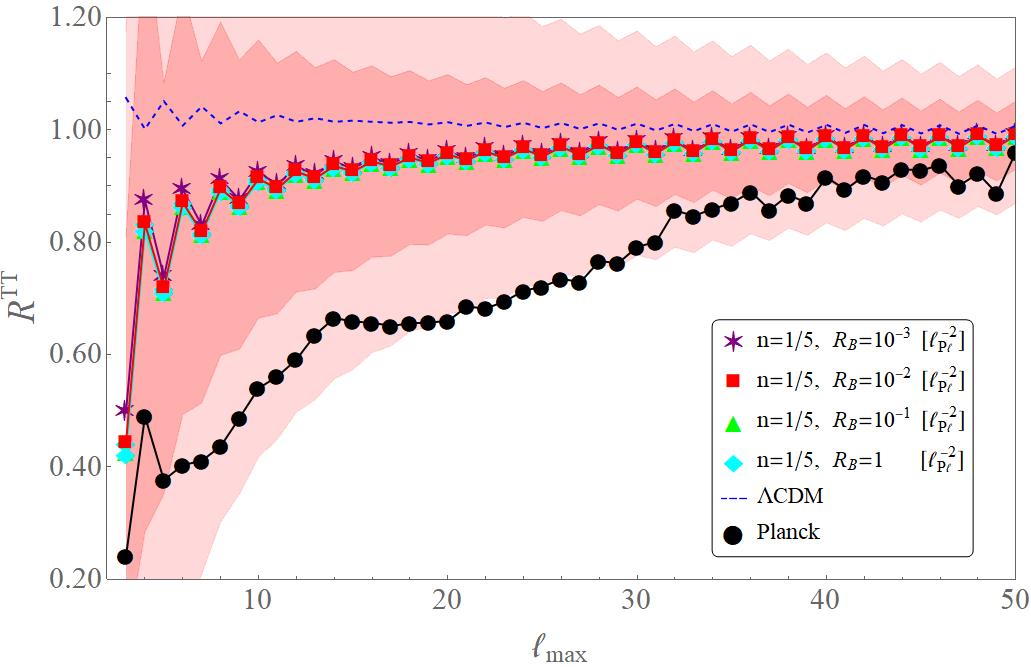}\quad
\medskip
\includegraphics[width=.45\textwidth]{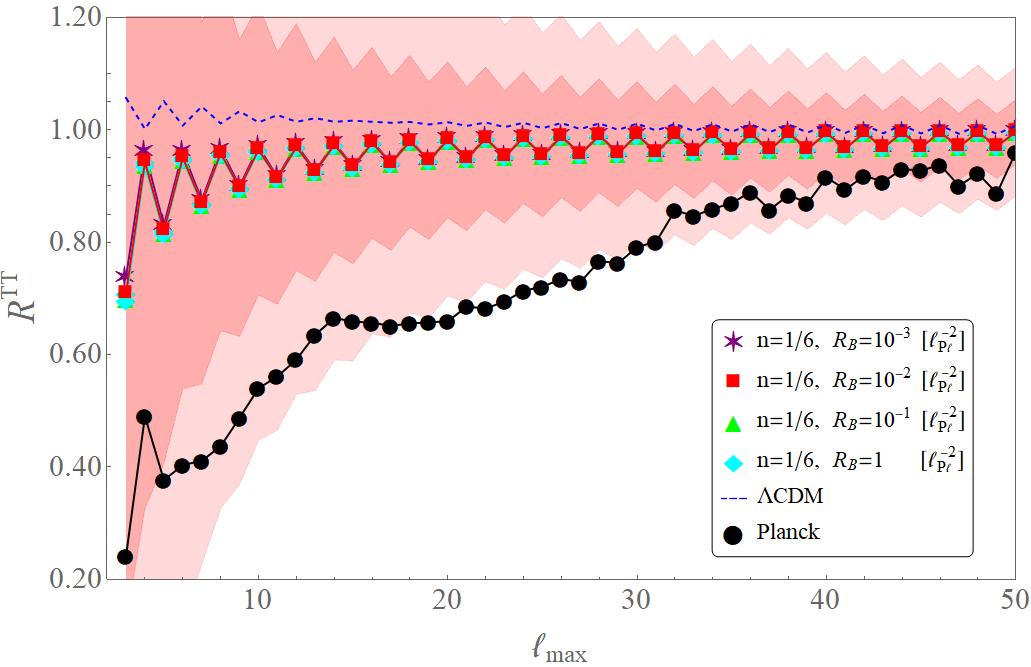}
\includegraphics[width=.45\textwidth]{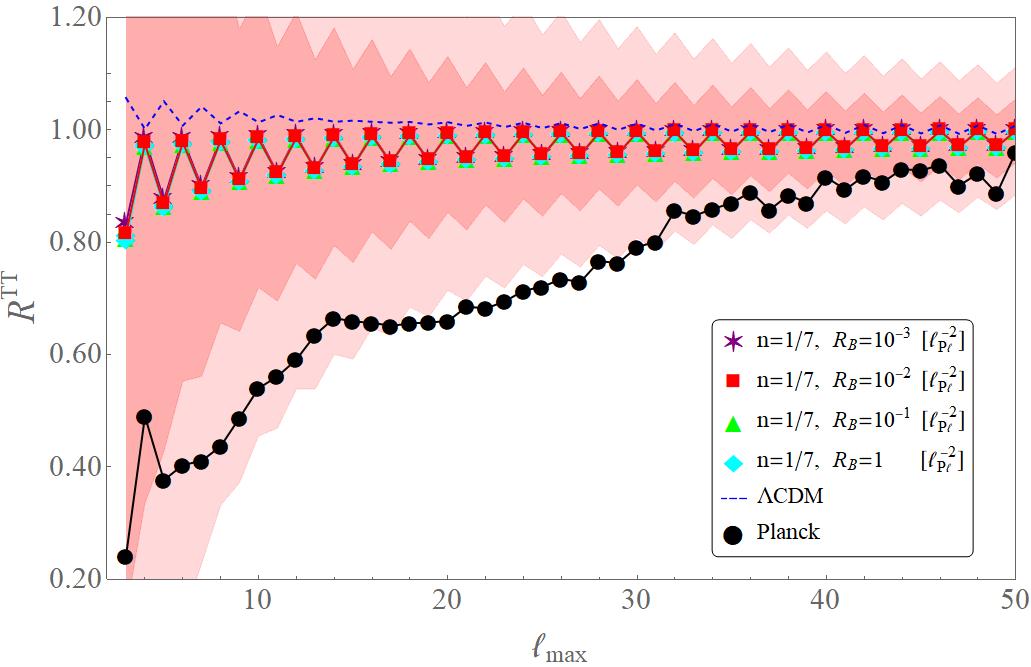}\quad
\caption{$R^{TT}$ vs $\ell_{max}$  predicted from our model for different values of $n$ and $R_B$. The blue-dashed line shows the result of  the $\Lambda$CDM model  with   almost scale invariant primordial perturbations, and the black dots correspond to data from Planck. We have also included the 1-$\sigma$ and $2-\sigma$  cosmic variance contours (shadowed regions) of our predictions for $R_B=10^{-2}\ell_{\rm Pl}^{-2}$. We observe that our model produces a clear preference for odd parity modes (i.e.\, $R^{TT}<1$), and closer to data for  large $n$.}
\label{Parity}
\end{figure}

We analyze in this section whether our model produces also an odd-parity asymmetry. As in the previous sections, we will evaluate this asymmetry for realizations of the primordial probability distribution for which the angular power spectrum contains a monopolar suppression that agrees with the observed value of  $S_{1/2}$. This can be done by simply inserting in equation (\ref{RTT}) the spectra $C^{\rm mod}_{\ell}$ computed above (and that is shown in Figure \ref{modCl}). We plot in Figure \ref{Parity} $R^{TT}(\ell_{\rm max})$ versus $\ell_{\rm max}$ in the range $[3,50]$ for different values of $n$ and $R_B$. The prediction of $\Lambda$CDM together with data from Planck, are also shown for comparison.  Overall, our model produces a clear preference for odd-parity correlations at low multipoles (i.e.\ $R^{TT}<1$), in contrast to the $\Lambda$CDM model. As in the previous subsections, we also observe that large values of $n$ do better, and in particular for $n=0.21$ and $n=1/5$ data remains within the 2-$\sigma$ region in the entire range of $\ell_{\rm max}$.

The origin of the odd-parity preference can be easily understood by looking at Figure \ref{modCl}. It is easy to understand from the definitions (\ref{Dpm})  
that a positive slope  of the curve $\frac{\ell (\ell+1)}{2\pi}C_{\ell}$  vs $\ell$  for small $\ell$'s gives rise to $D_{-}(\ell_{\rm max})>D_{+}(\ell_{\rm max})$ in that range. While this slope is negative in the $\Lambda$CDM theory for $\ell <30$, it is  positive  in our model. Hence, the odd-parity preference in our model is a consequence of  the suppression of the angular power spectrum for low $\ell's$.

\section{Dipolar modulation}\label{dipolar}

A dipolar asymmetry in the CMB was first reported in the early WMAP data releases \cite{Eriksen:2003db}, and since then it has been consistently found in all CMB maps, including  Planck  data  \cite{Ade:2013kta,Ade:2015hxq,Akrami:2019bkn} (and even found a posteriori in  data from COBE \cite{Eriksen:2003db}). The accumulated evidence makes it difficult to attribute this asymmetry to residual systematics or foregrounds, and at the present time there is little dispute about the fact that it is a real feature in the CMB. 
The observed asymmetry has a  peculiarity, shared by the power suppression discussed above: it is only observed in correlations involving large angular separations or, equivalently, low multipoles $\ell$. This scale-dependence facilitates to distinguish it from the Doppler-generated dipole caused by our relative velocity to the cosmological rest frame. But on the other hand, it implies that the physical mechanism causing it---assuming it is not a statistical fluke---affects only the lowest multipoles $\ell$. It has been proven difficult to come out with theoretical explanations able to accommodate this scale-dependence, while at the same time respecting the existing constrains on anisotropies coming from the quadrupolar modulation \cite{Akrami:2018odb} (see \cite{Dai:2013kfa} for a summary of some ideas in the literature). 
 
The Planck team has measured the dipolar asymmetry in several ways, and has quantified its amplitude and direction. The results are compatible with previous findings by WMAP. The amplitude can be conveniently parametrized by means of the BipoSH coefficients discussed above. A dipolar modulation contributes to $A_{\ell\ell'}^{LM}$ for $L=1$, which is different from zero only for $\ell'=\ell+1$ due to the   properties of the Clebsch-Gordan coefficients.  
Furthermore, one can write $A_{\ell\ell+1}^{1M}$ in terms of three coefficients $m_{1M}$, with $M=-1,0,1$, defined as 
\be \label{aa} A_{\ell\ell+1}^{1M}\equiv m_{1M}\, G^1_{\ell\ell+1},\, \hspace{1cm} {\rm where} \hspace{1cm}  G^1_{\ell\ell+1}\equiv (C_{\ell} +C_{\ell+1}) \sqrt{\frac{(2\ell+1)(2\ell+3)}{4\pi\, 3}}\, C^{10}_{\ell,0,\ell+1,0}\, .\ee
$G^1_{\ell\ell+1}$ is the so-called  form factor for a dipolar modulation (see e.g.\ \cite{Ade:2015hxq} and Appendix \ref{BipoSH} below).  Defined in this way, $m_{1M}$ is a function of  $\ell$. The Planck collaboration   has reported the amplitude of the dipolar modulation in terms of the value of the function  $A_1(\ell)$, defined as \cite{Ade:2015hxq}
\be \label{bb} A_1\equiv \frac{3}{2}\sqrt{\frac{1}{3\pi} (|m_{11}|^2+|m_{10}|^2+|m_{1-1}|^2)} \, .\ee
The modulation signal is reconstructed in \cite{Ade:2015hxq} in non-overlapping bins of width $\Delta \ell=64$, up to $\ell_{\rm max}=512$, and the observed amplitude $A_1$  deviates significantly ($\sim 3$ standard deviations) from what is expected from an isotropic distribution  only in the first bin $\ell \in [2,64]$. The signal is  compatible with zero within $2\sigma$ for all the higher $\ell$-bins. The observed value of $A_1$ in the lower $\ell$-bin is very similar for the four maps used by Planck, and its average value is $A^{\rm obs}_1=0.068\pm 0.023$. The direction of the dipole is also in agreement between the different component separation maps,  and   with previous observations, and it lies close to the ecliptic axis.  Other ways of determining the amplitude and direction of the dipole modulation provide compatible results.

 The simplest phenomenological parametrization of a dipolar modulation can be obtained by adding a dipole to an otherwise isotropic temperature distribution, $\delta T  (\hat n)=\delta T^{\rm iso} (\hat n)\, (1+d \, \hat n\cdot \hat p)$, where  $\delta T^{\rm iso} (\hat n)$ is a statistically isotropic distribution, $\hat p$ indicates the direction of the dipole, and  $d$ its amplitude. However, as we discuss in Appendix \ref{BipoSH}, this simple model predicts a modulation with uniform amplitude over all scales in the sky, i.e.\ scale-independent, in sharp contrast with data. A more elaborate model 
 is needed to account for the observations.

As discussed in section \ref{sec:NGMod}, our model cannot predict the direction of the dipole modulation, since it is randomly generated in the concrete realization that we observe. But we can make probabilistic statements about its amplitude. As discussed in the previous subsection, we focus on realizations producing the observed value of $S_{1/2}$, and investigate  whether these realizations also give rise to a dipolar modulation compatible with observations. This is a non-trivial demand, since the model needs to account both for the amplitude and the scale dependence of the dipole, with no additional parameters to play with.

The prediction for the dipole amplitude $A_1(\ell)$ is obtained  by first computing the expected  square value of the coefficients $m_{1M}$ from the BipoSH coefficients \be \langle |m_{1M}(\ell)|^2\rangle =\langle|A^{1 M}_{\ell \ell' }|^2\rangle/(G^1_{\ell\ell+1})^2 \, , \ee where $\langle|A^{1 M}_{\ell \ell' }|^2\rangle$  can be computed from  (\ref{varalpha}). The resulting expression for $A_1(\ell)$ is  
\be A_1(\ell)=\frac{3}{2}\sqrt{\frac{1}{3\pi} \sum_M  \langle |m_{ 1M}|^2\rangle}=\frac{3}{2}\frac{1}{\sqrt{\pi}}\frac{1}{C_{\ell}^{ \rm mod} +C_{\ell+1}^{ \rm mod}} \sqrt{\frac{1}{2\pi}\,  \int dq\, q^2 \, P_{\phi}(q) \, |\mathcal{C}_{\ell \ell+1}^1(q)|^2}\, ,\ee 
where $\mathcal{C}_{\ell \ell'}^L(q)$ was defined in (\ref{Ctilde}), and it contains the information about the primordial  non-Gaussianity. As emphasized in \cite{Ade:2013kta,Ade:2015hxq}, the form factors $G^1_{\ell\ell+1}$ must be computed by using the modulated power spectrum $C_{\ell}^{\rm mod}$ that we show in Figure \ref{modCl}, since we want to evaluate the amplitude of the dipole on the {\em observed} power spectrum. We observe that  the power suppression enhances the  relevance of the dipole. 

\begin{figure}[htp]
\centering
\includegraphics[width=.485\textwidth]{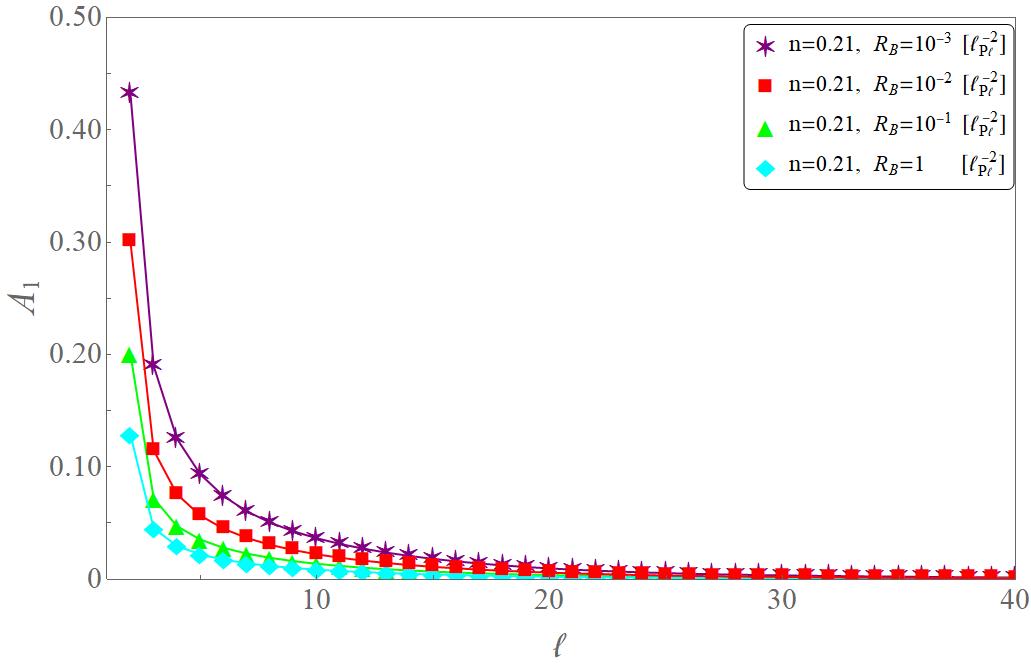}\quad
\includegraphics[width=.485\textwidth]{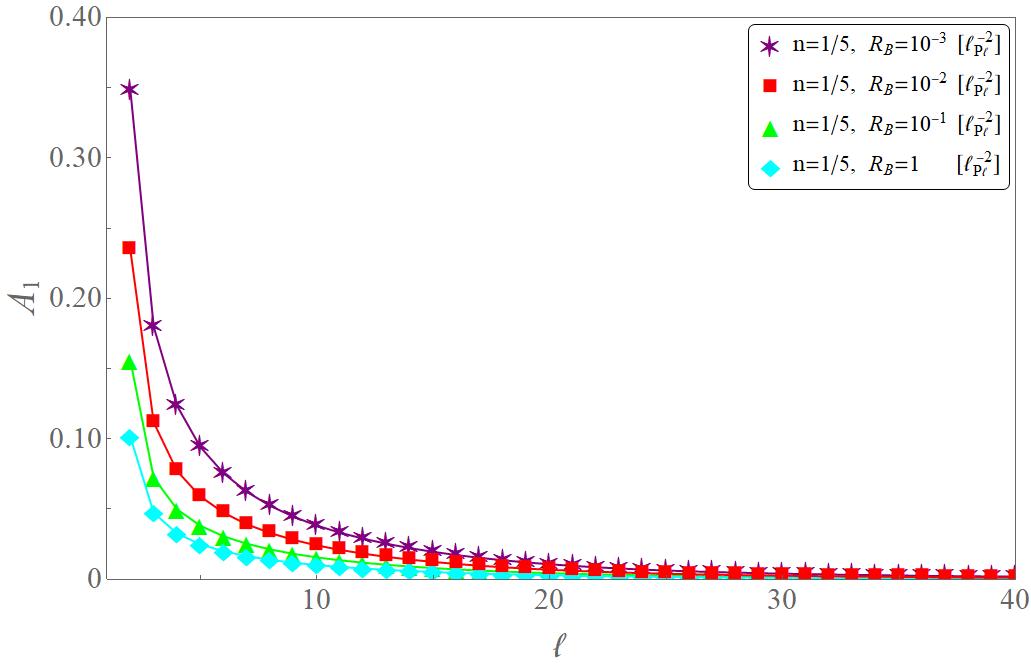}\quad
\medskip
\includegraphics[width=.485\textwidth]{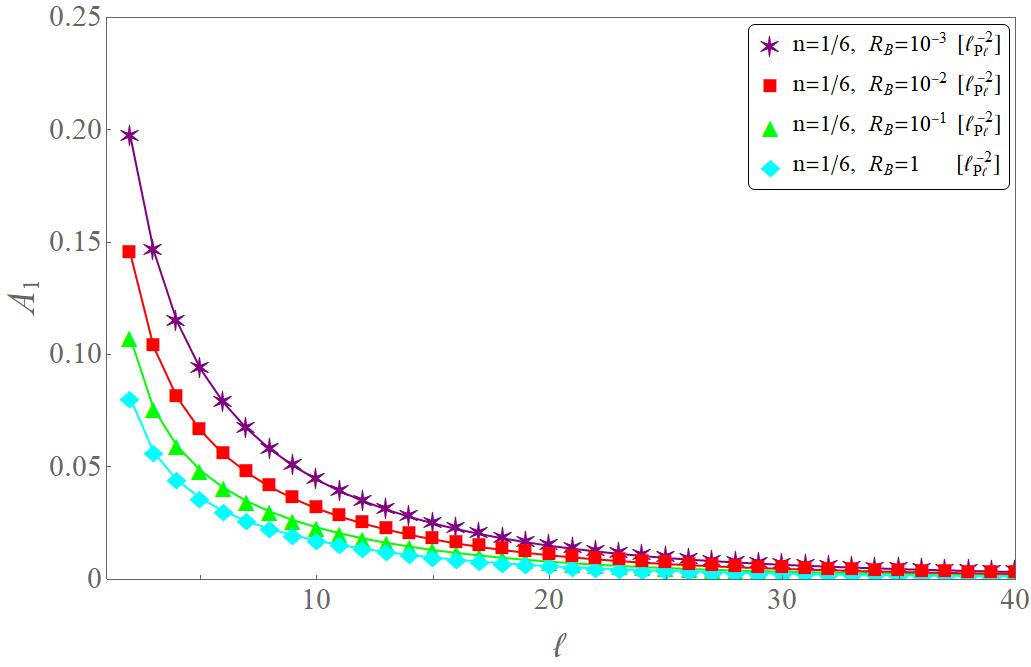}
\includegraphics[width=.485\textwidth]{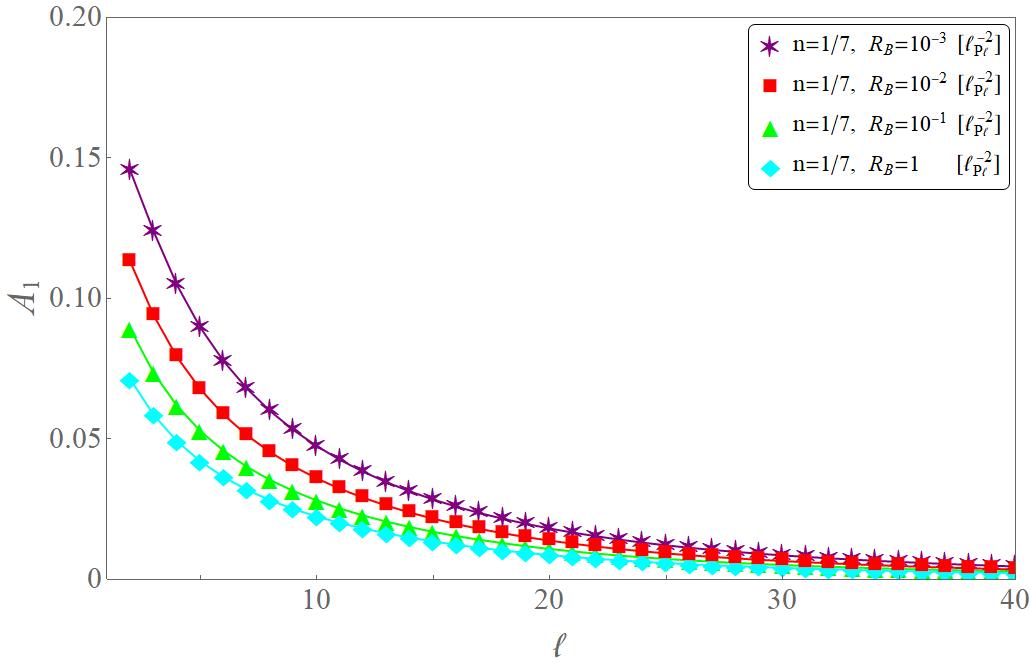}\quad

\caption{Amplitude of the dipolar modulation  $A_1(\ell)$  obtained from our model, for different values of $n$ and $R_B$, the latter in Planck units. Each plot shows the result for a fixed value of $n$.} 
\label{A1many}
\end{figure}

\begin{figure}[htp]
\centering
\includegraphics[width=.7\textwidth]{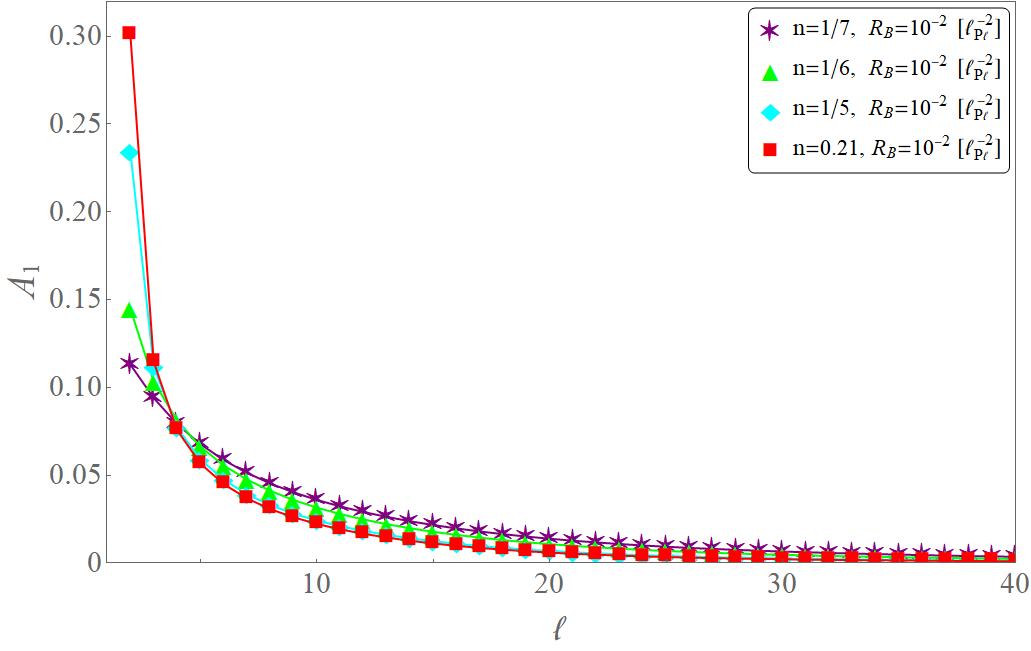}
\caption{Amplitude of the dipolar modulation  $A_1(\ell)$ for different values of $n$ and $R_B=10^{-2}$, the latter in Planck units.} 
\label{A1}
\end{figure}

We show in Figures \ref{A1many} and \ref{A1} our results for  $A_1(\ell)$ for different values of the parameters $n$ and $R_B$. 
As one could expect, the amplitude of the dipole $A_1(\ell)$ decreases with $\ell$, as a consequence of the scale-dependence that the bounce introduces in the primordial   power spectrum and non-Gaussianity. The value of $A_1(\ell)$ is significant only for low multipoles $\ell \lesssim 30$ and shows an amplitude and scale dependence in consonance with observations. As in previous sections, larger values of $n$ produce a mean amplitude that is closer to the value reported in  \cite{Ade:2015hxq},  $A^{\rm obs}_1=0.068\pm 0.023$ for $\ell<64$. However, the observations  \cite{Ade:2015hxq} do not resolve the way $A_1(\ell)$ varies inside the bin $\ell\in [2,64]$, and consequently a more quantitative comparison is not possible at present. Our model predicts that most of the contribution to $A^{\rm obs}_1$ comes from the lowest multipoles $\ell \lesssim 30$. \\

\section{Quadrupolar modulation}\label{quadrupolar}

We discuss in this section  the quadrupolar modulation that  comes together in our model with the monopolar and dipolar effects discussed in the previous two sections. The Planck satellite has looked for a quadrupolar modulation in the CMB and,  although they found a signal, it is not significant enough to claim the detection of new phsyics, but rather it is consistent with statistical fluctuations in an isotropic universe \cite{Akrami:2018odb}---i.e. the $p$-value of the observed quadrupole is large (see table 17 in  \cite{Akrami:2018odb}). These observation  impose constraints in our model:  the predicted amplitude for the quadrupolar modulation must not be larger than the signal found by Planck. We investigate in this section if  this is the case.%

First, we discuss the results from Planck.  Ref.\ \cite{Akrami:2018odb}  does not report the observed quadrupolar modulation in a model-independent manner, but it rather  assumes that the primordial power spectrum of comoving curvature perturbations contains a quadrupolar contribution of the form
\be \label{Pquadmod} P_{\mathcal{R}}(\vec k)=P^{0}_{\mathcal{R}}(k)\, \left( 1+\sum_{M=-2}^{2} g_{2M}\, (k/k_0)^{r}\, Y_{2M}(\hat k)\right)\, ,\ee
where $P^{0}_{\mathcal{R}}(k)$ is isotropic, and $k_0=0.05\, {\rm Mpc}^{-1}$ is a reference scale, where the $k$ dependence is restricted to a simple power law. Hence, this  model is parameterized  by the amplitudes $g_{2M}$ and the  tilt $r$. The analysis in \cite{Akrami:2018odb} obtains the best-fit value for  the total amplitude $g_2\equiv\sqrt{ \sum_M |g_{2M}|^2/5}$, for different values of $r$. The results for $r=-2,-1$ are  $g_2=3.30\times 10^{-5}$, and $4.34 \times 10^{-3}$, respectively.\footnote{Reference  \cite{Akrami:2018odb} also reports the value of  $g_2$ for scale-invariant and blue-tilted quadrupoles, i.e.\ zero and positive values of  $r$, respectively. However, these values  are not of interest to us, since our model produces instead a red-tilted quadrupole.} 

In order to compare these results with the predictions of our model, it is more convenient to translate the constraints on $g_2$ reported by Planck to the language of BipoSH coefficients. A quadrupolar modulation in the primordial power spectrum  affects only the BipoSH coefficient $A^{LM}_{\ell \ell'}$ with $L=2$ and $\ell'=\ell+2$. 
However, since our model {\it cannot be}  recast as a primordial spectrum of the form (\ref{Pquadmod}) with a power law scale-dependence, the comparison with Planck's results  is only qualitative.  Our  goal here is simply to compare orders of magnitude, rather than details. One would have to use ``raw'' data, not biased by the assumption (\ref{Pquadmod}), in order to make a quantitative comparison of our predictions with observations---but such analysis is beyond the scope of this work.

\begin{figure}[htp]
\centering
\includegraphics[width=.485\textwidth]{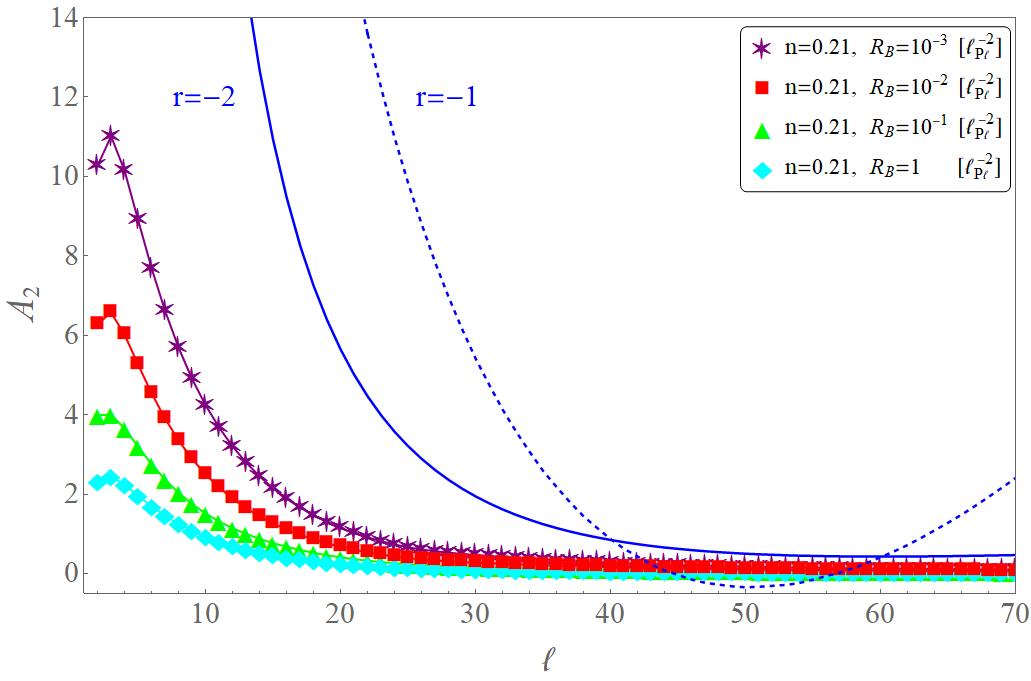}\quad
\includegraphics[width=.485\textwidth]{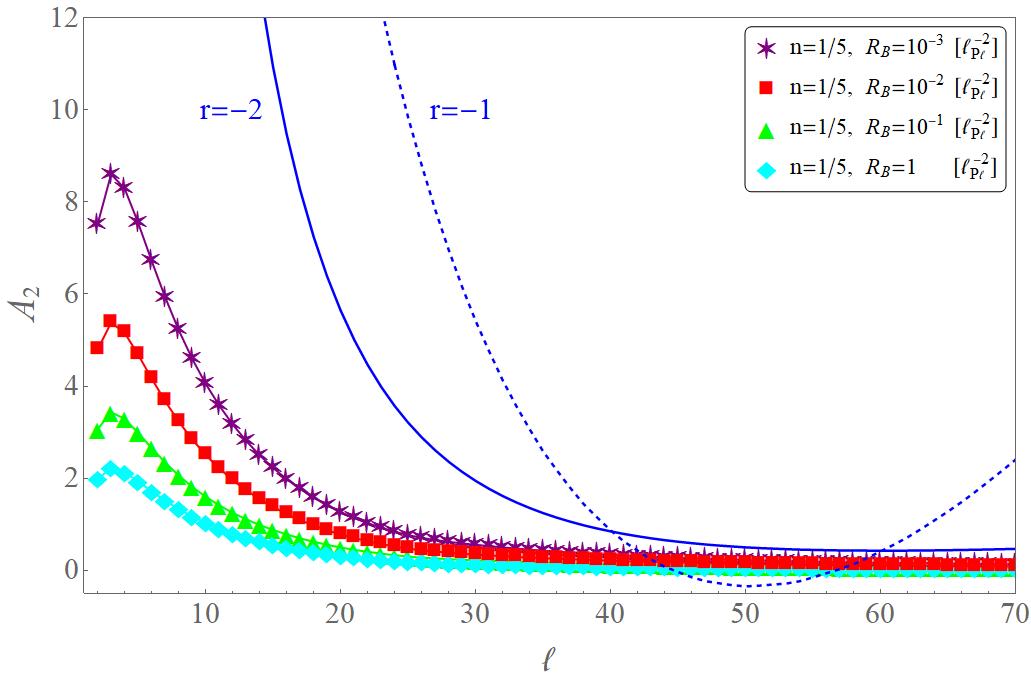}\quad
\medskip
\includegraphics[width=.485\textwidth]{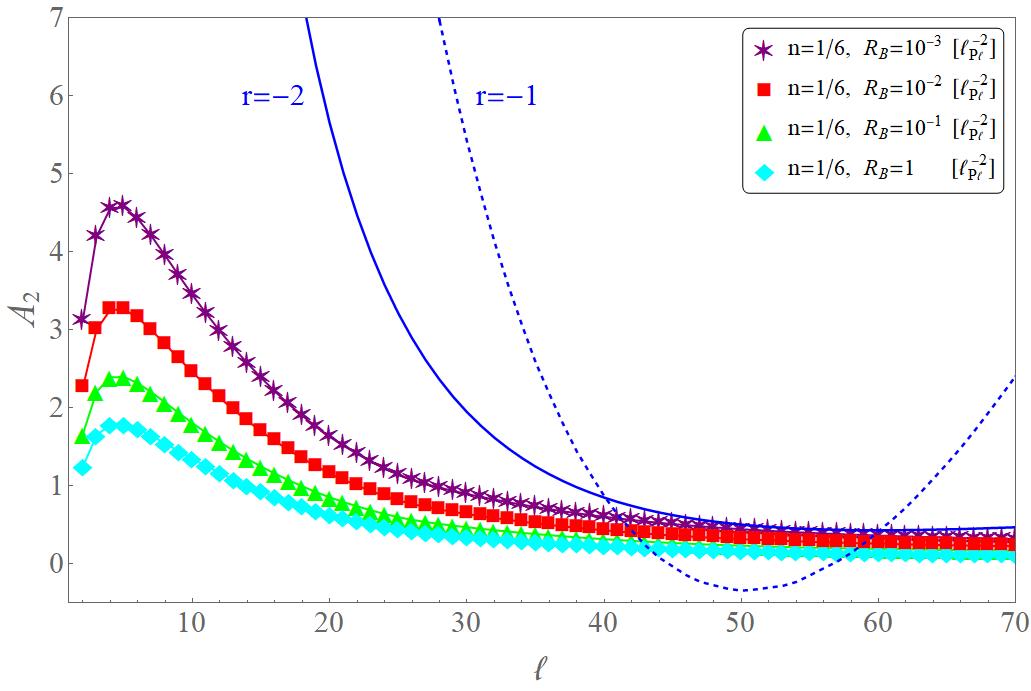}
\includegraphics[width=.485\textwidth]{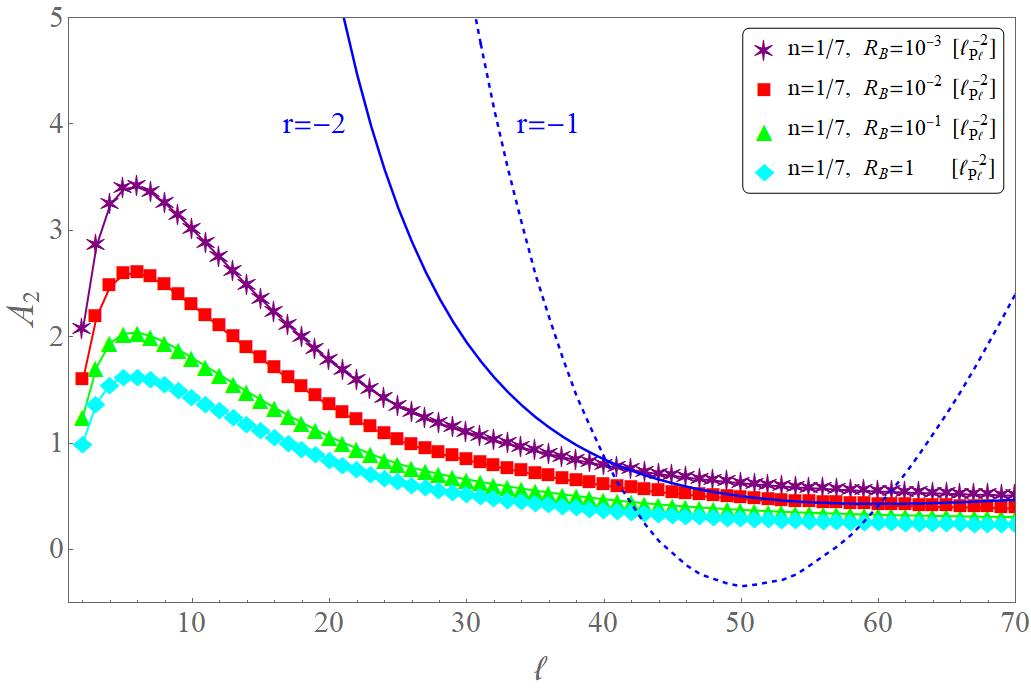}\quad
\caption{Results for the amplitude of the quadrupolar modulation  $A_2(\ell)$  predicted from our model, for different values of $n$ and $R_B$. The results from Planck using the two phenomenological models described in the main text for $r=-1$ and $r=-2$ are shown through the solid and dashed blue lines, respectively.}
\label{A2}
\end{figure}

The value for $A^{2M}_{\ell \ell+2}$ from the model (\ref{Pquadmod}) used by Planck can be written in terms of $g_{2M}$ and $r$ as (see Appendix \ref{BipoSH})
\be A^{2M}_{\ell \ell+2}=(3/5)^2 \,  C^{20}_{\ell 0 \ell+2 0}\,  \sqrt{\frac{(2\ell+1)(2\ell+5)}{4\pi\, 5}}\ \left[\frac{2}{\pi}\int dk_1  k_1^2 \,  \Delta_{\ell}(k_1)\Delta_{\ell+2}(k_1) \,P^{0}_{\mathcal{R}}(k_1)\, g_{2M}   \left(\frac{k_{\ma 1}}{k_0}\right)^{r}\,  \right] \, . \nonumber \ee
From this, and by assuming $g_{2M}\approx g_2$ for all $M$, the value of $A^{2M}_{\ell \ell+2}$ can be estimated  by simply replacing in the previous expression $g_{2M} $ by Planck's results for  $g_{2} $. 
For an easier comparison with our model, we define the quantity  $A_2(\ell)\equiv \sqrt{\frac{1}{5}\sum_M |A^{2M}_{\ell \ell+2}|^2} \times \Big( C^{20}_{\ell 0 \ell+2 0}\,  \sqrt{\frac{(2\ell+1)(2\ell+5)}{4\pi\, 5}}\Big)^{-1}$, and  plot the results from Planck for  $r=-2$ and $r=-1$ in  Figure \ref{A2}.  

On the other hand, the prediction of our model for  $A_2(\ell)$ can be directly obtained from expression (\ref{varalpha}) with $L=2$. As in  previous sections, we obtain the results for different values of the parameters $n$ and $R_B$, for realizations that produce a value of $S_{1/2}$ in agreement with observations.  We  also plot the results in Figure \ref{A2}. We see that for all values of $n$ and $R_B$ that we have considered, the predictions of our model are well below  Planck's observations for low multipoles $\ell$, and are of the same size for $\ell \sim 50$. For larger values of $\ell$ our predictions are also below the results from Planck for both for  $r=-2$ and $r=-1$. (We also notice that our predictions are above Planck's results for $r=-1$ around $\ell \sim 50$, although only  for a small window.) 
Given that, as discussed above, the models used by Planck are different from ours, the only conclusion we can extract from this qualitative analysis  is that the predictions from our model are not in conflict with Planck's constraints for the quadrupole, but rather reinforce the interpretation of the  small quadrupolar modulation in the CMB as a statistical fluke.   However, a comparison with unbiased data  would be needed in order to derive more precise conclusions. It is possible that our model can also provide a physical origin for the observed quadrupole.  This is an exciting possibility that we will further investigate in future work.

We have also checked that the BipoSH coefficient with $L=3,4,\cdots$ in our model are all smaller than for $L=2$, and therefore no observable higher order modulations are predicted, in consonance with observations.

\section{Validity of the perturbative expansion}\label{validitypert}

This paper rests on the idea that non-Gaussianity can affect significantly the form of the power spectrum at large angular scales. We have restricted ourselves to leading order non-Gaussianity, which means that we have used perturbation theory at the  next-to-leading-order.  
A natural question is whether this truncation is  justified or, on the contrary, perturbation theory breaks down in our model. The goal of this section is to address this point, and to show that perturbation theory remains under control.  

We will separate the analysis into two steps. In the first one, we will not consider the non-Gaussian modulation in the way we did in section \ref{sec:NGMod}, but rather we will consider the quantum theory of the primordial perturbations and ask whether the large values of $\mathfrak{f}_{_{\rm NL}}\sim 10^3$ advocated in the previous section for super-horizon modes jeopardizes the validity of the perturbative expansion. We will address this question by computing the corrections to the two-point function of comoving curvature perturbations originated from  next-to-leading-order terms. These corrections can be evaluated following the analysis of \cite{Maldacena:2002vr} (see also \cite{Agullo:2017eyh}), and they can be codified in a correction  $\Delta \mathcal{P}_{\mathcal{R}}(k)$ to the primordial power spectrum $\mathcal{P}_{\mathcal{R}}(k)$, defined as 
\bea \label{PDP}
\langle 0|\h{\mathcal{R}}_{\vec k_1} \h{\mathcal{R}}_{\vec k_2}|0\rangle &=& (2\pi)^3 \delta^{(3)}(\vec k_1+\vec k_2)\, \f{2\pi^2}{k_1^3}  \,\left[ \mathcal{P}_{\mathcal{R}}(k_1)+ \, \Delta \mathcal{P}_{\mathcal{R}}(k_1)\right] \, .\eea
Explicit calculations produce (see Appendix C, and \cite{Agullo:2017eyh} for further details)
\bea \label{DeltaP} \Delta \mathcal{P}_{\mathcal{R}}(k_1)&=&\f{k_1^3}{\pi^2}\, \Bigg[  \l(-\f{a}{z}\r)^3\, \left[-\f{3}{2}+3\f{V_{\varphi}\, a^5}{\kappa\, \pp\, \pi_a}+\f{\kappa}{4}\f{z^2}{a^2}\right] \int\f{\d^3p}{(2\pi)^3}  \, B_{\delta\varphi}(\vec{k}_1,\v p, -\v k_1-\v p)\,  \nonumber \\ &+&  \l(-\f{a}{z}\r)^4\, \left[-\f{3}{2}+3\f{V_{\varphi}\, a^5}{\kappa\, \pp\, \pi_a}+\f{\kappa}{4}\f{z^2}{a^2}\right]^2 \,   \int\f{\d^3p}{(2\pi)^3}\,  \left|\frac{v_p}{a}\right|^2 \, \left|\frac{v_{| \v k_1-\v p|}}{a}\right|^2\Bigg],
\eea
where  all quantities are evaluated at the end of inflation. $B_{\delta\varphi}(\vec{k}_1,\v k_2, \v k_3)$ is the Bispectrum for inflaton perturbations in the comoving gauge, written in  Appendix \ref{pertsham}, and $p_{\varphi}=a^3 \dot \varphi $ and $\pi_a=-6\ \kappa\, a\, \dot a$ are the conjugate momenta of the scalar field $\varphi$ and the scale factor $a$, respectively. The two-point function (\ref{PDP}) is proportional to $\delta^{(3)}(\vec k_1+\vec k_2)$ as a consequence of the underlying homogeneity. The difference with the calculation done in section \ref{sec:NGMod}, is that we are not fixing here any spectator mode, but applying quantum averages on all modes. 

The value of $\Delta \mathcal{P}_{\mathcal{R}}$ in our model is  dominated  by 
the first line in equation (\ref{DeltaP}), which contains the contribution of the non-Gaussianity generated by the bounce. The second term is the so-called ``field redefinition term'' \cite{Maldacena:2002vr}. We have numerically evaluated  $\Delta \mathcal{P}_{\mathcal{R}}$ in our model by using the form of the non-Gaussianity written in  (\ref{fNL}). As a representative sample of our result,  we show in Figure \ref{fig:vopt} the ratio $|\Delta \mathcal{P}_{\mathcal{R}}/\mathcal{P}_{\mathcal{R}}|$ for a bounce characterized by $n =0.21$ and $R_B =1$ in  Planck units. This calculation is done by using a unit amplitude for the non-Gaussiaity, $\mathfrak{f}_{_{\rm NL}}\,=\, 1$---hence, to obtain the actual  result we need to multiply by the values of $\mathfrak{f}_{_{\rm NL}}$ given in Table \ref{T1}. Figure \ref{fig:vopt} shows that $|\Delta \mathcal{P}_{\mathcal{R}}/\mathcal{P}_{\mathcal{R}}|$ is of order $10^{-7}$, so it remains much less than one even after multiplying by  $\mathfrak{f}_{_{\rm NL}}\sim 10^3$.

The values in Figure \ref{fig:vopt} can be qualitatively understood as follows.  The first line in (\ref{DeltaP}) can be approximated by noticing that the terms in the square bracket are proportional to the slow-roll parameters during inflation  (that we generically denote by $\epsilon$) and that the Bispectrum is of the order of $\mathfrak{f}_{_{\rm NL}}\,\mathcal{P}_{\mathcal{R}}^2 $.  Hence, the first line is proportional to $\epsilon\,\mathfrak{f}_{_{\rm NL}}\,\mathcal{P}_{\mathcal{R}}^2 $, where $\mathcal{P}_{\mathcal{R}}$ should be understood as the average value of the power spectrum among all wavenumbers involved in the calculation. 
The second line in (\ref{DeltaP}), on the other hand, is of the order of $\epsilon^2\,\mathcal{P}_{\mathcal{R}}^2$. 
For $\mathfrak{f}_{_{\rm NL}}\ge 1$ and $\epsilon \sim 10^{-2}$, it is obvious that the first line in (\ref{DeltaP}) dominates. Hence, $\Delta \mathcal{P}_{\mathcal{R}}(k_1)/\mathcal{P}_{\mathcal{R}}(k_1) \sim \epsilon\, \mathcal{P}_{\mathcal{R}}$ for  $\mathfrak{f}_{_{\rm NL}}=1$. 
Higher order contributions introduce additional powers of $\mathcal{P}_{\mathcal{R}}$, and since this quantity is always much smaller than one in our model, these contributions are negligibly small. In this sense,  one can intuitively think about $\mathcal{P}_{\mathcal{R}}$ as the small ``parameter'' that makes the perturbative expansion well defined. 

\bfig
 \ig[width=.7\textwidth]{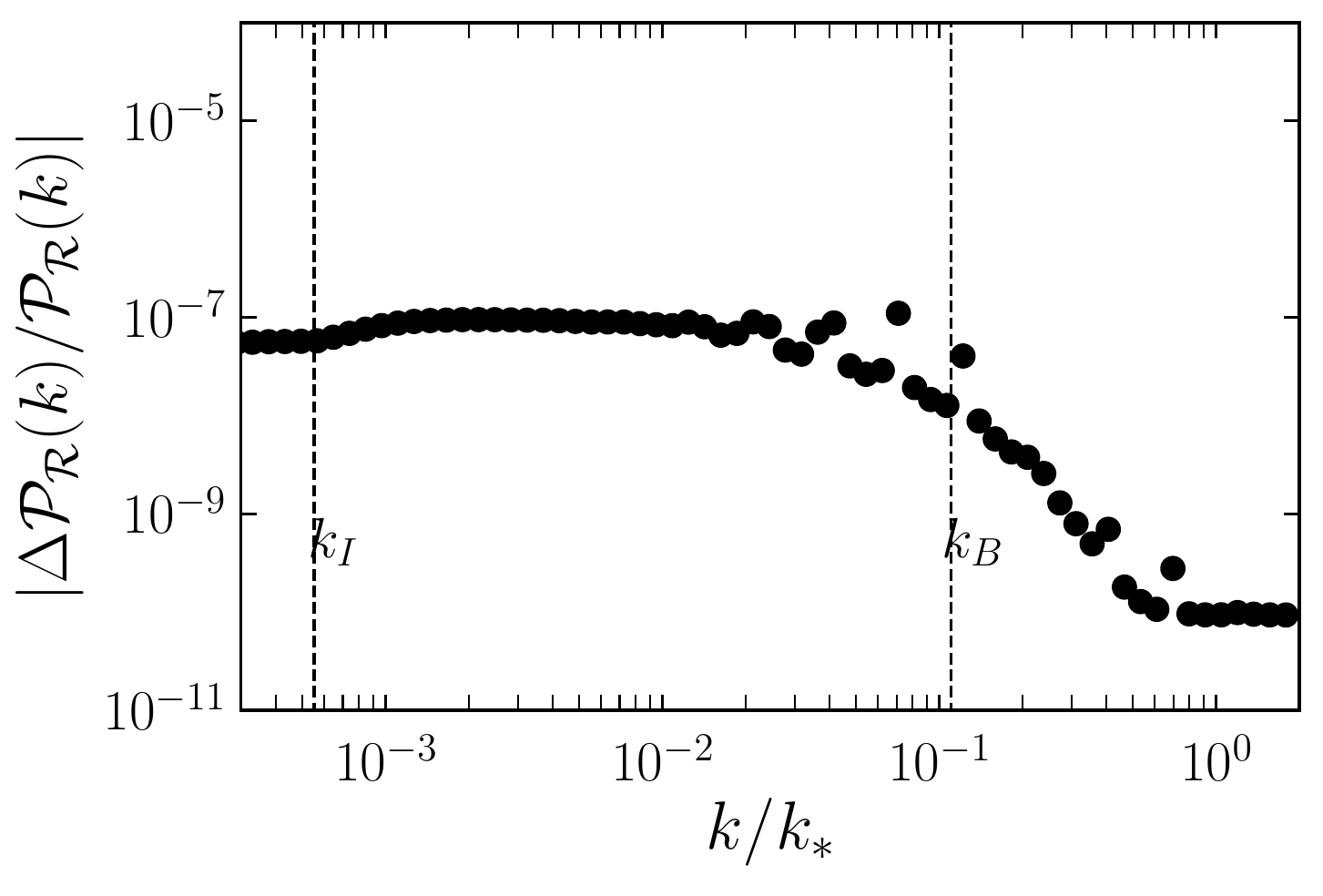}
\caption{Numerical estimate of the first-order correction to the power spectrum in the bouncing model (\ref{an}) with $n\, =\, 0.21$ and $R_B\, =\, 1 $ in Planck units. In obtaining this figure, we have worked with $\mathfrak{f}_{_{\rm NL}}\,=\,1$. The result  grows linearly with  $\mathfrak{f}_{_{\rm NL}}$.We have checked that the other values of $n$ and $R_B$ used in this paper all produce smaller values for this ratio.} \label{fig:vopt}
\efig

The second question we want to analyze concerns the contribution of higher order perturbative terms to the non-Gaussian modulation of the power spectrum. This is not captured by the previous calculation, since there we have computed the average value of $\Delta \mathcal{P}_{\mathcal{R}}(k_1)$ and not its ``variance'', which contains information about the expected  deviations  from the mean in typical realizations. The largest effect of the modulation occurs for the monopolar modulation. Indeed, we observe that the correction to the ``bare'' angular power spectrum in equation (\ref{Cobs}) is not small, and it is in fact a significant fraction of the final result---although the relative contribution is always smaller than one in all our calculations.  This means that the non-Gaussian modulation affects the observed power spectrum significantly. The order of magnitude of the correction can be understood by looking, for instance,  at the derivation in subsection \ref{modPsp}, and in particular to equation (\ref{ngmod}). This equation reveals that the relative contribution of the non-Gaussian modulation to individual realizations is of order $ \mathfrak{f}_{_{\rm NL}} \sqrt{\mathcal{P}_{\mathcal{R}}}$. For a primordial power spectrum whose average is of the order of $10^{-8}- 10^{-7}$, and $\mathfrak{f}_{_{\rm NL}}\sim 10^3$, this quantity is smaller than, but close to one.  We argue, however, that this does not signal any problem with the perturbative expansion.

The non-Gaussian modulation originates from second order perturbations, so the validity of the perturbative expansion must be evaluated by comparing this result with higher order contributions. But 
 higher orders introduce additional powers of $\mathcal{P}_{\mathcal{R}}\ll 1$. So the next-to-leading-order contribution to the non-Gaussian modulation is of order  $ \mathfrak{f}_{_{\rm NL}} (\mathcal{P}_{\mathcal{R}})^{3/2}$, which is sub-leading by virtue of  the smallness of $\mathcal{P}_{\mathcal{R}}$. Therefore, we see again that it is the fact that $\mathcal{P}_{\mathcal{R}}$ remains much smaller than one in our model that makes the results robust under the addition of higher perturbative corrections.

\section{Discussion and Conclusions}\label{conclusions}

The anomalies at large angular scales in the CMB have generated   a significant amount of  interest and new developments. 
Although it is possible that the observed signals are  the result of a statistical excursion of the $\Lambda$CDM model, the possibility that some of them originated from new physics is indeed fascinating. As emphasized by the  Planck collaboration in \cite{Akrami:2018vks}, it is worth exploring  new ideas, since having a theoretical model based on a few free parameters  would  allow the development  of novel probes on similar angular scales,  which could  increase the significance of existing signals.  In  this paper we have proposed a model able to relate multiple anomalies in a simple manner. Our ideas were inspired by studying the concrete theory of the early universe arising in loop quantum cosmology \cite{Agullo:2012sh,Agullo:2012fc,Agullo:2013ai,Agullo:2015tca,Agullo:2015aba,Agullo:2017eyh}, where  non-perturbative quantum gravity effects replace the big bang singularity by a  cosmic bounce. A detailed study revealed  that the mechanism generating the  large scale anomalies does not depend on the fundamental principles and peculiarities upon which this theory rests, and that this mechanism can also be materialized in other theories.  We have presented here a general phenomenological model that describes the minimum ingredients needed for the mechanism that accounts for the anomalies to work.  The key ingredient is  the existence of a cosmic bounce followed by a phase of slow-roll inflation. We have parameterized the bounce in a simple manner, by means of two numbers $n$ and $R_B$, and have studied under what conditions the model can collectively account for the anomalies discussed above. The main assumptions on which our model rests  are: (i) There was a cosmic bounce in the early universe followed by a phase of slow-roll inflation that started when the spacetime curvature was of the order $10^{-10}-10^{-11}$ in Planck units.  (ii) The new physics that causes the bounce loses relative relevance soon after it, and the matter content becomes dominated by a scalar field that is responsible for
inflation. (iii) The amount of expansion accumulated after the bounce is such that the wavenumber-scale characteristic of the bounce $k_B$ is red-shifted to scales that today are of the order of $k_*=0.002\, {\rm Mpc}^{-1}$. This is equivalent to requiring a number of $e$-folds of expansion between the bounce and the end of inflation $\sim 70$. This also implies that the potential energy of the scalar field is small relative to its kinetic energy at the time of the bounce. 
(iv) We have assumed that the form of the equations that describe comoving curvature perturbations is not drastically modified by new physics, except for the fact that the scale factor and the rest of background quantities describe a  bouncing universe. 
With these assumptions, and by  using general arguments about the  expected form of the non-Gaussianty in bouncing models with a phase of slow-roll inflation, we have concluded that, if the amplitude $ \mathfrak{f}_{_{\rm NL}}$ of the Bispectrum is of order $ \mathfrak{f}_{_{\rm NL}}\sim 10^3$, the large scale anomalies are expected features in the CMB.   They are traces left by the non-Gaussian correlations between the longest wavelengths we can observe and super-horizon modes. We find interesting that these imprints can account for a quite  diverse set of anomalies. A collective explanation is something  that has remained elusive so far.

It is important to emphasize that, as  explained in section \ref{sec:NGMod},  our model accounts  for the observed features in the sense that they are significantly more likely to be found than in the $\Lambda$CDM model, and consequently they should not be considered anomalous. In other words,  the anomalies in the CMB arise in our model as the result of a statistical excursion, but the required  excursion is much more probable  than in the  $\Lambda$CDM model. 

 We have found that the typical effects of a pre-inflationary bounce on the CMB can be described in the form of a modulation of the primordial power spectrum with angular multipolar dependence $L=0,1,2,\cdots$. The dominant multipole is $L=0$, and the amplitude of higher multipoles $L$ decreases rapidly with $L$. All modulations are scale dependent, and are large only for large angular scales in the CMB. The spherically symmetric monopolar modulation  produces a power suppression relative to the mean value in a large fraction of realizations. We have found that this power suppression induces other effects, such as a preference for odd parity correlations, and a decrease in the lensing amplitude  $A_L$. Furthermore, these effects come together with an anisotropic dipolar modulation, with amplitude and scale dependence  in consonance with observations. We have analyzed the details of  these effects for different types of bounces, that we have parameterized in a simple manner, and have contrasted the predictions with data, finding for some values of these parameters a remarkably good agreement with data, as measured by the $\chi^2$ parameter. 
In particular, we have found that models with larger value of $n$, namely $n=0.21$ and  $n=1/5$, fit  data better, and  produce a dipolar modulation and a parity asymmetry closer to the observed values.
 Future investigations will focus on studying whether this model can also account for other observed features, such as the details of the  quadrupolar modulation observed by Planck and discussed in section \ref{quadrupolar}. Another exciting possibility is that the dipolar asymmetry in our model can generate the alignment of the $\ell=2$ and $\ell=3$ multipoles observed in the CMB  \cite{deOliveira-Costa:2003utu,Schwarz:2004gk,Land:2005dq}. This alignment constitutes another anomaly in the CMB (see \cite{Schwarz:2015cma} for a pedagogical summary), and results in \cite{Chang:2013lxa} show that it can in fact be originated from  the dipolar asymmetry, since the later induces correlations between multipoles $\ell$ and $\ell+1$ that make the observed alignment significantly more probable than in $\Lambda$CDM.

It is  our view that materializing our ideas in a  general phenomenological model that contains just a few free parameters offers many benefits, particularly  in order to contrast with observations and in designing new ways of testing it. A phenomenological model may do not be so attractive to more mathematically minded cosmologists, due to the lack of fundamental ideas supporting it. But as  argued above, loop quantum cosmology provides a concrete example where our model  emerges from first principles \cite{Ashtekar:2006rx,Ashtekar:2006wn,Ashtekar:2011ni,Agullo:2016tjh,Agullo:2013dla,Agullo:2013ai,Agullo:2015tca,Agullo:2017eyh}.   Other examples include braneworld bouncing scenario discussed in \cite{Shtanov:2002mb} and 
the higher-derivative scalar-tensor  theory introduced in \cite{Chamseddine:2016uef}, which was proven in \cite{Liu:2017puc} to contain  bouncing solutions  that produce the similar FLRW spacetimes as found in loop quantum cosmology. These examples provide a proof of-concept for our ideas.

 Our  conclusions indicate  that it may be premature to  dismiss all the CMB anomalies as simple fluctuations of  a universe ruled by the $\Lambda$CDM model complemented with almost scale invariant primordial density perturbations. We rather propose that 
a simple modification of the form of the primordial  perturbations make the observed features compatible with the $\Lambda$CDM model, while respecting other constraints.  The almost scale invariant perturbations are normally accounted for by appealing to an early phase of inflation. Our  initial perturbations are rather justified by adding a cosmic bounce prior to the inflationary era. It is indeed a fascinating possibility that the observed anomalies may carry information about such a remote era. It is of interest to extend our results to include tensor modes, in order to find new ways of testing the ideas proposed here.

\section*{Acknowledgments}
We specially thank Boris Bolliet for many discussions, inputs, and initial collaboration in this project. We have  benefited from  discussions with A.  Ashtekar,  B.  Gupt, J. Olmedo, J. Pullin, and P. Singh. We  thank B. Gupt for assistance with Planck data. This work is supported by the NSF CAREER grant PHY-1552603, and from the Hearne Institute for Theoretical Physics. V.S. was supported by Louisiana State University and Inter-University Centre for Astronomy and Astrophysics during earlier stages of this work. This research was conducted with high performance computing resources provided by Louisiana State University (http://www.hpc.lsu.edu). This paper is based on observations obtained from Planck (http://www.esa.int/Planck), an ESA science mission with instruments and contributions directly funded by ESA Member States, NASA, and Canada.

\appendix

\section{Bipolar Spherical Harmonic coefficients}\label{BipoSH}

 The  bipolar spherical harmonics (BipoSH) are a convenient basis to characterize deviations from statistical isotropy in the CMB. They have been extensively used in the literature, in particular by the Planck collaboration to report the amplitude of the dipolar anomaly. In this section, we briefly summarize the definition of these coefficients.  See e.g.\ \cite{Hajian_2003, Joshi:2009mj} for additional details.  
 
 The two-point correlation function of  temperature anisotropies in the CMB, $\langle \delta T(\hat n_1) \delta T(\hat n_2)\rangle$, is a function of two directions, $\hat n_1$ and $\hat n_2$. A basis for functions of two directions is given by the familiar product of two spherical harmonics,  $Y_{\ell_1 m_1}(\hat n_1)Y_{\ell_2 m_2}(\hat n_2)$; the coefficients of the expansion of the temperature two-point function in this basis are   the elements of the covariance matrix  $\langle a_{\ell_1 m_1}a_{\ell_2 m_2}\rangle$:
\be \langle \delta T(\hat n_1) \delta T(\hat n_2)\rangle=\sum_{\ell_1\ell_2}\sum_{m_1m_2} \langle a_{\ell_1 m_1}a_{\ell_2 m_2}\rangle \, Y_{\ell_1 m_1}(\hat n_1)Y_{\ell_2 m_2}(\hat n_2) \, . \ee
The inverse transformation reads
\be \langle a_{\ell_1 m_1}a_{\ell_2 m_2}\rangle=\int d\Omega_1 d\Omega_2  \, \langle \delta T(\hat n_1) \delta T(\hat n_2)\rangle\, Y^*_{\ell_1 m_1}(\hat n_1)Y^*_{\ell_2 m_2}(\hat n_2) \, .\ee 
The  BipoSH, commonly denoted by $\{Y_{\ell_1}(\hat n_1)\otimes Y_{\ell_2}(\hat n_2)\}_{LM}$, is  another basis of functions of two directions. It is different, but closely related to the product of two spherical harmonics. They are related by
\be \label{change} \{Y_{\ell_1}(\hat n_1)\otimes Y_{\ell_2 }(\hat n_2)\}_{LM}\equiv \sum_{m_1,m_2}\, C^{LM}_{\ell_1,m_1,\ell_2,m_2}\, Y_{\ell_1 m_1}(\hat n_1)Y_{\ell_2 m_2}(\hat n_2)\, ,\ee
where $C^{LM}_{\ell_1,m_1,\ell_2,m_2}$ are  Clebsch-Gordan coefficients (recall that the Clebsch-Gordan coefficients  are zero unless $\ell_1+\ell_2\geq L\geq|\ell_1-\ell_2|$, $M=m_1+m_2$).  The BipoSH can be heuristically understood as the decomposition of the product of two spherical harmonics in functions with ``well-defined total angular momentum''. Note that the product of two spherical harmonics is labeled by four numbers, namely $\ell_1,\ell_2,m_1,m_2$. The BipoSH are also labeled by four numbers, but they are instead $\ell_1,\ell_2,L,M$. See \cite{Joshi:2009mj}  for a list of properties of the BipoSH. 

The expansion of  $\langle \delta T(\hat n_1) \delta T(\hat n_2)\rangle$ in the BipoSH's, 
\be \langle \delta T(\hat n_1) \delta T(\hat n_2)\rangle=\sum_{\ell_1\ell_2}\sum_{LM}\, A^{LM}_{\ell_1\ell_2} \ \{Y_{\ell_1}(\hat n_1)\otimes Y_{\ell_2 }(\hat n_2)\}_{LM} \, . \ee
define the BipoSH coefficients, denoted by $A^{LM}_{\ell_1\ell_2}$. The inverse transformation reads
\be A^{LM}_{\ell_1\ell_2}=\int d\Omega_1 d\Omega_2  \, \langle \delta T(\hat n_1) \delta T(\hat n_2)\rangle\ \{Y_{\ell_1}(\hat n_1)\otimes Y_{\ell_2 }(\hat n_2)\}^*_{LM} \, .\ee 
It should be obvious from the expressions above that the BipoSH coefficients $A^{LM}_{\ell_1\ell_2}$ and  $ \langle a_{\ell_1 m_1}a_{\ell_2 m_2}\rangle$ are related by means of the Clebsch-Gordan coefficients. It is a simple exercise to show that 
\be  \langle a_{\ell_1 m_1}a^{\star}_{\ell_2 m_2}\rangle=(-1)^{m_2}\, \sum_{LM}\, A_{\ell_1\ell_2}^{LM}\, C^{LM}_{\ell_1,m_1,\ell_2,-m_2}\, .\ee
Similarly, the inverse relation is 
\be \label{atoBipoSH}  A_{\ell_1\ell_2}^{LM}=\sum_{m_1,m_2}\ \langle a_{\ell_1 m_1}a^{\star}_{\ell_2 m_2}\rangle\, \, (-1)^{m_2}\, C^{LM}_{\ell_1,m_1,\ell_2,-m_2}\, .\ee

We end this summary by considering two simple examples that will help us to better understand the  information encoded in the BipoSH coefficient. Let us start by considering a statistically homogenous and isotropic CMB, for which   $\langle a_{\ell_1 m_1}a^{\star}_{\ell_2 m_2}\rangle=\, C_{\ell_1}\, \delta_{\ell_1,\ell_2}\, \delta_{m_1,m_2} $, where $C_{\ell_1}$ is the standard angular power spectrum. 
Then, using equation (\ref{atoBipoSH}) we obtain
%
\be A_{\ell_1\ell_2}^{LM}=(-1)^{\ell_1}\, \sqrt{2\ell_1+1}\, C_{\ell_1}\, \delta_{L,0}\, \delta_{M,0}\, \delta_{\ell_1,\ell_2}\, , \ee
where we have used the following property of the Clebsh-Gordan coefficients, $\sum_b (-1)^{a-b}\, C^{c0}_{aba-b}=\sqrt{2a+1}\, \delta_{c0} $. 
Therefore, for an isotropic CMB the only non-zero BipoSH coefficients are $A_{\ell\ell}^{00}$, and  the angular power spectrum is given by $C_{\ell}=(-1)^{\ell}\, A_{\ell\ell}^{00}/{\sqrt{2\ell+1}}$. In other words, all the information about the  isotropic part of the power spectrum is encoded in $A_{\ell\ell}^{00}$. The advantage of the BipoSH coefficients is that a non zero value of any  BipoSH coefficient for $L>0$ unambiguously indicates a departure from statistical isotropy.

As a second example, let us consider a temperature  distribution in the CMB $\delta T(\hat n)$ which is statistically isotropic except for a modulation with angular distribution given by a combination of  spherical harmonics:
\be \label{isomod}  \delta T(\hat n)=\delta T^{\rm iso}(\hat n) \left(1+\sum_{LM} g_{LM}\,Y_{LM}(\hat n)\right )\, \ee
where $\delta T^{\rm iso}(\hat n)$  indicates the statistically isotropic part and $g_{LM}$  the amplitude of the modulation. For $L=1$ we have a dipolar modulation, for $L=2$ quadrupolar, etc. The BipoSH coefficients for this example turn out to be
\be A^{LM}_{\ell \ell'}=g_{LM}\, (C_{\ell}+C_{\ell'})\,  \sqrt{\frac{(2\ell+1)(2\ell'+1)}{4\pi\, (2L+1)}}\,  C^{L 0}_{\ell 0 \ell' 0}\ \, . \ee
As one could expect,  they are proportional to the amplitude of the modulation $g_{LM}$. This simple example shows that BipoSH coefficients are a sharp tool to look for deviations of statistical isotropy in the CMB, and to characterize their angular distribution. Note also that in this example all the dependence in $\ell$ and $\ell'$ of the BipoSH coefficients $A^{LM}_{\ell \ell'}$ are in the so-called form factors $(C_{\ell}+C_{\ell'})\,  \sqrt{\frac{(2\ell+1)(2\ell'+1)}{4\pi\, (2L+1)}}\,  C^{L 0}_{\ell 0 \ell' 0}$, i.e., if one factors out these form factors, the remaining amplitudes  $g_{LM}$ are independent of $\ell$. One then says that the modulation is scale-independent.  
 Interestingly enough, the anomalies that have been observed in the CMB only appear for low $\ell$'s, and therefore are {\em scale-dependent}. We then  conclude that the simple model  (\ref{isomod})  is insufficient to describe the observed anomalies. We need a more sophisticated model that can accommodate scale-dependence.  This is indeed  the situation for the non-Gaussian modulation discussed in this paper.
 
 \begin{figure}[htp]
\centering
\includegraphics[width=.49\textwidth]{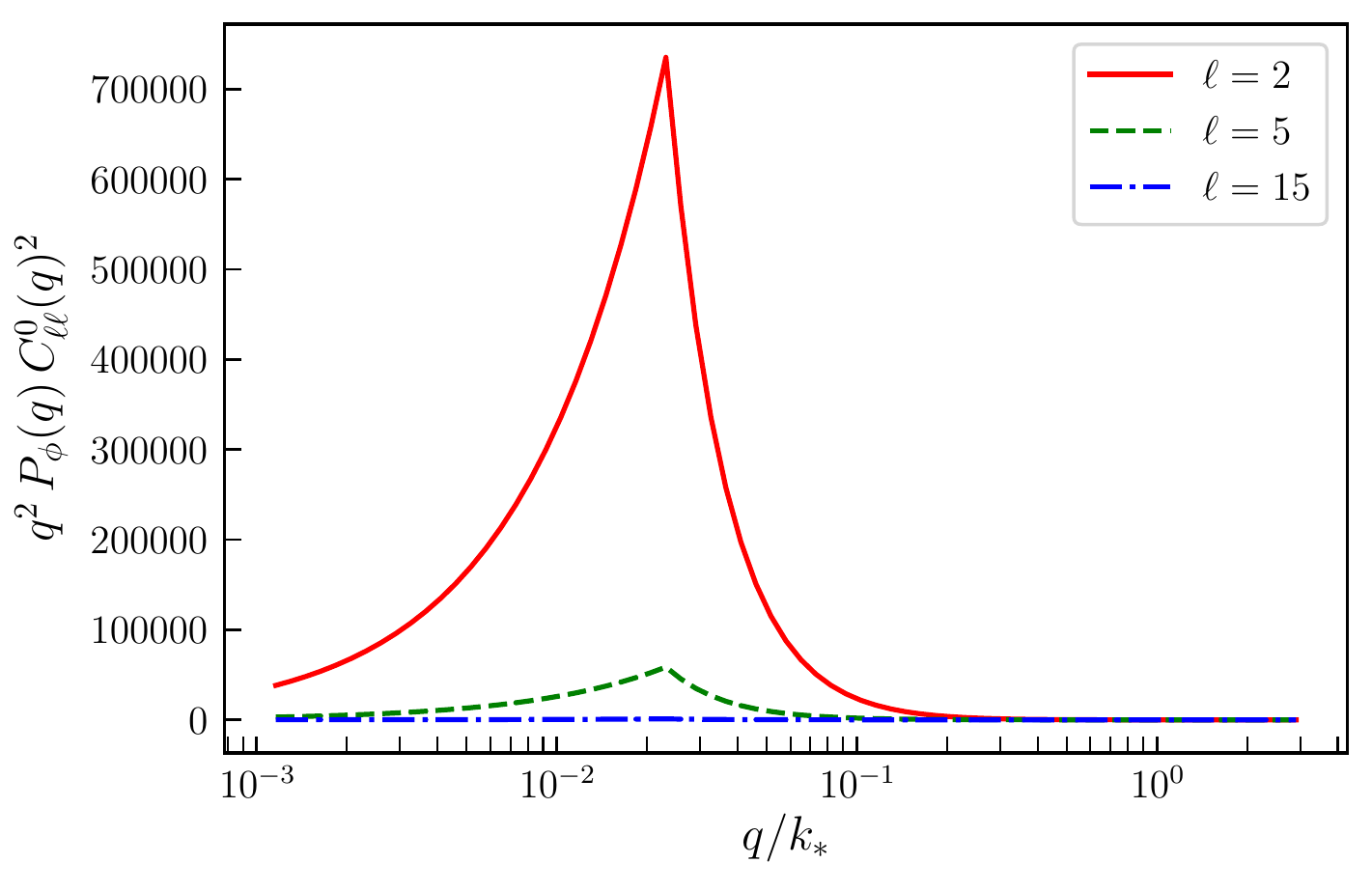}\quad
\includegraphics[width=.45\textwidth]{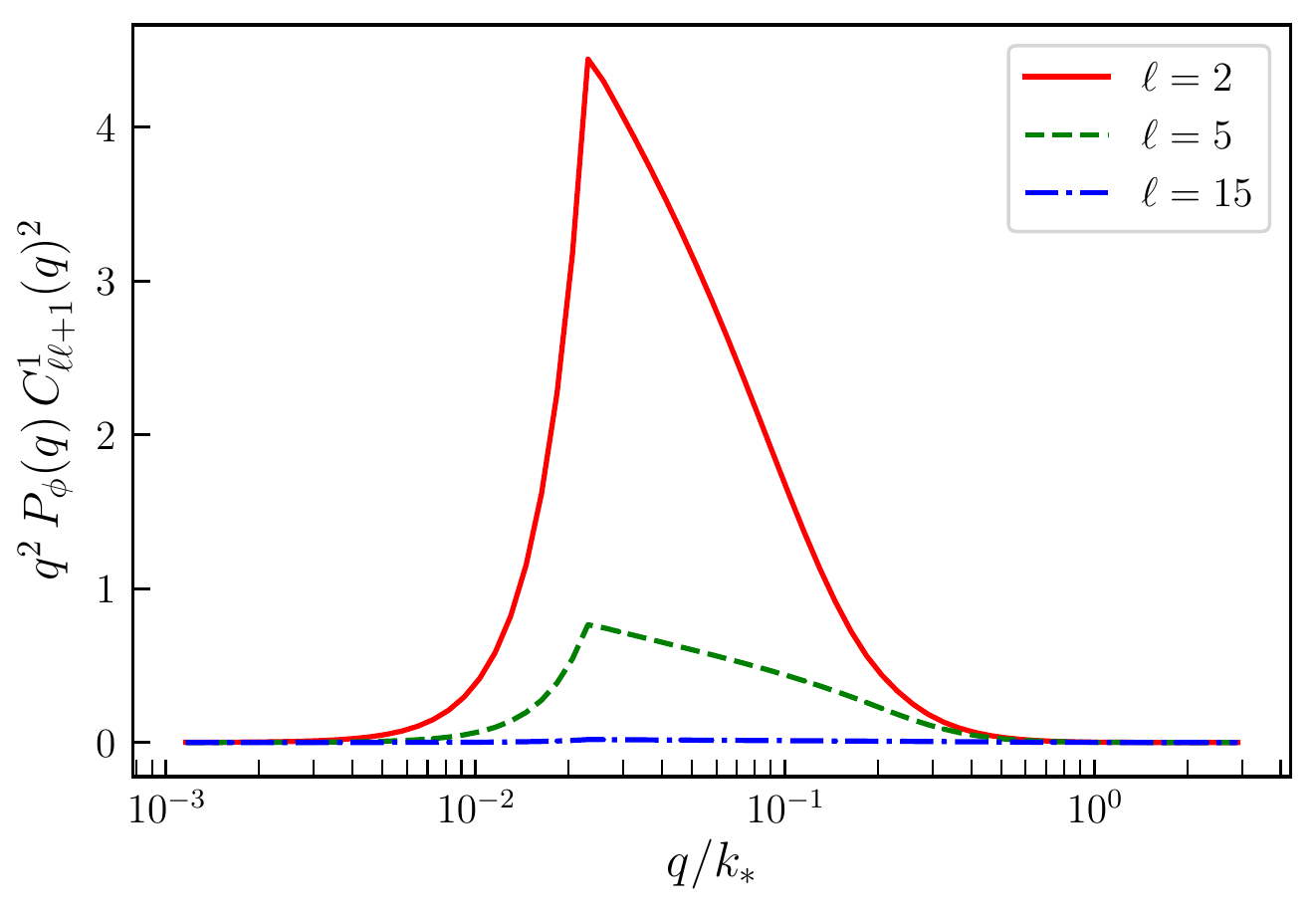}
\medskip
\includegraphics[width=.5\textwidth]{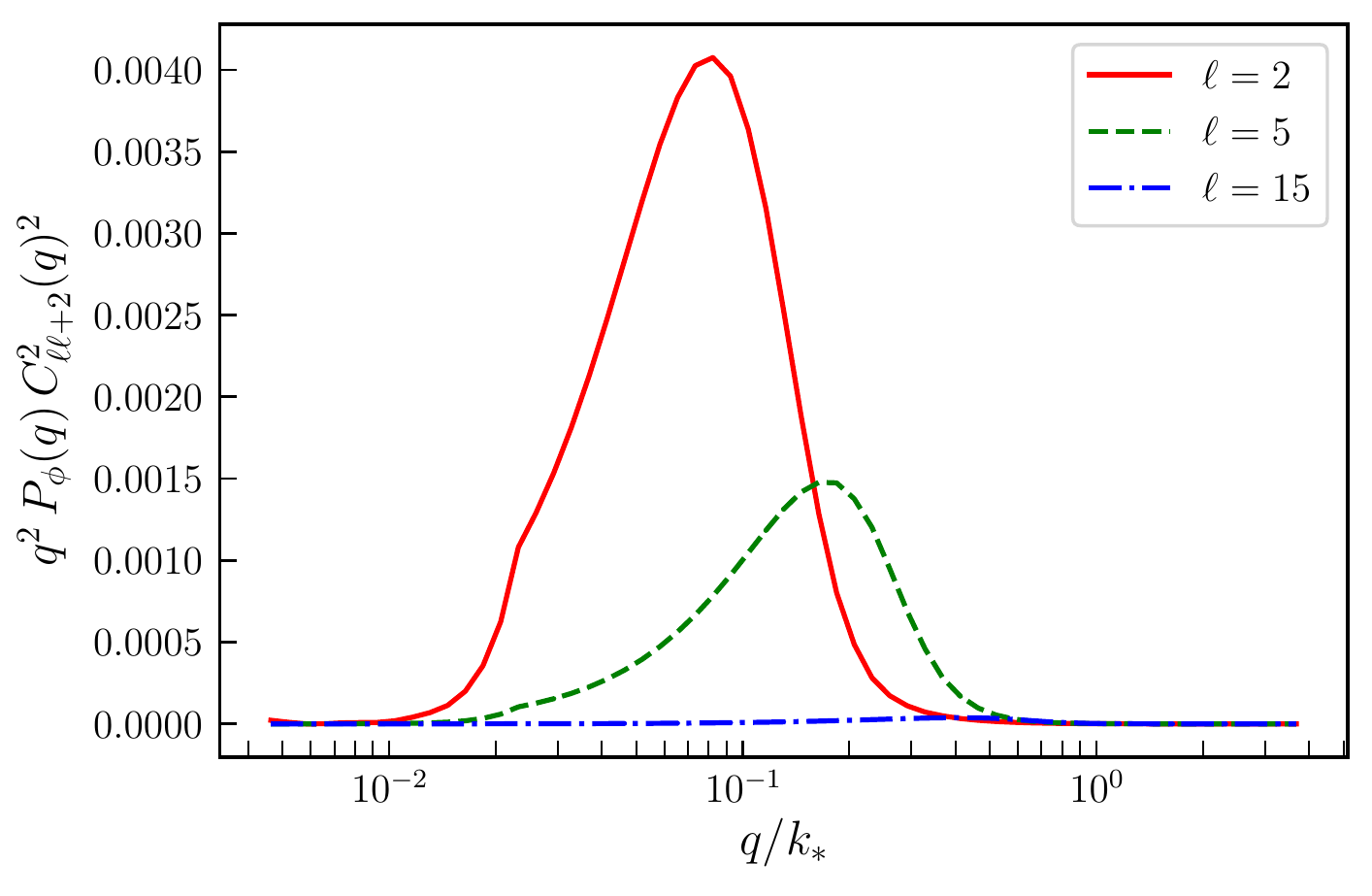}
\caption{Integrand in Eqn.\ (\ref{varalpha}),  i.e $q^2\,P_\phi(q)\,|\mathcal{C}^L_{\ell\,\ell'}|^2$ versus $q/k_*$, for $L = 0$ (upper left), $L = 1$ (upper right) and $L = 2$ (bottom panel). }
\label{integrand}
\end{figure}

 \section{Check of the validity of the approximation used in section \ref{covmaxmod}}\label{approx}
 This appendix provides a consistency check for the assumption made in section \ref{covmaxmod}, under which CMB modes $\vec k_1$ are more strongly correlated with modes $\vec q$ with norms $q$ significantly smaller than $k_1$. The more direct way of checking whether this assumption is satisfied in our model is by plotting  the integrand in equation (\ref{varalpha}), in order to see what values of $q$ contribute the most to the amplitudes of the non-Gaussian modulation. We plot in Figure \ref{integrand} these integrands for $L=0,1$ and $2$, and for three representative values of $\ell$, namely $\ell=2,5$ and $15$. We see first that the integrand decreases when we increase either $L$ or $\ell$. This shows, on the one hand, that there is a hierarchy in the amplitudes of the modulation, being the largest  for the monopolar one ($L=0$), then the dipolar ($L=1$), etc. And on the other hand, that the amplitude of the the modulation is larger for small values of $\ell$ (i.e.\ large angular scales). 
 
 Furthermore, these figures also show that the values of $q$ contributing the most correspond to Fourier modes with wavelengths a bit larger than our Hubble radius today. The mode whose wavelength equals the Hubble radius today is $k_{\rm min}\approx k_*/8.9\approx 0.1\, k_*$, and Figure \ref{integrand} indicates that the dominant  contribution to the modulation comes from $q\lesssim k_{\rm min}$, hence justifying our approximation. We note, however, that the approximation is better satisfied for the monopolar and dipolar modulations.

\section{Some details about the calculation of the primordial non-Gaussianity}\label{pertsham}

This appendix provides further details about the calculations summarized in section \ref{bispectrum}, and illustrates them by using the Hamiltonian for scalar perturbations derived from general relativity (see \cite{Agullo:2017eyh} for additional details). We work in the Hamiltonian or canonical formalism, where the phase space of the homogeneous degrees of freedom is four dimensional $\Gamma_{FLRW}=\{a,\varphi,\pi_a,p_{\varphi}\}$, where $\pi_a$ and $p_{\varphi}$ are the  momenta conjugate to $a$ and $\varphi$, respectively. For perturbations, we work in the spatially flat gauge, and describe scalar perturbations by the inflaton field perturbations $\delta \varphi(\v x)$ and the conjugate momentum $\delta p_{\varphi}(\v x)$. We don't consider tensor modes in this appendix. The relation of $\varphi$ and the Bardeen potential $\Phi$ used in the main body of the paper is 

\be \label{zetadph} \Phi(\v x)=\f{3}{5}\f{a}{z} \, \dph(\v x)-\f{3}{5}\left[-\f{3}{2}+3\f{V_{\varphi}\, a^5}{\kappa\, \pp\, \pi_a}+\f{\kappa}{4}\f{z^2}{a^2}\right]  \left(\f{a}{z} \, \dph(\v x)\right)^2\, \cdots \ee
where $z=-\f{6}{\kappa}\f{p_{\varphi}}{\pi_a}$, and the dots indicate terms proportional  to spatial and time derivatives of $\delta \varphi$; they produce sub-leading contributions to observable quantities  when evaluated at the end of inflation. From this we have (now in Fourier space)
\bea \label{3pz}
& & \langle 0|\h{\Phi}_{{\vec k}_1} \h{\Phi}_{{\vec k}_2} \h{\Phi}_{{\vec k}_3}|0\rangle=\left(\f{3}{5}\f{a}{z}\right)^3  \langle 0|\h{\dph}_{{\vec k}_1} \h{\dph}_{{\vec k}_2} \h{\dph}_{{\vec k}_3}|0\rangle\nonumber \\ &-& \left(\f{3}{5}\right)^3 \left(-\f{3}{2}+3\f{V_{\phi}\, a^5}{\kappa\, \pp\, \pi_a}+\f{{\kappa}}{4}\f{z^2}{a^2}\right)\, \left(-\f{a}{z}\right)^4\, \Big[\int \f{d^3p}{(2\pi)^3} \, \langle 0|\h{\dph}_{{\vec k}_1} \h{\dph}_{{\vec k}_2}  \h{\dph}_{{\vec p}}\,  \h{\dph}_{{\vec k}_3-\v p}|0\rangle + (\v k_1 \leftrightarrow \v k_3)+ (\v k_2 \leftrightarrow \v k_3) \, \nonumber \\
&+&\cdots\Big] \, ,  \eea
where $(\v k_i \leftrightarrow \v k_j)$ denotes terms obtained after interchanging $\v k_i$ and $\v k_j$. Dots  indicate higher order contributions. The four-point functions in the second line can be computed by expanding $\h{\dph}^{\rm I}_{{\vec k}}$ in creation and annihilation operators 

\be \h{\dph}^{\rm I}_{{\vec k}}=  \left(\hat A_{\vec k} \, \varphi_k(\eta) + \hat A^\dagger_{-\vec k}\, \varphi_{k}^*(\eta)\right) \, , \ee
where the Fourier modes $\varphi_k(\eta)$ are related to the variable $v_k$ used in section \ref{powspec} by $\varphi_k(\eta)=v_k/a$. From this equation we obtain
\be \label{4p2} \int \f{\d^3p}{(2\pi)^3} \, \langle 0|\h{\dph}_{{\vec k}_1} \h{\dph}_{{\vec k}_2}  \h{\dph}_{{\vec p}}\,  \h{\dph}_{{\vec k}_3-\v p}|0\rangle = (2\pi)^3 \delta^{(3)}(\v{k}_1+\v{k}_2+\v{k}_3)  \,2 \,  |\varphi_{k_1}|^2 |\varphi_{k_2}|^2\,  \, .\ee

On the other hand, the leading order contribution to the three-point function in (\ref{3pz}) requires knowledge about the Hamiltonian describing self-interaction of scalar perturbations $\delta \varphi$. To illustrate the calculation in a concrete scenario, we will use in this appendix the Hamiltonian for $\delta\varphi$ as predicted by general relativity.   
 At next to leading order in perturbations, it is given by \cite{Agullo:2017eyh}
\be \mathcal{H}_{\delta \phi}=\mathcal{H}^{(2)}+\mathcal{H}_{\rm Int}+\mathcal{O}(\delta\varphi^4)\, \ee
where 
\begin{eqnarray} \label{hams}
 \mathcal{H}^{(2)}\,=\, N\f{1}{2}\,\int \d^3 x \,  \biggl[\, \f{1}{\,a^3}\, \dpp^2\, +\, a^3\, (\v \partial \delta\varphi)^2\, 
 +\,a^{3}\,  \u\, \dph^2\biggr]\, ,
\end{eqnarray}
where $\u$ was written below equation (\ref{chieq}), and
\begin{eqnarray}\label{eq:H3}
 \mathcal{H}^{(3)}\, 
 &=&
 \, N\,\int\, \d^3  x\, \biggl[ 
\left(  \frac{9\,\kappa\,p_{\varphi}^3}{4\,a^4\,\pi_a}-\f{27 \, \pp^5}{2\, a^6\pi_a^3} -\, \frac{3\,a^2\,p_{\varphi}\,V_{\varphi\varphi}}{2\,\pi_a}\, 
 +\frac{a^3\,V_{\varphi\varphi\varphi}}{6} \right) \,\delta\varphi^3\, \nonumber \\
 &&-\, \frac{3\,p_{\varphi}}{2\,a^4\,\pi_a}\,\delta p_{\varphi}^2\, \delta\varphi\,-\f{9 \, \pp^3}{ a^5 \pi_a^2} \, \dpp\dph^2
-\, \frac{3\,a^2\, p_{\varphi}}{2\, \pi_a}\delta\varphi\, (\vec{\partial}\delta\varphi)^2 +\, \frac{3\,p_{\varphi}^2}{N\, a\,\pi_a}\,\delta\varphi^2 \partial^2\chi\, +\, \frac{3}{2}\f{a^2\,p_{\varphi}}{N^2\,\kappa\,\pi_a}\,\delta\varphi\,\partial^2\chi\,\partial^2\chi \nonumber \\
 &&+\,3\, \f{\pp^2}{N\, a\,\pi_a}\, \delta\varphi\, \partial^i\chi\partial_i\delta \varphi +\f{1}{N}\, \delta p_{\varphi}\,\partial_i \delta\varphi\, \partial^i\chi\,
 -\, \frac{3}{2}\f{a^2\,p_{\varphi}}{N^2\,\kappa\,\pi_a}\, \delta\varphi\, \partial_i\partial_j\chi\, \partial^i\partial^j\chi\,
 \biggr] \, ,
\end{eqnarray}
where $N$ is the lapse function that specifies the time variable one is using, a sub-index $\varphi$ in the potential $V(\varphi)$ indicates derivative with respect to $\varphi$, and 
\be \chi=\,N\,  \f{ \sqrt{9}\,\kappa}{a^3}\, \partial^{-2} \biggl[ \biggl(\,\f{\pp}{2} -\, \f{a^5\, V_{\varphi}}{\kappa\,\pi_a} \biggr)\delta {\varphi}\, -\, \f{\pp}{\kappa\, a\, \pi_a}\delta p_{\varphi}\, \biggr]\ . \ee
By performing a Legendre transformation, it can be checked that these expressions agree with the third-order Lagrangian derived in \cite{Maldacena:2002vr}. From this Hamiltonian we obtain

\bea \label{bispdph} & &\langle 0|\h{\dph}_{{\vec k}_1}(\eta) \h{\dph}_{{\vec k}_2}(\eta) \h{\dph}_{{\vec k}_3}(\eta)|0\rangle =(2\pi)^3\delta^{(3)}( \vec{k}_1+\vec{k}_2+\vec{k}_3)\, B_{\delta\varphi}(k_1,k_2,k_3) \, ,\eea
where
\bea \label{Bphi} && B_{\delta\varphi}(k_1,k_2,k_3)=2\,  \, {\rm Im} \Big[\varphi_{\vec{k}_1}( \eta)\varphi_{\vec{k}_2}( \eta)\varphi_{\vec{k}_3}( \eta)\nonumber \\
&\times& \int_{ \eta_0}^{\eta} \d\eta'\, \Big( f_1(\eta')\, \varphi^{\star}_{{k}_1}( \eta')\varphi^{\star}_{{k}_2}(\eta')\varphi^{\star}_{{k}_3}( \eta') 
+f_2( \eta')\, \varphi^{\star}_{{k}_1}( \eta')\varphi^{\star}_{{k}_2}( \eta'){\varphi'}_{{k}_3}^{\star}(\eta') 
+ f_3( \eta')\, \varphi_{{k}_1}^{\star}(\eta'){\varphi'}_{{k}_2}^{\star}( \eta'){\varphi'}_{{k}_3}^{\star}( \eta') \nonumber \\
&+&(\v k_1 \leftrightarrow \v k_3)+ (\v k_2 \leftrightarrow \v k_3) \Big)
 \Big ]+\mathcal{O}(\mathcal H^2_{\rm int}) \, ,\eea
where the functions $f_1( \eta)$, $f_2(\eta)$ and $f_3(\eta)$ are combinations of background functions  and wavenumbers, given by

\bea
f_1(\eta)\, &=&\, a\,\biggl[\, 2\, \biggl(\, \f{243\,\pp^7}{2\,\kappa\,a^8\,\pi_a^5}\, 
-\, \f{81\,\pp^5}{2\,a^6\,\pi_a^3}\, +\, \f{27\,\kappa\,\pp^3}{8\,a^4\,\pi_a}\, 
+\, \f{81\,\pp^4\,V_{\varphi}}{\kappa\,a\,\pi_a^4} - \f{27\,a\,\pp^2\,V_{\varphi}}{2\,\pi_a^2}\,+\, \f{27\,a^6\,\pp\,V_{\varphi}^2}{2\,\kappa\,\pi_a^3}\biggr)\,\nonumber\\
&&\, \times\biggl(\, 1\, -\ \f{(\v k_1 \cdot \v k_2)^2}{k_1^2\,k_2^2}\,\biggr)\, 
+\, \f{3\,a^2\,\pp}{\pi_a}\,\v k_1 \cdot \v k_2\,+\, \f{9\,a\,\pp^2\,V_{\varphi}}{\pi_a^2}\, -\, 
\f{3\,a^2\,\pp\,V_{\varphi\varphi}}{\pi_a}\, +\, \f{a^3\,V_{\varphi\varphi\varphi}}{3}\,\biggr]\,, \\
f_2(\eta)\, &=&\, a^3\,\biggl[\, \biggl(\, \f{81\,\pp^5}{\kappa\,a^7\,\pi_a^4}\, -\, \f{27\,\pp^3}{2\,a^5\,\pi_a^2}\, 
+\,\f{27\,\pp^2\,V_{\varphi}}{\kappa\,\pi_a^3}\,\biggr)\biggl(\, 2\, -\, \f{(\v k_1 \cdot \v k_3)^2}{k_1^2\,k_3^2}\, 
-\, \f{(\v k_2 \cdot \v k_3)^2}{k_2^2\,k_3^2}\,\biggr)\, -\, \f{9\,\pp^3}{a^5\,\pi_a^2}\, \nonumber\\
&&\, +\, \biggl(\, \f{-3\, \kappa\,\pp}{2\,a^3}\, +\, \f{9\,\pp^3}{a^5\,\pi_a^2}\, +\, \f{3\,a^2\,V_{\varphi}}{\pi_a}\,\biggr)\,
\biggl(\, \f{\v k_1\cdot \v k_2}{k_1^2}\, +\, \f{\v k_1\cdot \v k_2}{k_2^2}\,\biggr)\,\biggr]\, ,\\
f_3(\eta)\, &=&\, a^5\,\biggl[\, \f{27\,\pp^3}{\kappa\,a^6\,\pi_a^3}\,\biggl(\,1\, -\, \f{(\v k_2\cdot \v k_3)^2}{k_2^2\,k_3^2}\,\biggr)\, 
-\, \f{3\,\pp}{a^4\,\pi_a}\, +\, \f{3\,\pp}{a^4\,\pi_a}\,\biggl(\, \f{\v k_1\cdot\v k_3}{k_3^2}\, +\, \f{\v k_1 \cdot \v k_2}{k_2^2}\biggr)\, \biggr]\, .
\eea

Putting everything together, the three-point function of the Bardeen potential is 
\be \langle 0|\h{\Phi}_{{\vec k}_1} \h{\Phi}_{{\vec k}_2} \h{\Phi}_{{\vec k}_3}|0\rangle=(2\pi)^3\delta^{(3)}( \vec{k}_1+\vec{k}_2+\vec{k}_3)\, B_{\Phi}(k_1,k_2,k_3) \, ,\ee
where the primordial bispectrum is 
\bea  \label{BR} B_{\Phi}(k_1,k_2,k_3)&=&\left(\f{3}{5}\f{a}{z}\right)^3\, B_{\delta\varphi}(k_1,k_2,k_3) \\ &-&  {\left(\f{3}{5}\right)^3}
\left[-\f{3}{2}+3\f{V_{\varphi}\, a^2}{\kappa\, \pp\, \pi_a}+\f{\sqrt{\kappa}}{4}\f{z^2}{a^2}\right]  \left(\f{a}{z}\right)^4\, 
2\,   \big(|\varphi_{k_1}|^2 |\varphi_{k_2}|^2+|\varphi_{k_1}|^2 |\varphi_{k_2}|^2+|\varphi_{k_2}|^2 |\varphi_{k_3}|^2\big) \nonumber \, , \eea
with all quantities  evaluated towards the end of inflation. Note that the second line of this expression does not contain any integral in time, and it is proportional to the slow-roll parameters. In the standard inflationary paradigm, the first term in this equation is of the same order as the second one. However, in the model studied in this paper, where a cosmic bounce takes place before inflation, the first term provides the leading order contribution, since it is this term that carries information about the pre-inflationary evolution of the perturbations.

 \bibliography{Refs}

\end{document}